\documentclass[]{aa}
\usepackage{graphicx}
\usepackage{txfonts}

\begin{document}

\title{Modelling the spectral energy distribution of ULIRGs II: The energetic environment and the dense interstellar
medium\thanks{Data and SED models of Fig. 1 are available in
electronic form. See also Tables B.1, B.2 and B.3. }}

\author{O. Vega\inst{1,2}
       \and
       M.S. Clemens\inst{2}
       \and
       A. Bressan\inst{1,2,3}
       \and
       G.L. Granato\inst{2}
       \and
       L. Silva\inst{4}
       \and
       P. Panuzzo\inst{2,5}}

\offprints{O. Vega, \email{olga.vega@oapd.inaf.it}}

\institute{ INAOE, Luis Enrique Erro 1, 72840 Tonantzintla, Puebla,
Mexico
           \and
            INAF, Osservatorio Astronomico di Padova, Vicolo dell'Osservatorio,
            5, 35122 Padova, Italy
           \and
            SISSA, Strada Costiera, I-34131 Trieste, Italy
           \and
           INAF, Osservatorio Astronomico di Trieste, Via Tiepolo 11, I-34131 Trieste, Italy
           \and
           Laboratoire AIM, CEA/DSM - CNRS - Universit\'{e} Paris Diderot, DAPNIA/Service
d'Astrophysique, B\^{a}t. 709, CEA-Saclay, F-91191 Gif-sur-Yvette
C\'{e}dex, France}

\date{Received ????; accepted ????}

\abstract {} {By using the spectral energy distribution (SED) from
the near-infrared to the radio of a statistically significant number
of luminous infrared galaxies we determine important physical
parameters for this population of objects. In particular we
constrain the optical depth towards the luminosity source, the star
formation rate, the star formation efficiency and the AGN fraction.}
{We fit the near-infrared to radio spectral energy distributions of
a sample of 30 luminous and ultra-luminous infrared galaxies with
pure starburst models or models that include  both  starburst and
AGN components.} {We find that although about half of our sample
have best-fit models that include an AGN component, only 30\% (9/30)
have an AGN which accounts for more than 10\% of the infrared
luminosity from 8 to $1000\;\rm \mu m$, whereas all have an
energetically dominant  starburst. Our derived AGN fractions are
generally in good agreement other measurements based in the
mid-infrared line ratios, Ne[V]/Ne[II]
 and O[IV]/Ne[II] measured by \emph{Spitzer} IRS, but much lower than
 those derived from PAH equivalent widths or
the mid-infrared spectral slope. Our models determine the mass of
dense molecular gas within which active star formation takes place
via the extinction required to reproduce the infrared part of the
SED. Assuming that this mass is that traced by the HCN molecule, we
reproduce the observed linear relation between HCN flux and infrared
luminosity found by Gao \& Solomon, 2004a. We also find that the
star formation efficiency, as defined by the current star formation
rate per unit molecular gas mass, falls as the starburst ages.} {If
the evolution of ULIRGs includes a phase in which an AGN contributes
an important fraction to the infrared luminosity, this phase should
last an order of magnitude less time than the starburst phase.
However, we find no convincing evidence that an energetically
important AGN is associated with a particular phase of the
starburst. Because the mass of dense molecular gas which we derive
is consistent with observations of the HCN molecule, it should be
possible to estimate the mass of dense, star-forming molecular gas
in such objects when molecular line data are not available.}


\maketitle

\section{Introduction}

With total infrared luminosities between $10^{11}- 10^{12}$
L$_\odot$ and $\geq 10^{12}$ L$_\odot$, respectively, Luminous and
Ultraluminous Infrared galaxies, (U)LIRGs, are the most luminous
objects in the local universe. Although scarce at low redshift, they
may account for the bulk of all star formation activity at $z > 2 -
3$ and dominate the far-infrared background (e.g. Blain et al.
2002).

Many of them are found in merging systems (e.g. Sanders et al. 1988),
suggesting  that dynamical interaction has
driven gas towards the nucleus, fueling a massive
starburst (SB) and/or the central massive black hole (Mihos
\& Hernquist 1996).

Despite extensive investigation over the last decades, there is
still considerable uncertainty as to the nature of their power
source.

The observation that ULIRGs obey the same FIR-radio correlation of
normal star forming galaxies (e.g. Sopp, Alexander \& Riley, 1990)
and display a tight correlation between dense molecular gas mass and
far-infrared luminosity (Gao \& Solomon, 2004a), constitute strong
evidence that these sources are powered predominantly by star
formation. However, ULIRGs may commonly host an AGN and the evidence
that AGN and starburst luminosities are correlated over a wide range
of IR luminosities (e.g. Farrah et al. 2003) render the above issue
more intriguing.

Many studies have been devoted to the quantification of the relative
contributions of AGN and starburst (SB), but they are hindered by
the large and uncertain extinction that, usually, is not negligible
even at NIR wavelengths (Goldader et al. 1995, Silva et al. 1998,
Murphy et al. 2001, Vald\'es et al. 2005).

In order to analyze the power mechanism in these sources, tracers
that do not suffer large extinction have been recently preferred,
such as hard X-rays, and MIR to radio diagnostics. Rigopoulou et al.
(1996) argue that the weakness of the hard X-ray luminosity seen in
most ULIRGs, reveals a real lack of AGN activity, because the gas
column densities required to absorb its emission exceed those
estimated from CO data. Similar conclusions have been found by
Risaliti et al. (2006) who concluded that energetically important
AGN are present in only 20\% of their ULIRG sample.

After the advent of ISO and \emph{Spitzer}, MIR diagnostics such as
the strengths of PAH emission and the 9.7${\mu}$m absorption
features, the high ionization lines, and the MIR continuum slope,
have been largely used to quantify the AGN contribution (Genzel et
al. 1998, Dudley 1999, Imanishi et al. 2007, Farrah et al. 2007,
Armus et al. 2007). However, in spite of a general consensus that
the starburst is dominant in about 80\% of ULIRGs, there are strong
discrepancies in the results obtained with the different methods.
This is in line with the recent finding that the strength of PAHs
and the shape of the mid-infrared continuum cannot be safely used to
disentangle AGN and starburst contributions because, on one hand,
there is evidence that PAHs can be destroyed in strong star forming
regions and, on the other, that MIR radiation can be self absorbed
in a highly obscured environment (Vega et al. 2005, Weedman et al.
2005).

In the radio domain, the most direct way of distinguishing between
AGN and starburst power sources in ULIRGs is to search for very
compact radio continuum emission towards the nuclei. However radio
sources of similar \emph{physical} sizes have been identified as
both AGN or a compact starburst, depending on the resolution of the
observations. Nagar et al. (2003), by using 15~GHz radio continuum
data with a resolution of 150 mas, have concluded that most of the
83 ULIRGs of their sample are AGN powered because of the compactness
of the radio sources detected. However, the resolution of their data
corresponds to 420 pc at the median redshift of their sample and, as
a comparison,  the SNe detected in the NW nucleus of Arp~220 by
Smith et al. (1998) are within a region $0\, \farcs 2 \times 0\,
\farcs 4$ ($75 \times 150 \;\rm pc$). Smith et al. find that no AGN
is necessary to explain the IR luminosity in this source. As long as
the brightness temperature does not exceed $10^{7} \;\rm K$ the
compactness of radio nuclei alone does not support an AGN
hypothesis. Smith, Lonsdale \& Lonsdale (1998) find that 7/11 ULIRGs
observed at 18 cm by VLBI could be modelled as pure starbursts,
which would be consistent with their finding that the molecular gas
mass is correlated with the radio flux on small scales.

In this paper we follow a different approach to examine the
energetic environment in (U)LIRGs. Instead of considering a single
spectral region, we base our study on the thorough analysis of the
panchromatic  spectral energy distribution (SED). Thus, we first
determine the NIR - radio SED of a sample of 30 local (U)LIRGs by
using archival IR photometry and spectroscopy, and new radio data by
Clemens et al. (2007). Then, these data are compared with suitable
models with SB and AGN components. Similar approaches have been
adopted by Klaas et al. (2001) and by Farrah et al. (2003). They
used starburst and AGN templates to model the SED from the NIR-submm
range, and from the optical-submm range, respectively. We prefer to
use models built on well calibrated star forming complexes instead
of templates re-scaled to different luminosities for two reasons.
Firstly, it avoids the dubious  process of "re-scaling" a "template"
to another luminosity. This point is well exemplified by the link
between lack of PAH emission and dominance of the molecular cloud
component with respect to the cirrus component (Vega et al. 2005).
Secondly, the use of models may give access to a more sound physical
picture of the environment within which such objects evolve. We are
not only interested in the nature of the power mechanism, but also
in its environment and evolution. This, in turn, can be used for the
study of even more extreme phases such as those likely found at high
redshift, which is one of the ultimate goals of the present
investigation.

Another major difference with previous works is the inclusion of
radio data in the analysis. Radio data constitute an independent
constraint on the starburst strength and age. Indeed, Bressan, Silva
\& Granato (2002) showed that deviations from the FIR/radio
correlation could be used to infer the evolutionary status of a
compact starburst. Such deviations are expected in bursts of short
duration because, at early times (a few $\times 10^6\;\rm yr$),
there is a depletion of synchrotron emission due to the low rate of
production of core collapse Supernovae. Such young sources are
characterized by a FIR/radio ratio that is larger than the average
of star forming objects, and by a radio slope that is flatter than
the average synchrotron slope.

The power of considering also radio data was illustrated by Bressan
et al. (2002) and Prouton et al. (2004). While Bressan et al.
limited their comparative analysis to the case of pure starbursts,
Prouton et al. (2004) have attempted a full panchromatic analysis by
including also the possible contribution from the AGN. The latter
authors pointed out the existence of a certain degree of degeneracy
between the AGN contribution and dust optical depth of the molecular
clouds, mainly arising from a  poor sampling of the SED in the MIR
spectral region.

The present work supersedes our previous ones for the following
reasons. From the observational side, we have greatly enlarged the
sample from the original 7 of Prouton et al. to 30 objects. We have
also reconsidered and largely increased the number of observational
data points, and consequently, our conclusions rely on much better
sampled SEDs. This is particularly true for the MIR range, where for
about one third of the sample we have also included the
\emph{Spitzer} IRS spectrum. From the theoretical side, we make use
of the most recent improvements in our GRASIL SB models (see Sect.
3). These models adopt a better treatment of PAH emission and
include results based on IR observations so that they allow a more
realistic rendering of the emission in the mid-infrared.

The paper is organized as follows. In Section \ref{sec:sample} we
describe our  sample selection and present the observed SEDs of our
galaxies. Section \ref{sec:models} describes the models and the
method used to obtain the best fit to the SEDs. Observed SEDs and
best fit models are published in electronic form, see also Tables
B.1, B.2 and B.3. The results of the fits are presented in Section
\ref{sec:results} and they are discussed in Section
\ref{sec:discussion}. Our conclusions are drawn in Section
\ref{sec:conclusion}, while notes on individual objects are provided
in Appendix A.
\begin{table*}
\caption{Global properties of our sample of (U)LIRGs:
Log(L$_{\rm{IR}}$) is the total (8 - 1000 $\mu$m) IR luminosity
calculated from the 4 IRAS bands following Sanders \& Mirabel
(1996). $D_{\rm{L}}$ is the luminosity distance of the galaxy,
$12/25$ and  $60/100$ are the IRAS colours and J-K is the observed
NIR colour. Galaxies marked with an $^*$ have warm IR colors,
$f_{25}/f_{60}>0.2$. Galaxies marked with a $^\dag$ have NIR colours
J-K$>2$, and those marked with a $^\S$ have available \emph{
Spitzer} IRS spectra. $q_{\rm{1.4GHz}}$ is the FIR to radio
luminosity ratio, and $\alpha_{22}^{1.4}$ is the radio spectral
index between 22 and 1.4 GHz. Galaxies are ordered by decreasing the
total IR luminosity} \label{tab:geninfo}
\begin{tabular}{lccccccclcl} \hline \hline
(1)&(2)&(3)&(4)&(5)&(6)&(7)&(8)&(9)&(10)&(11)\\
NAME &Log(L$_{\rm{IR}}$)&$D_{\rm{L}}$&12/25&60/100&J-K&$q_{\rm{1.4GHz}}$&$\alpha_{22}^{1.4}$&Spectral&Radio&References\\
&(L$_\odot$)&(Mpc)&&&&&&type&size (\arcsec)\\
 \hline
UGC~8058$^{*,\dag,\S}$&12.49&174.0&0.21&1.06&2.22&2.14&-0.25&Sy1&$\le 0.07\times0.06$&a,b, c, d, e, f, g, h, i, j, l, n\\
IR~14348-1447$^{\dag}$&12.30&351.3&0.19&0.97&1.71&2.36&-0.75&L&$0.17\times0.14$&a,  b, j, n\\
IR~12112+0305&12.28&309.3 &0.17&0.85&1.22&2.66&-0.51&HII&$0.19\times0.10$&a, b, j, m, n, o\\
IR~05189-2524$^{*,\dag,\S}$&12.11&175.7&0.21&1.20&2.17&2.74&-&Sy2&$0.20\times 0.17$&a,  b, j, n, p\\
UGC~9913$^\S$&12.11&73.5&0.06&0.91&1.18&2.59&-0.46&HII/L&$0.32\times 0.19$&a,  b, j, l, m, n, o, q\\
IR~08572+3915$^{*,\S}$&12.09&243.6&0.19&1.53&1.90&3.27&-0.11&L/Sy2&$0.09\times 0.07$&a,  j, n, o\\
UGC~8696$^\S$&12.09&155.3&0.10&1.01&1.30&2.25&-0.83&Sy2/L&$0.32\times 0.18$&a, b, e, j, k, l, n, \\
IR~15250+3609$^\S$&12.01&229.7&0.15&1.17&1.16&2.76&-0.45&L&$\le 0.06\times 0.05$&a,  b, j, m, n, o\\
IR~10565+2448$^\S$& 11.98&177.9&0.16&0.80&1.09&2.44&-&HII&$\le 0.25 \times 0.25 $&a,  m, n, o, r\\
UGC~5101$^\S$&11.90&162.2&0.24&0.57&1.80&1.99&-0.81&L/Sy1.5&$0.14\times 0.11$&a,  j, k, n, s\\
IZW~107 &11.83&165.6&0.15&0.86&1.11&2.36&-0.95&HII&$0.28\times 0.25$&a,  m, o, t\\
IR~01364-1042&11.78&200.4&0.13&0.96&1.19&2.71&-0.50&L&$0.19\times0.08$&a  \\
IR~10173+0828&11.74&198.9   &0.13&1.07&1.16&2.84&-0.44&-&$\le 0.08\times0.05$&a, m, o\\
Arp~299$^*$&11.72&41.5&0.16&0.97&1.29&2.09&-0.80&-&$0.38\times 0.28$&a, q, u, v\\
UGC~4881&11.65&161.8&0.21&0.69&1.15&2.37&-0.85&HII&$0.20\times 0.15$&a, o\\
CGCG~436-30$^\S$&11.64&127.8&0.14&1.11&1.31&2.40&-0.63&L/HII&$0.45\times 0.30$&a, o, w\\
IC~1623&11.63&81.4&0.19&0.74&1.41&2.08&-0.88&-&$\le 0.40 \times 0.25$&a, t, x, y \\
NGC~1614$^*$&11.60&64.5 &0.20&0.99&1.28&2.45&-0.68&HII/Sy2&$-$&a, k, m, o, s, z, aa, ab, ac\\
UGC~8387$^\S$&11.58&95.0&0.19&0.61&0.96&2.32&-0.95&L/HII&$0.52\times 0.33$&a, m, o, r, ab\\
NGC~7469$^*$&11.58&66.0&0.23&0.77&1.26&2.29&-0.84&Sy1.2&$0.16\times 0.09$&a, h, k, o, ad, ae, af\\
UGC~2369 &11.57&127.7 &0.17&0.79&1.12&2.33&-0.80&HII&$0.13\times 0.13$&a, m, o, t, ag\\
IIIZW~35&11.56&112.0 &0.10&0.94&1.27&2.58&-0.52&L/HII&$0.18\times 0.14$&a,  o\\
IC~5298$^*$&11.54&111.9&0.17&0.75&1.29&2.53&-0.79&Sy2/L/HII&$0.12\times 0.09$&a, o\\
Arp~148 &11.51&141.6&0.33&0.55&1.37&2.37&-0.64&HII&$0.22\times 0.10$&a, m, o\\
NGC~2623$^\S$&11.49&74.8&0.12&0.83&1.15&2.49&-0.60&L/Sy2&$0.38\times 0.26$&a,  k, m,  ac\\
Mrk~331&11.44&74.9&0.19&0.81&1.24&2.52&-0.70&HII/Sy2&$0.12\times 0.09$&a,  k, o, r, ab\\
NGC~34&11.44&79.6&0.15&1.01&1.18&2.49&-0.79&Sy2&$0.39\times 0.29$&a, u, ah \\
NGC~5256&11.44&113.7&0.33&0.66&1.47&1.90&-0.77&Sy2/HII&$0.25\times 0.21$&a, b, o, ac, ad, ai \\
UGC~6436&11.41&140.1&0.44&0.57&1.05&2.65&-0.70&L/HII&$\le 0.30\times 0.30$&a, o \\
NGC~6286&11.16&74.4&0.50&0.44&1.36&1.94&-0.95&HII/L&$\le 0.25\times 0.25$&a, l, r, w \\
\hline
\end{tabular}

\textsc{References:} (a) NED, IRSA, Clemens et al. (2007), (b) Klaas
et al. (2001), (c) Roche \& Chandler (1993), (d) Joyce \& Simon
(1976), (e) Rieke (1978),
 (f) Rieke (1976), (g) Joyce et al. (1975), (h) Weedman et al. (2005), (k) Benford (1999), (j) Rigopoulou et al. (1996), (l) Carico et al. (1992),
  (m) Dunne \& Eales (2001), (n) Farrah et al. (2003), (o) Dunne et al. (2000), (p) Maiolino \& Rieke (1995), (q) Chini, Kruegel \& Lemke  (1986),
  (r) Dale et al. (2000), (s) Carico et al. (1988), (t) Spinoglio et al. (2002), (u) Gallais et al. (2004), (v) Charmandaris, Stacey \& Gull (2002),
  (w) Stickel et al. (2004), (x) Frayer et al. (1999), (y) Le Floc'h et al. (2002), (z) Rieke \& Low (1972), (aa) Lebofsky \& Rieke (1979),
  (ab) Soifer et al. (2001), (ac) Brandl et al. (2006), (ad) P\'erez-Garc\'{\i}a \& Rodr\'{\i}guez- Espinosa (2001), (ae) Alonso-Herrero et al. (2001),
  (af) Papadopoulos \& Allen (2000), (ag) Hwang et al. (1999), (ah) Albrecht, Kr\"{u}gel \& Chini (2007), (ai) Alonso-Herrero et al. (2003)\\

\end{table*}

\section{Sample}
\label{sec:sample}

We study the sample  of 30 (U)LIRGs presented by Clemens et al.
(2007, hereafter Paper I). The selection criteria were specificated
in that work and are summarized as follows: (a) The galaxies are
compact (U)LIRGs from  Condon et al. (1991a) (i.e. brighter than
$5.25\;\rm Jy$ at $60\;\rm \mu m$) with   radio data at 3 or more
frequencies, including the new measurements at 22~GHz (Paper I). The
well-sampled radio spectra will allow us to put strong constraints
on the age of the burst of star formation (Bressan et al. 2002). (b)
We also require the galaxies to have a well-sampled infrared SED,
with data in the NIR, MIR and FIR spectral ranges. We have excluded
from the analysis data at wavelengths shorter than the NIR because,
due to the large extinction, they are likely to be dominated by the
underlying old stellar population and not by the starburst (e.g.
Surace et al. 2000, Farrah et al. 2001). Thus, the inclusion of
optical data would introduce new free parameters, all related with
the geometry, ages, and metallicities of this unobscured stellar
population. To a similar extent, also the NIR continuum may be
dominated by the old population, but its inclusion is needed because
a  J-K colour excess over that expected from a pure stellar
component
 is a strong indication of the presence of an AGN (e.g. Berta
et al. 2003).

The original  sample presented in paper I is of 31 galaxies, however we exclude
one of the galaxies (IRAS~03359+1523) from this analysis because of
the confusion with a nearby source of similar strength in the
far-infrared.

In Table \ref{tab:geninfo} we list some general information for our
sample, useful for the analysis performed here. Column (1) displays
the name of the galaxy. In Col. (2) we list the logarithm of the
total (8 - 1000 $\mu$m) IR luminosity (Sanders \& Mirabel 1996) for
each galaxy. In  Col. (3) we list the luminosity distances,
$D_{\rm{L}}$, of the galaxies, calculated by assuming
$H_o=75$~Mpc/km/s, $\Omega_M=0.3$ and $\Omega_\Lambda=0.7$, and the
radial velocities given in NED\footnote{The NASA/IPAC Extragalactic
Database is operated by the Jet Propulsion Laboratory, California
Institute of Technology, under contract with the National
Aeronautics and Space Administration.}.  Columns (4) and (5) list
the 12/25 and 60/100 IRAS colours respectively, where all the IR
data were taken from NED. Column (6) displays the J-K colour. NIR
data were taken from IRSA\footnote{The NASA/IPAC Infrared Science
Archive is operated by the Jet Propulsion Laboratory, California
Institute of Technology, under contract with the National
Aeronautics and Space Administration.}. In Col. (7) we list the
value of the FIR to radio luminosity ratio, $q_{\rm{1.4GHz}}=
\log\left(\frac{FIR}{3.75 \times 10^{12} \ W \
m^{-2}}\right)-\log\left(\frac{S_{1.4 \ GHz}}{W \
m^{-2}Hz^{-1}}\right) $, and in Col. (8) the radio spectral index
between 22 and 1.4 GHz, $\alpha_{22}^{1.4}= -\frac{
log(S_{1.4GHz}/S_{22 GHz})} {log ({1.4GHz}/22 GHz)}$. In both cases,
radio data were taken from Paper I. Column (9) displays the optical
spectral classification of each galaxy from NED. Column (10)
displays the size of the radio emitting region at 1.4 GHz from
Condon et al. (1991a).

\subsection{The NIR to radio data}
For all the sources of our sample we collected NIR data  from the
2MASS All-Sky Extended Source Catalog and IRAS data from  NED. Radio
data were collected from Paper I. Additional IR and sub-mm
broad-band data were taken from the literature. The references to
those works are quoted in the last column of Table
\ref{tab:geninfo}. In order to clarify which  NIR to radio data were
 effectively used in our SED modelling, we also provide, in
electronic form, the SED data for each galaxy.

We  searched for  \emph{Spitzer} IRS spectra of our ULIRG sample in
the SSC archive and found them for 11 galaxies (projects ID 105, and
14, P.I.: J. R. Houck, and project ID 30323, P.I.: L. Armus) for
eleven galaxies of our sample, UGC~8058, IRAS~05189-2524, UGC~9913,
IRAS~08572+3915, UGC~8696, IRAS~15250+3609, IRAS~10565+2448,
UGC~5101, UGC~8387, CGCG~436-30, and NGC~2623. In order to include
them in the fitting procedure, we calculated from the \emph{Spitzer}
spectra the values of the specific fluxes at 5.8 and 10 $\mu$m,
those in the LW2 and LW3 ISO bands and those in the IRAS 12 and 25
microns bands, by taking into account the suitable filter responses.
We excluded the fluxes in the IRAC 3 and IRAC 4 bands because these
are too  dominated by molecular absorption bands (e.g. Lahuis et al.
2007), which are not included in our models.

In the compilation of fluxes care was taken to avoid confusion
between different close sources and, in the case of interferometric
radio data, losses due to high resolution mapping.

Our objects are compact, with the radio emitting
regions  ranging from $290 \times 239$ pc for IR~14348-1447 to 43 $\times$
33 pc for Mrk~331 (see Col. 10 in table \ref{tab:geninfo}).
They are also isolated so that we are confident that our
fluxes sample the entire starburst in all  cases.

\section{Models}
\label{sec:models}

In this section we describe the  model that will be used to
interpret the SEDs of our galaxy sample.

As already mentioned, there is considerable evidence that the
starburst component is always present and possibly dominates the
bolometric luminosity in luminous and ultra-luminous IRAS galaxies.
Therefore, we adopt the following approach to find the best fit
model. We first test the null-hypothesis that our sample sources are
pure starbursts. Then, if the observed SED shows an excess emission
over the best fit in those wavelength regions that are more affected
by the  presence of an embedded AGN, typically in the NIR-FIR range,
we reject the null-hypothesis and add an AGN component to the model.

The starburst models (SB) are selected from a library generated with
GRASIL\footnote{GRASIL can be run via the WEB interface
web.pd.astro.it/galsynth/ which is maintained at OAPD, INAF by L.
Paoletti, A. Petrella \& D. Selvestrel, and can also be downloaded
from the web page http://adlibitum.oats.inaf.it/silva/default.html}
(Silva et al. 1998) and involve seven main free parameters that will
be discussed below.

The best fit is obtained by minimizing a merit function calculated
as
\begin{equation}\label{equ:chi}
\chi
=\frac{1}{N}\sum_{i=1}^{N}\left(\frac{F_{mod}(i)-F_{obs}(i)}{Err(i)}\right)^2
\end{equation}
where $F_{mod}(i)$, $F_{obs}(i)$, and $Err(i)$ are the model flux
values,  the observed fluxes and observational errors respectively.
$N$ is the number of passbands used for the fit and it is always
greater than the number of parameters of the starburst model (i.e.
7, see Sec. 3.2). Thus, the best fit SB model is always well
constrained. We consider that a satisfactory best fit starburst
model is reached when, in all passbands, the residuals between
observations and models are less than the typical observational
error of the passband.

However, in some cases, the residuals between the observations and the SB model
in one or more passbands, mainly in the NIR/MIR range,
are larger than the typical error of the corresponding flux. Then we reject our
null-hypothesis and consider the possibility that an AGN component,
characterized by a hotter dust emission, is present. The AGN model
is selected from a library generated with the radiative transfer
code developed by Granato \& Danese (1994) and added to the SB model
in such a way that the total luminosity is
\begin{equation}
L_{\lambda}^{\rm{tot}}= (1-f)\times L_{\lambda}^{\rm{SB}}+f \times
L_{\lambda}^{\rm{AGN}} \  {\rm{ , with }} \ 0\leq f \leq 1.
\end{equation}
The best fit is again searched by using the merit function given by
equation  \ref{equ:chi}. Since the AGN models are characterized by
other seven free parameters, one could question that the fit is not
more well constrained because the total number of free parameters
(SB + AGN) may exceed the number of observed passbands. Actually,
anticipating our results, this is the case for one object, IRAS
01364-1042, where the number of adopted passbands is 11 and the
estimated contribution of the AGN amounts to $\sim$10\%. However, as
will be discussed below, we could have simply estimated the AGN
contribution as the flux excess over the best fit SB model.

A more critical point is concerned with the radio emission of the
AGN which is not considered in the model  by Granato and Danese
(1994).  Following Prouton et al. (2004), we assume that the
contribution from the AGN to the radio emission is negligible so
that the radio emission is entirely ascribed, and actually becomes a
strong constrain, to the SB component. This assumption is supported
by VLBI observations of the majority of our sources, indicating an
average value of $S_{\rm{VLBI}}/S_{\rm{Tot}}=0.12$ at 1.6 GHz, thus
suggesting that the contribution of the AGN to the radio can be
neglected (Smith et al. 1998).

In the followings subsections, we briefly describe the library of SB
and AGN models.

\subsection{Starburst models: a brief description of GRASIL}
GRASIL is a population synthesis code which predicts the SED of
galaxies from  far-UV to radio, including state-of-the-art treatment
of dust reprocessing (Silva et al. 1998, Silva 1999, Granato et al.
2000), production of radio photons by thermal and non-thermal
processes (Bressan et al. 2002), nebular lines emission (Panuzzo et
al. 2003) and an updated treatment of PAHs emission (Vega et al.
2005). We refer the reader to the original papers for details, but
we summarize here, for convenience, the features most relevant to
the present work.

The star formation history of a galaxy with a burst of star
formation is  computed in the GRASIL chemical evolution code by
adopting
 a Schmidt-type law,
$SFR(t) =\nu_{\rm{Sch}} M_{\rm{gas}}(t)^k$, of efficiency
$\nu_{\rm{Sch}}$, and exponent $k$. The burst of star formation is
simulated by superposing on this quiescent component
 a starburst
phase, characterized by an exponentially decreasing star formation
rate, with e-folding times "$t_{\rm{b}}$" and age "$age_{\rm{b}}$".
In order to quantify the number of massive stars, the code  adopts a
 initial mass function (IMF). The chemical evolution code
determines, as a function of the galaxy age ($T_{\rm{gal}}$), the
star forming rate ($SFR$), the supernova rate (SNR), the number of
stars of a given mass and metallicity, the remaining mass of gas
($M_{\rm{gas}}$), and the total baryonic mass of the galaxy
($M_{\rm{T}}$).

Then, GRASIL computes the interaction between the stellar radiation
and dust using a relatively realistic and flexible geometry for both
stars and dust. In general, the system is described as a
superposition of an exponential disk component and a bulge
component, the latter modelled by a King profile. However, for the
starburst models, we consider that the spherically symmetric King
profile alone, with core radii corresponding to the radial scale
lengths for the star and dust components,  $r_\star ,
r_{\rm{dust}}$, and identical spatial dependence, i.e.  $r_\star =
r_{\rm{dust}}=r$, is a reasonably good approximation (Silva et al.
1998).

One of the most novel features of GRASIL is that the ISM is divided
into two phases, the diffuse ISM, corresponding to cirrus dust, and
the much denser molecular clouds component (MCs). The code performs
the radiative transfer of starlight through the MCs and diffuse ISM,
by taking into account that new stars are born inside MCs and
progressively get rid of them (either because they escape or because
the clouds are destroyed). This process leads to a {\it
age-dependent extinction} of stellar generations, in the sense that
the younger stars are more affected by dust obscuration in MCs. This
is approximately described in GRASIL as follows. If $t_{\rm{esc}}$
is the timescale for the process, the fraction of starlight radiated
inside the clouds at time $t$ after they formed is given by
\begin{equation}
F(t)=\left\{\begin{array}{ll} 1&(t<t_{esc})\\
2-\frac{t}{t_{esc}}&(t_{esc}<t<2t_{esc})\\
0&(t>t_{esc})
\end{array}\right.
\end{equation}
In practice, 100\% of the stars younger than $t_{\rm{esc}}$ are
considered to radiate inside the MCs, and this percentage decreases
linearly to 0\% in $2 t_{\rm{esc}}$. The timescale $t_{\rm{esc}}$ is
a fundamental parameter, which is found to be larger in starburst
than in normal disk-like galaxies (Silva et al. 1998). The mass of
gas involved in each component is specified by the the parameter
$f_{\rm{mc}}$, which represents the relative fraction of gas in the
molecular cloud phase. Then, the total molecular mass,
$M_{\rm{mol}}= f_{\rm{mc}}\times M_{\rm{gas}}$, is subdivided into
spherical clouds whose optical depth, $\tau_1$, is another important
free parameter. In general, GRASIL computes the mass of dust by
using a Gas-to-Dust mass ratio, $G/D$. This approach assumes that
gas and dust are well mixed and have the same spatial distribution.
As for the intrinsic dust properties, GRASIL adopts a mixture of
graphite and silicate grains and PAHs with continuous size
distribution. Here as in other previous works (see Silva et al. 1998
and Vega et al. 2005 for more details),  the dust parameters are set
to  match the extinction and emissivity properties of the local ISM.

The simple stellar populations (SSPs) included in GRASIL are based
on the Padova stellar models, and cover a wide range of ages and
metallicities. Starlight reprocessing from dust in the envelopes of
AGB stars is included directly into the SSPs, as described by
Bressan, Granato \& Silva (1998).
\subsection{The library of starburst SEDs}

Our starburst library spans a wide range of parameters: star
formation history, obscuration times, dust opacities, etc. The
underlying quiescent star formation history is fixed using a
Schmidt-type law for star formation with an efficiency of
$\nu_{\rm{Sch}}=0.5 \ Gyr^{-1}$ and an exponent $k=1$, with an
infall of primordial gas on a timescale of $t_{\rm{inf}}=9$ Gyr. For
the IMF, we have adopted a Salpeter IMF with  slope $x=1.35$ and
$M_{\rm{low}}=0.1$ $M_\odot$ to $M_{\rm{up}}=100 $ $M_\odot$. To
mimic the starburst, we superimposed, at a late time (11.95 Gyr), an
exponential burst with different e-folding times, $t_{\rm{b}}$. The
strength of the burst was fixed by imposing the constraint that the
mass of stars formed during the burst is  $10 \%$ of the gas mass at
that epoch. We then follow the evolution of the SED models at
different ages ($age_{\rm{b}}$) from the early starburst phase to
the post-starburst and quiescent-normal phase (see Vega et al. 2005
for a description of those phases). For the computation of
non-thermal radio emission we assumed that the lower limit of
initial mass giving rise to a type II SN  is 8 $M_\odot$ (Portinari,
Chiosi \& Bressan 1998, Bressan et al. 2002b).

Among the GRASIL parameters that mostly affect the IR-radio range we
recall the optical depth of MCs (parameterized via $\tau_{\rm{1}}$
i.e. at $1 \mu$m), the mass-fraction of dense molecular gas to total
gas mass ($f_{\rm{mol}}$), the escape time of newly born stars from
their parent MCs ($t_{\rm{esc}}$), and the gas-to-dust ratio
($G/D$). Since, there is evidence that the metallicity in this kind
of galaxies is  slightly super-solar (Rupke, Veilleux \& Baker
2007), we assumed a conservative value for the metallicity of the
gas, $Z\sim 1.1 Z_\odot$, and a $G/D \propto 1/Z$, with the
proportionality constant  chosen to have $G/D = 110$ for $Z=Z_\odot$
(Dwek 1998).

For the spatial distribution of stars and dust, we adopted a
spherically symmetric King profile (Silva et al. 1998) where the the
King profile core radius $r$ was allowed to vary by the typical
sizes of compact starburst regions.  The last free parameter that we
considered in our models was the sub-mm dust emissivity index,
$\beta$. Theoretical studies indicate $\beta$ between 1.5 to 2 but
infrared SEDs fitted with single or double modified black bodies,
require values of $\beta$ as low as $\beta\simeq$0.7 (e.g. Dunne et
al. 2000). Instead when the dust temperature distribution is
properly computed as in GRASIL, fits to similar sources require
$\beta>1.5$, consistent with theoretical predictions.

In summary our starburst models are characterized by the following
free parameters, whose ranges of values are listed in Tables
\ref{tab:chevo} and \ref{tab:gra}: the e-folding time and age of the
burst $t_{\rm{b}}$ and $age_{\rm{b}}$, the sub-mm dust spectral
index $\beta$, the escape time $t_{\rm{esc}}$, the MC optical depth
at 1 $\mu$m, $\tau_{1}$, the fraction of gas in MCs $f_{\rm{mc}}$,
and the core radius of the King profile for the dust-star
distributions $r$.
\begin{figure*}
\centerline{
\includegraphics[scale = 0.43]{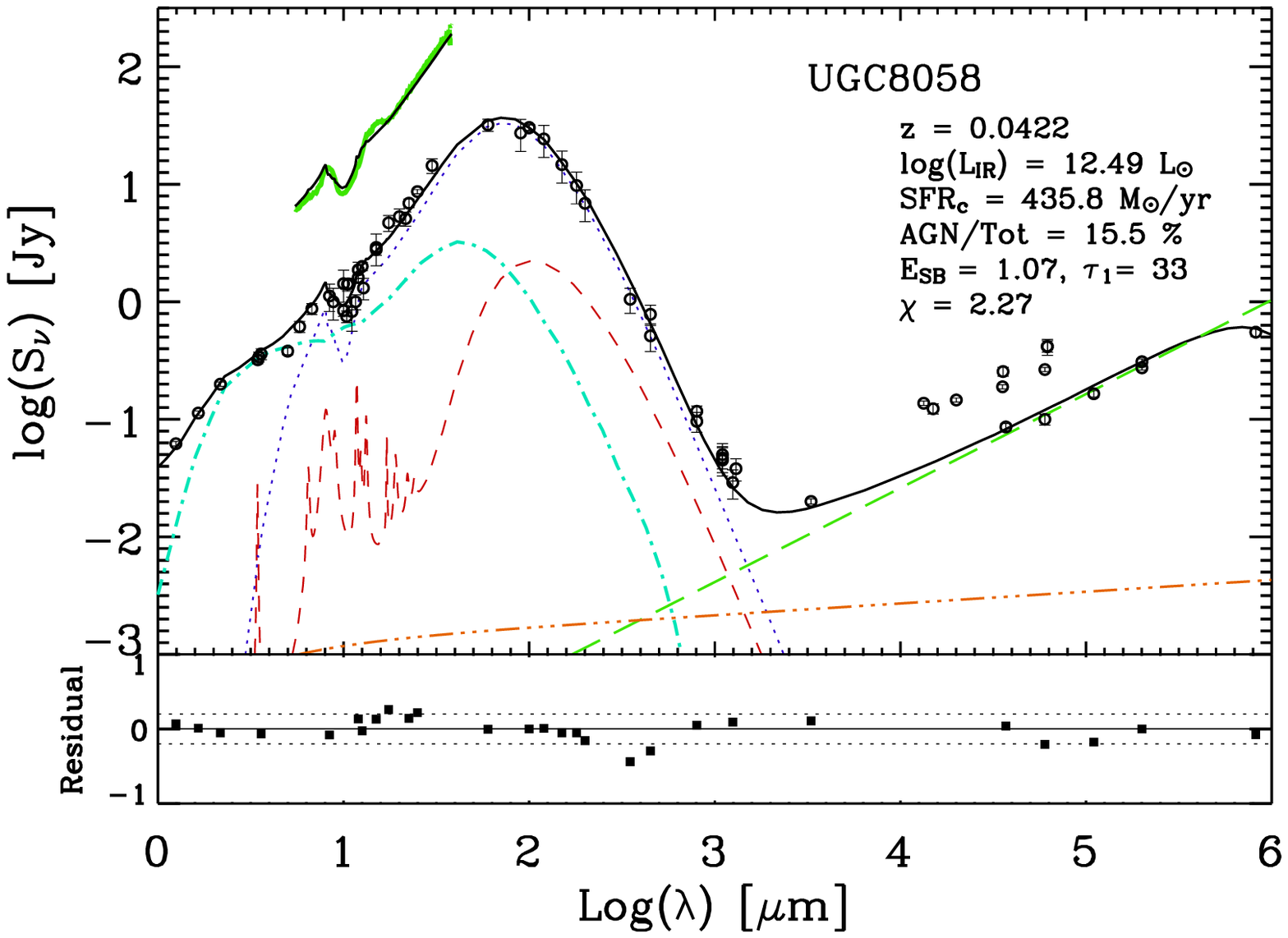}
\includegraphics[scale = 0.43]{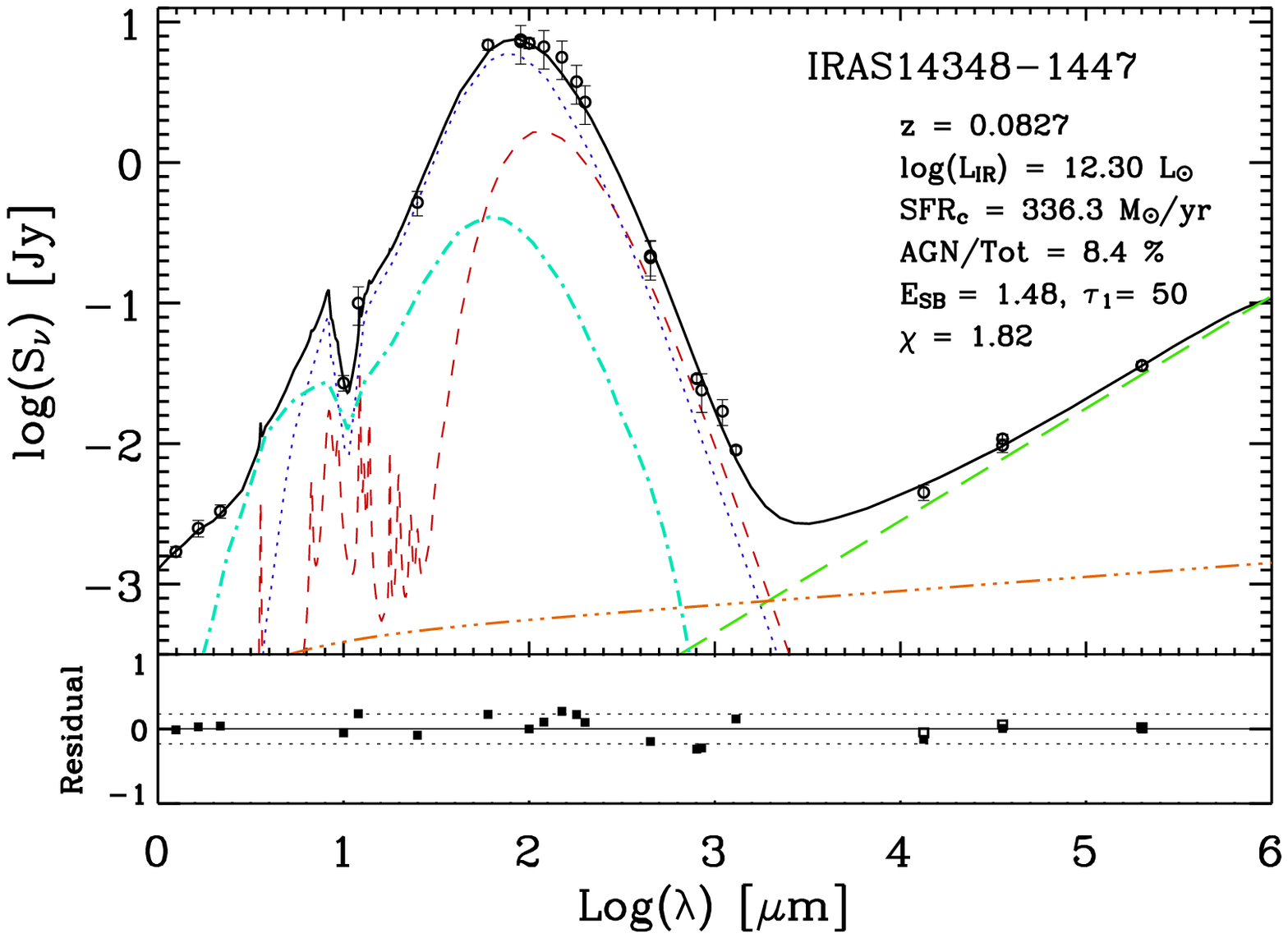}
} \centerline{
\includegraphics[scale = 0.43]{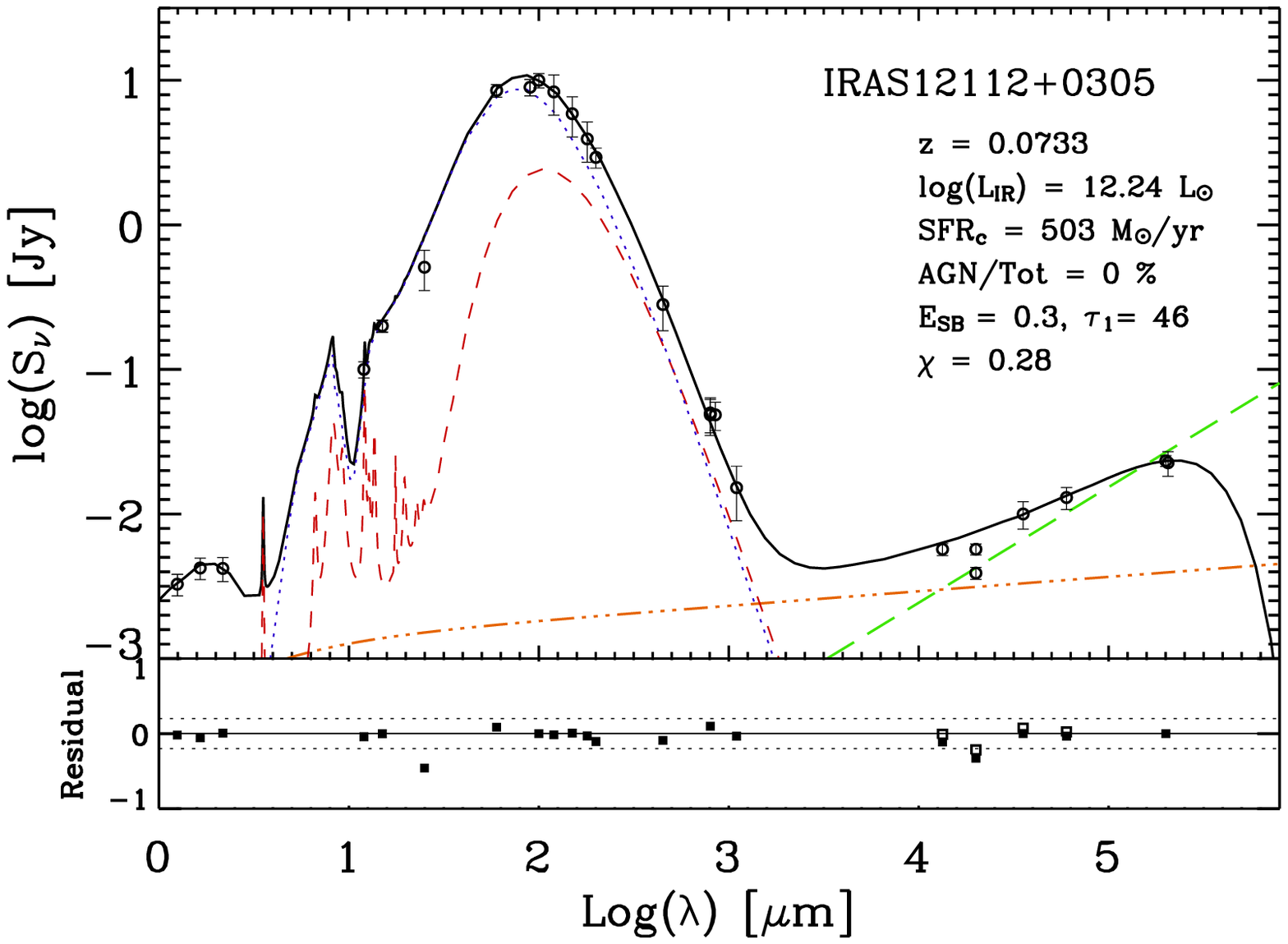}
\includegraphics[scale = 0.43]{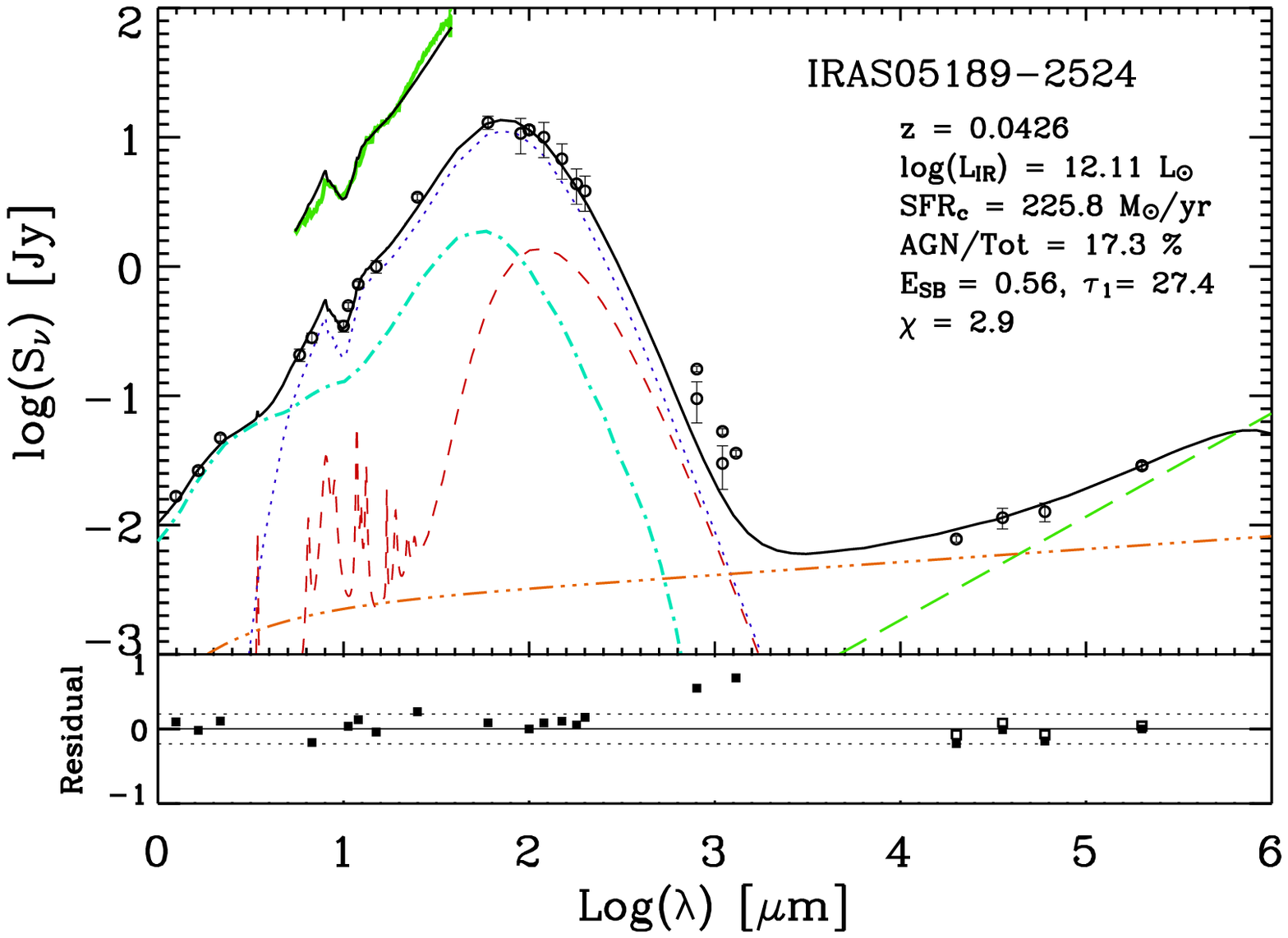}
} \centerline{
\includegraphics[scale = 0.43]{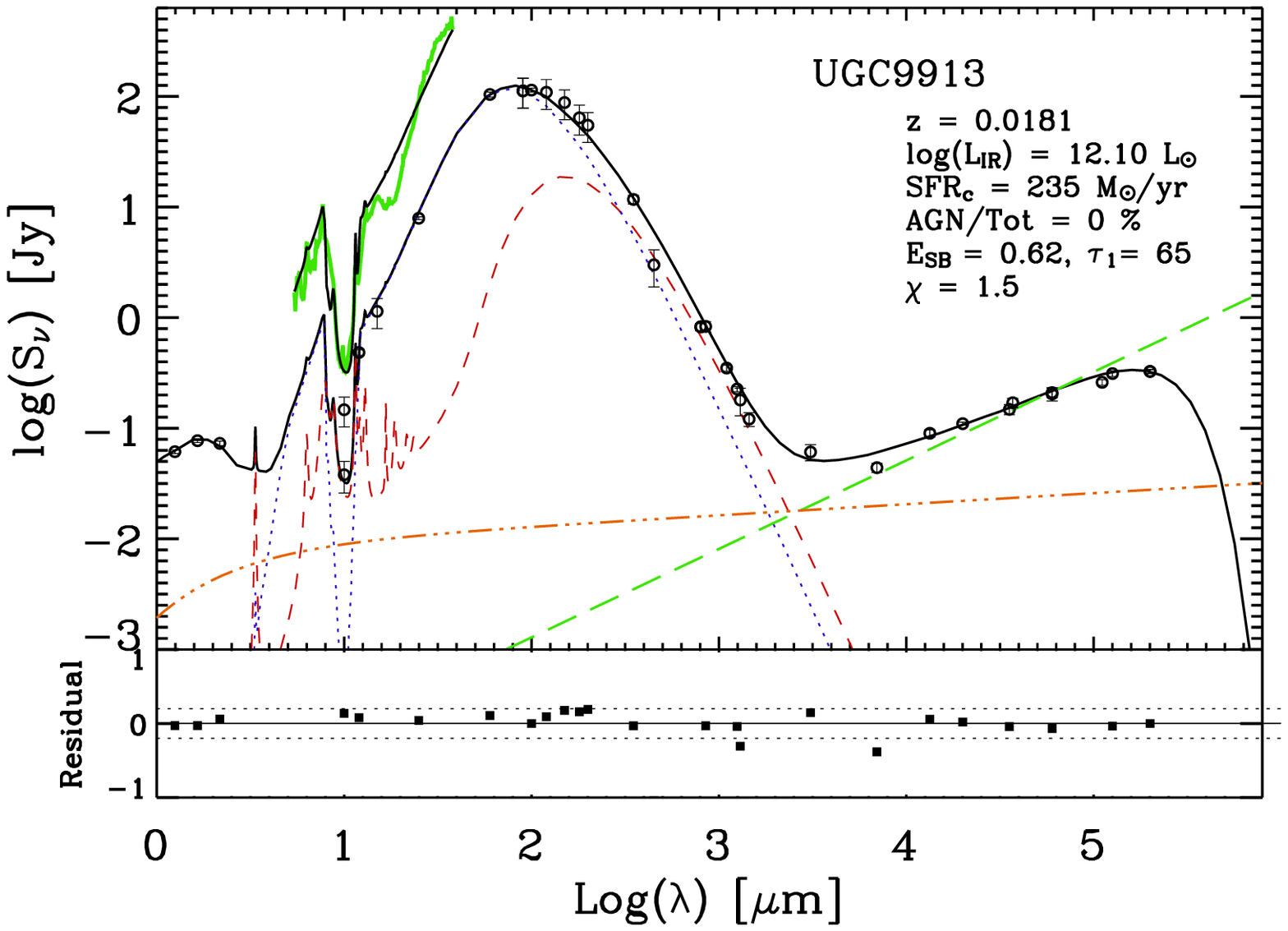}
 \includegraphics[scale = 0.43]{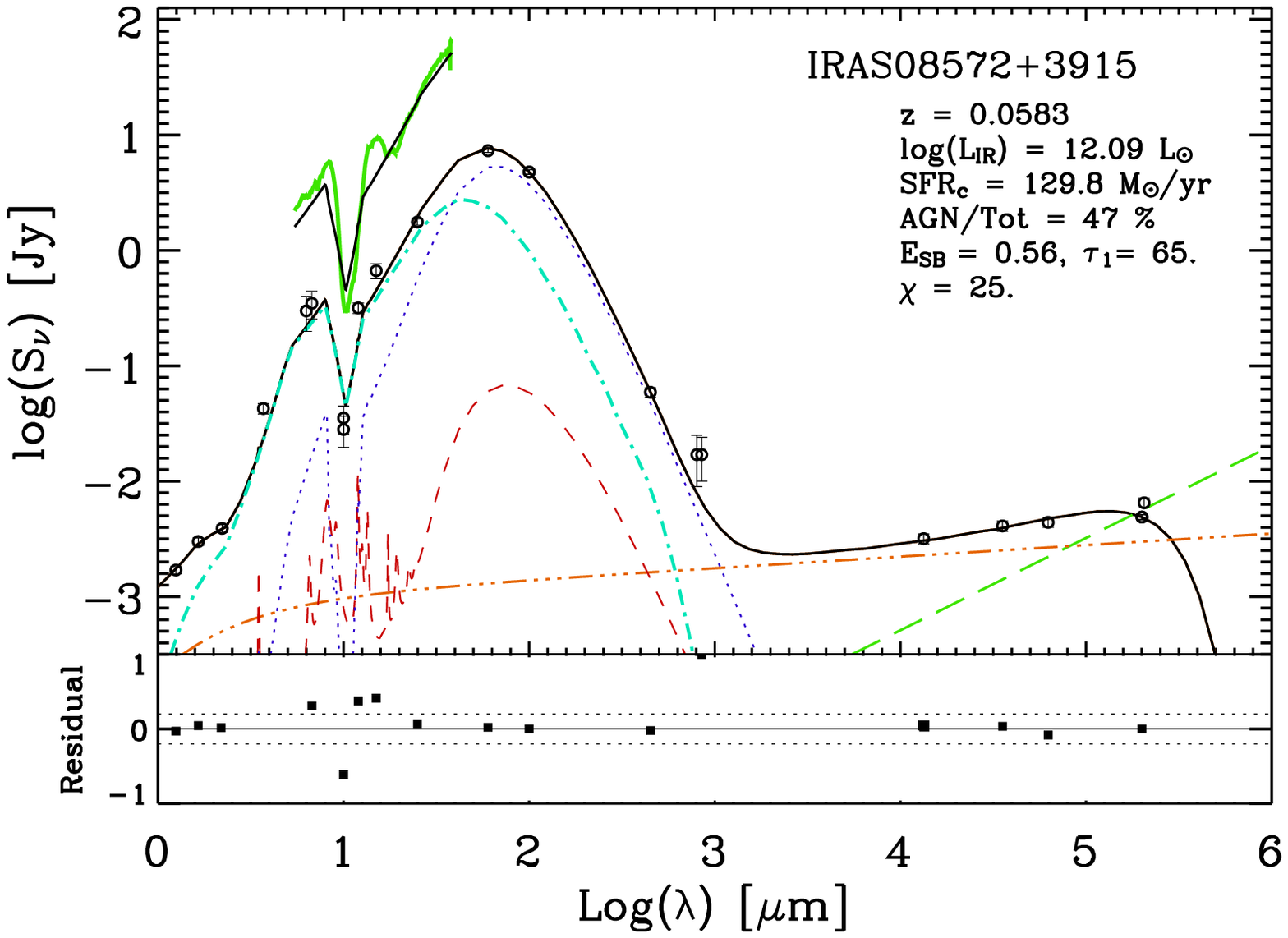}
 }\caption{Comparison between the broad band SED (open circles)
 and the best fit model (thick solid line).
The different emission components of the starburst are: the diffuse
medium (short dashed line), the molecular clouds (dotted line),
thermal radio emission (three dots-dashed line), and the non-thermal
radio emission (long dashed line). The additional emission from the
AGN, when needed, is indicated by a dot-dashed line. When available,
the IRS low resolution spectrum (solid thick line) and the
corresponding model SED (solid thin line) are shown upward displaced
by one dex. The residuals between models and data are shown in the
bottom panels as $(f_{\rm{data}}-f_{\rm{model}})/f_{\rm{data}}$. The
thin dotted lines in the residual plots represent  $\pm 20\%$ of the
residual values. The filled squares represent the residuals and
their wavelengths indicate the data used in the minimization
procedure. The value of the $\chi$ for the best fit is displayed in
each plot. See Sect. \ref{sec:results} and tables \ref{tab:fitpara}
and \ref{tab:derived}, for more details.} \label{Fig:sbsed}
\end{figure*}

\begin{figure*}
\addtocounter{figure}{-1}   \centerline{
\includegraphics[scale = 0.42]{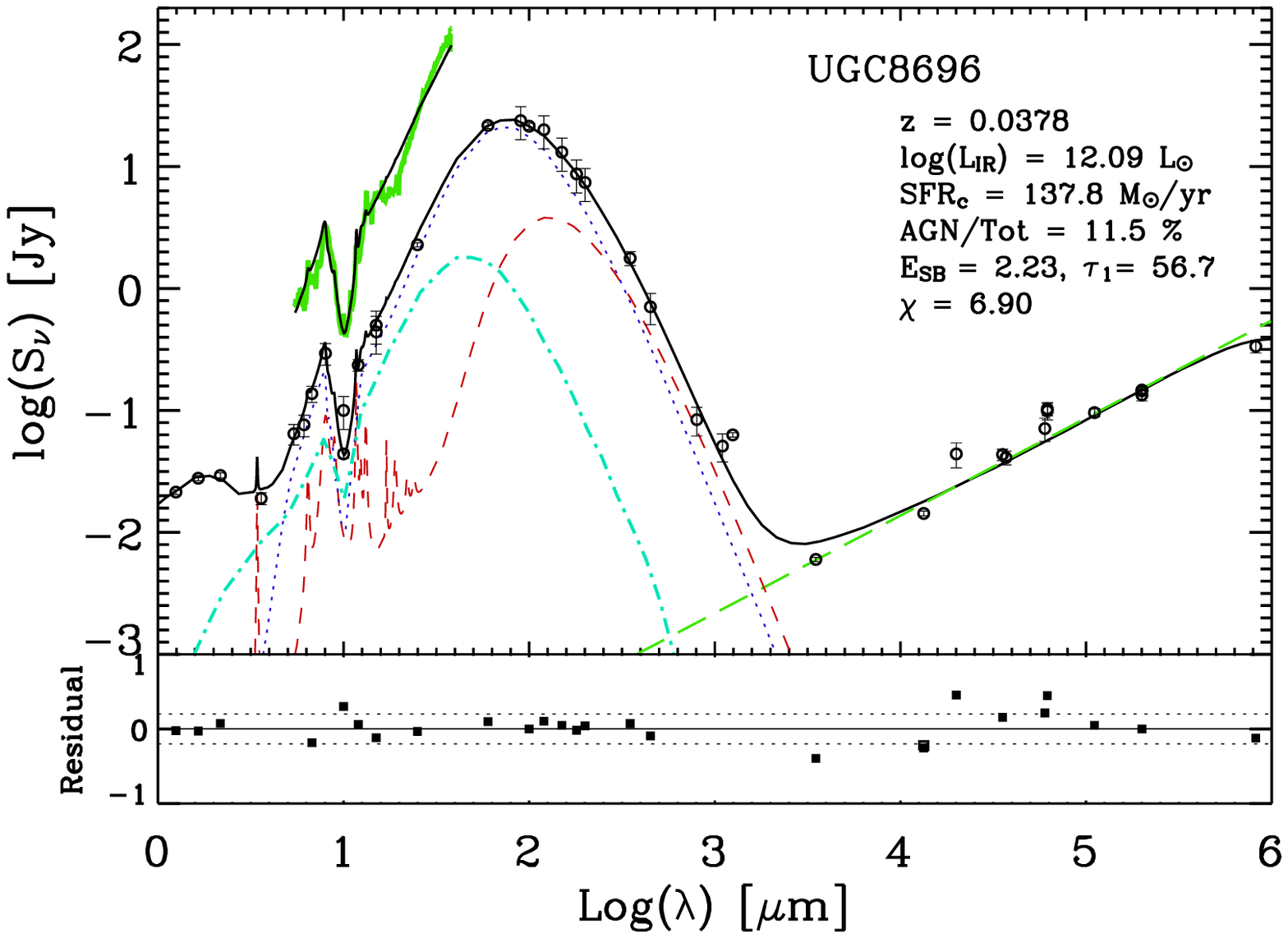}
\includegraphics[scale = 0.42]{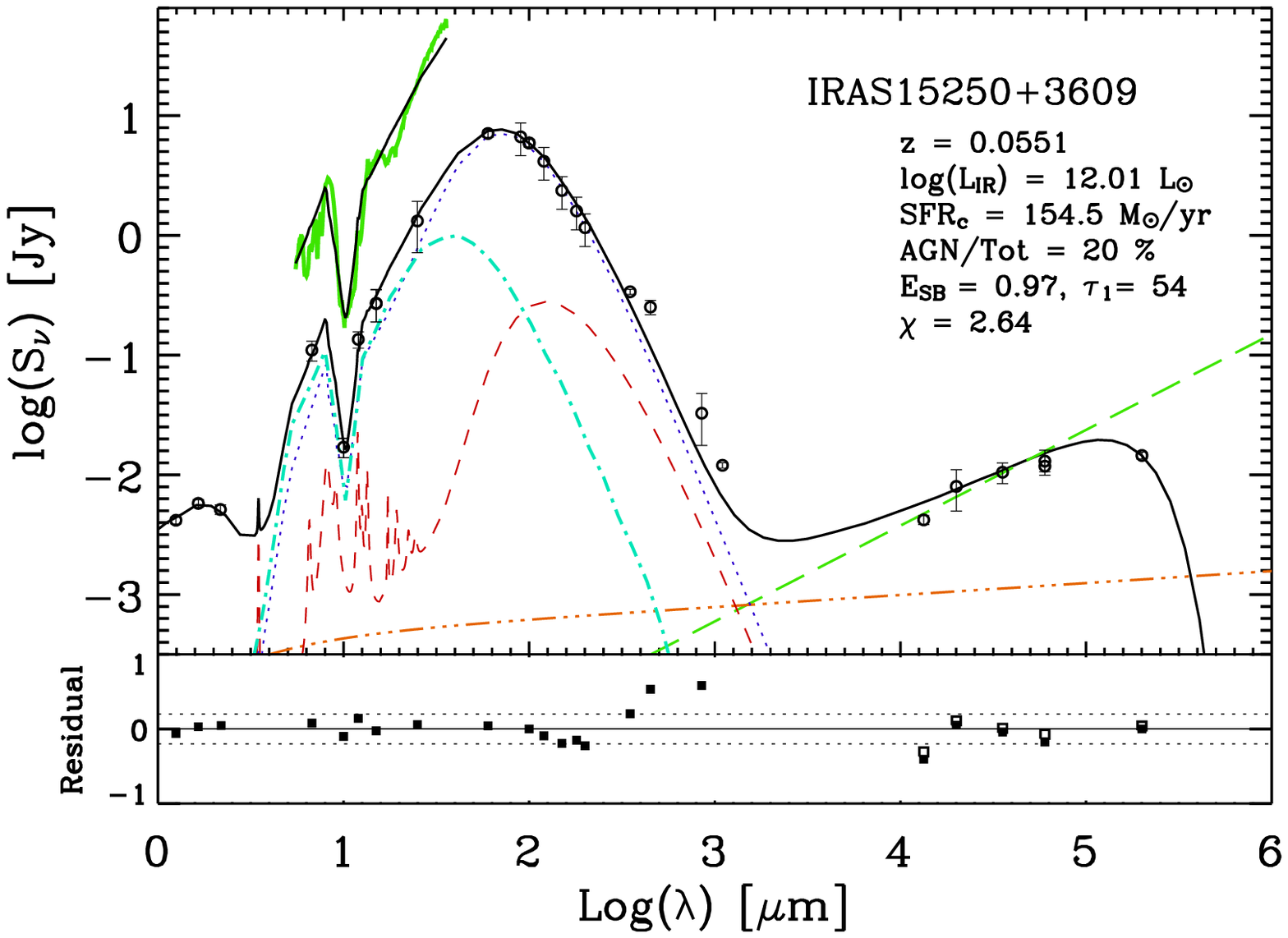}
} \centerline{
\includegraphics[scale = 0.42]{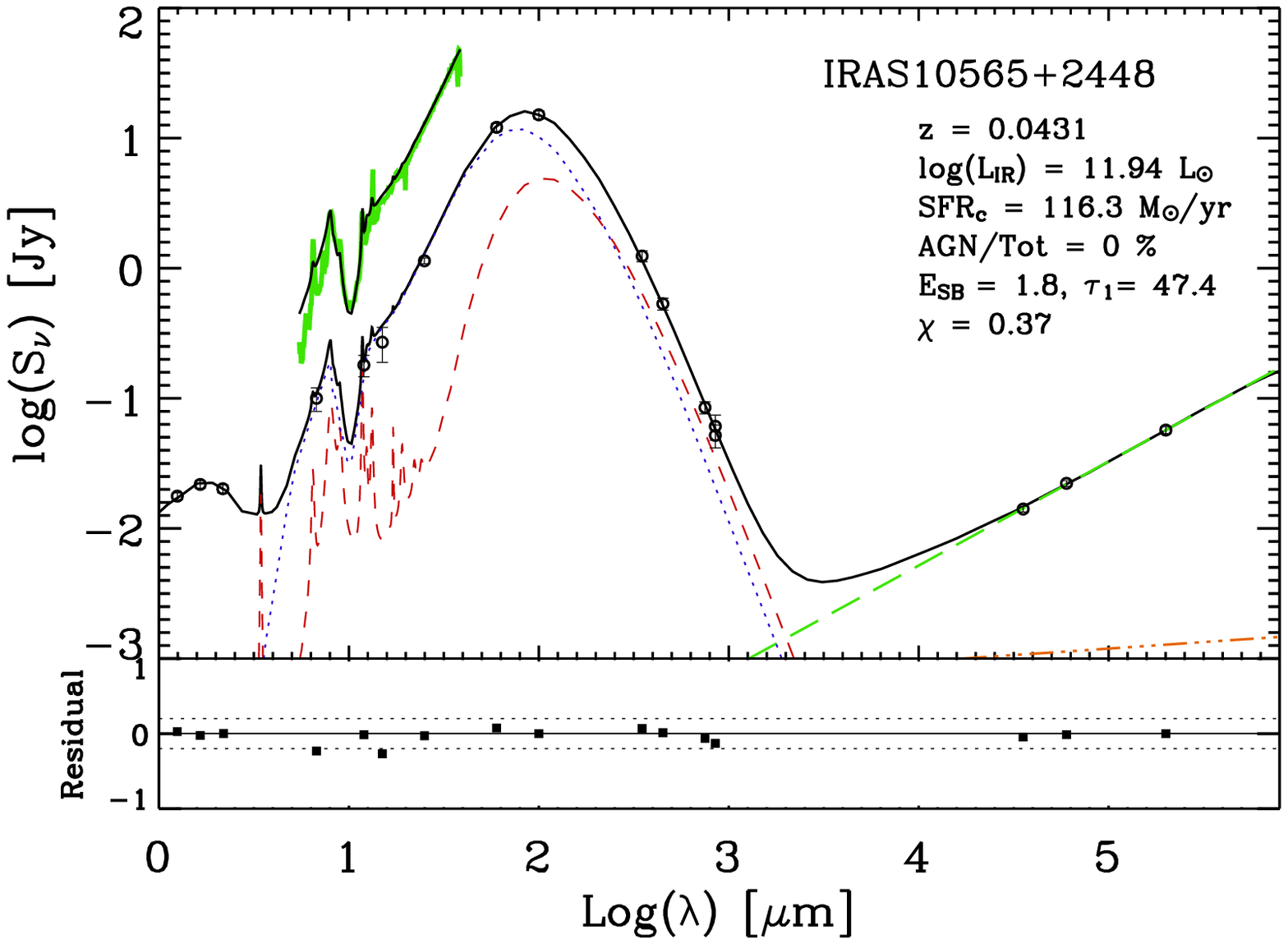}
\includegraphics[scale = 0.42]{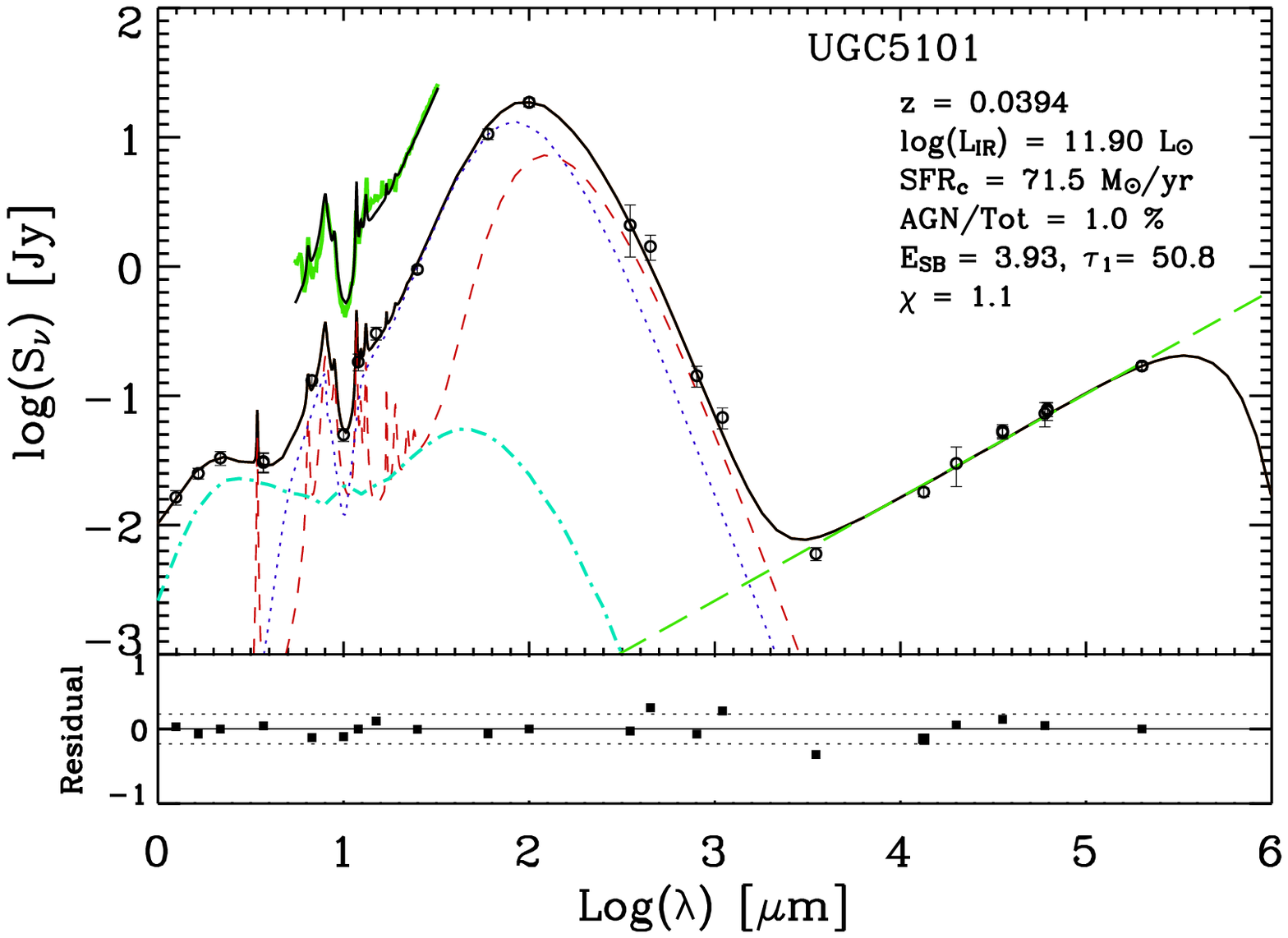}
} \centerline{
\includegraphics[scale = 0.42]{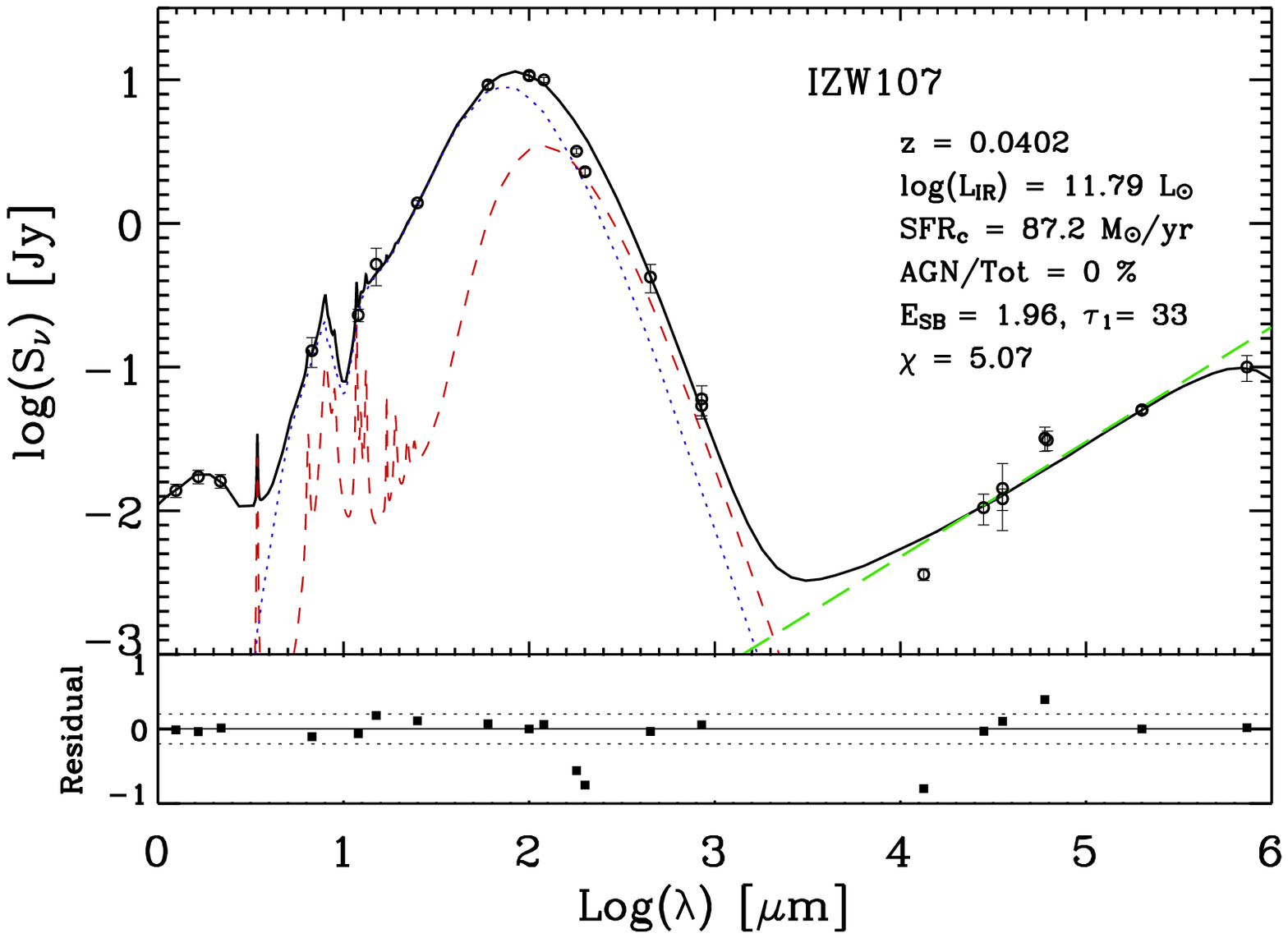}
\includegraphics[scale = 0.42]{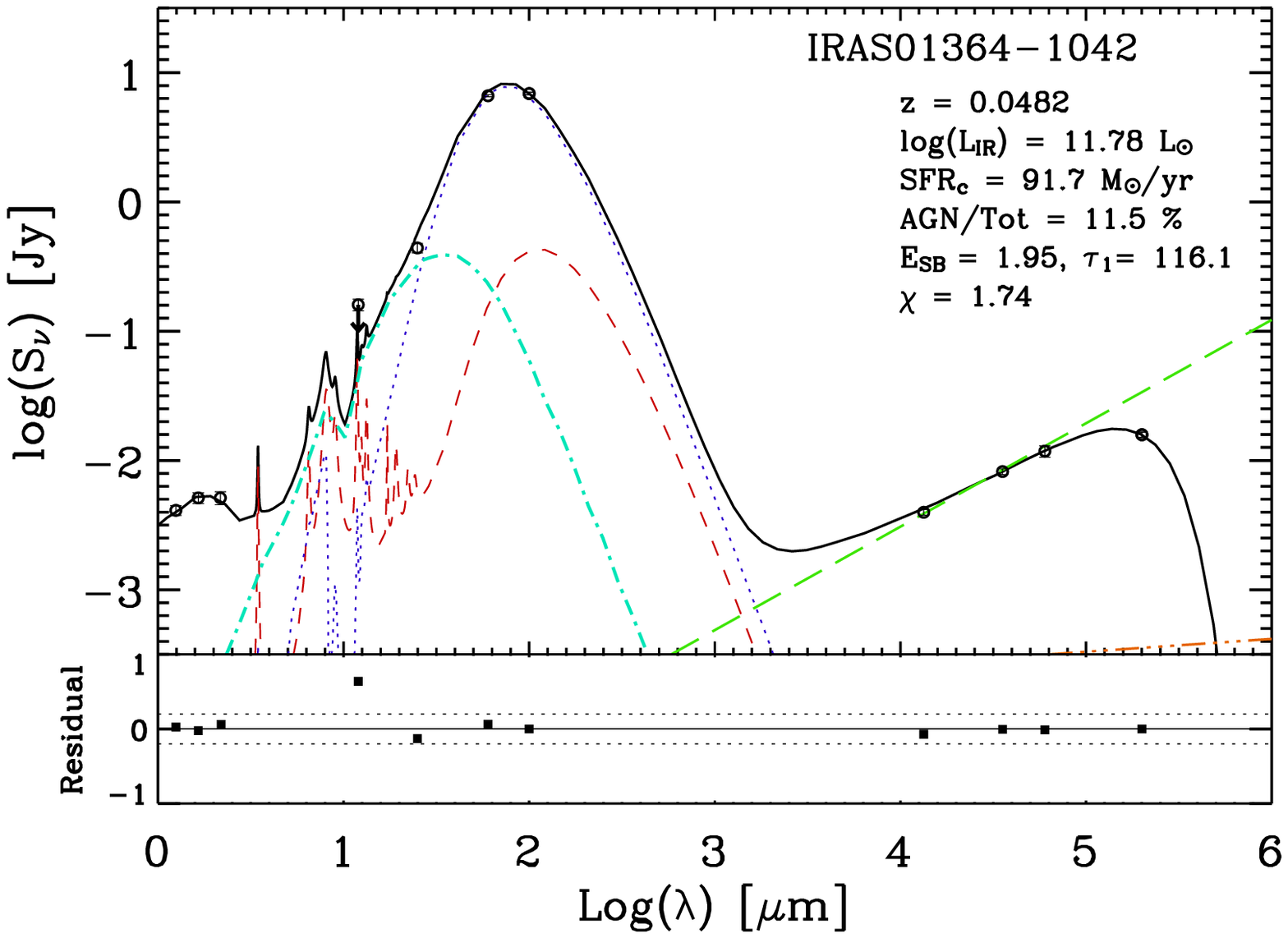}
} \centerline{
\includegraphics[scale = 0.42]{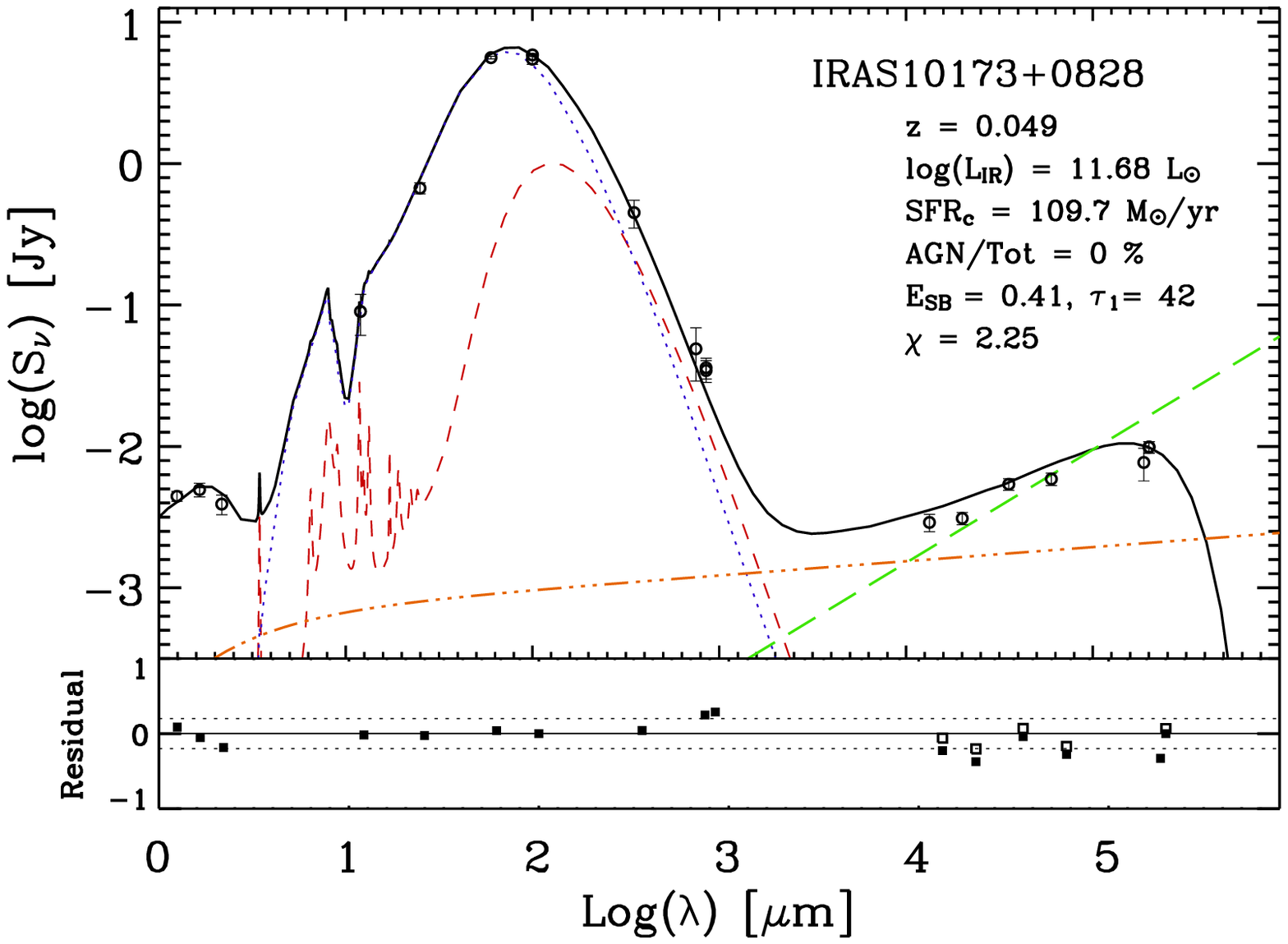}
\includegraphics[scale = 0.42]{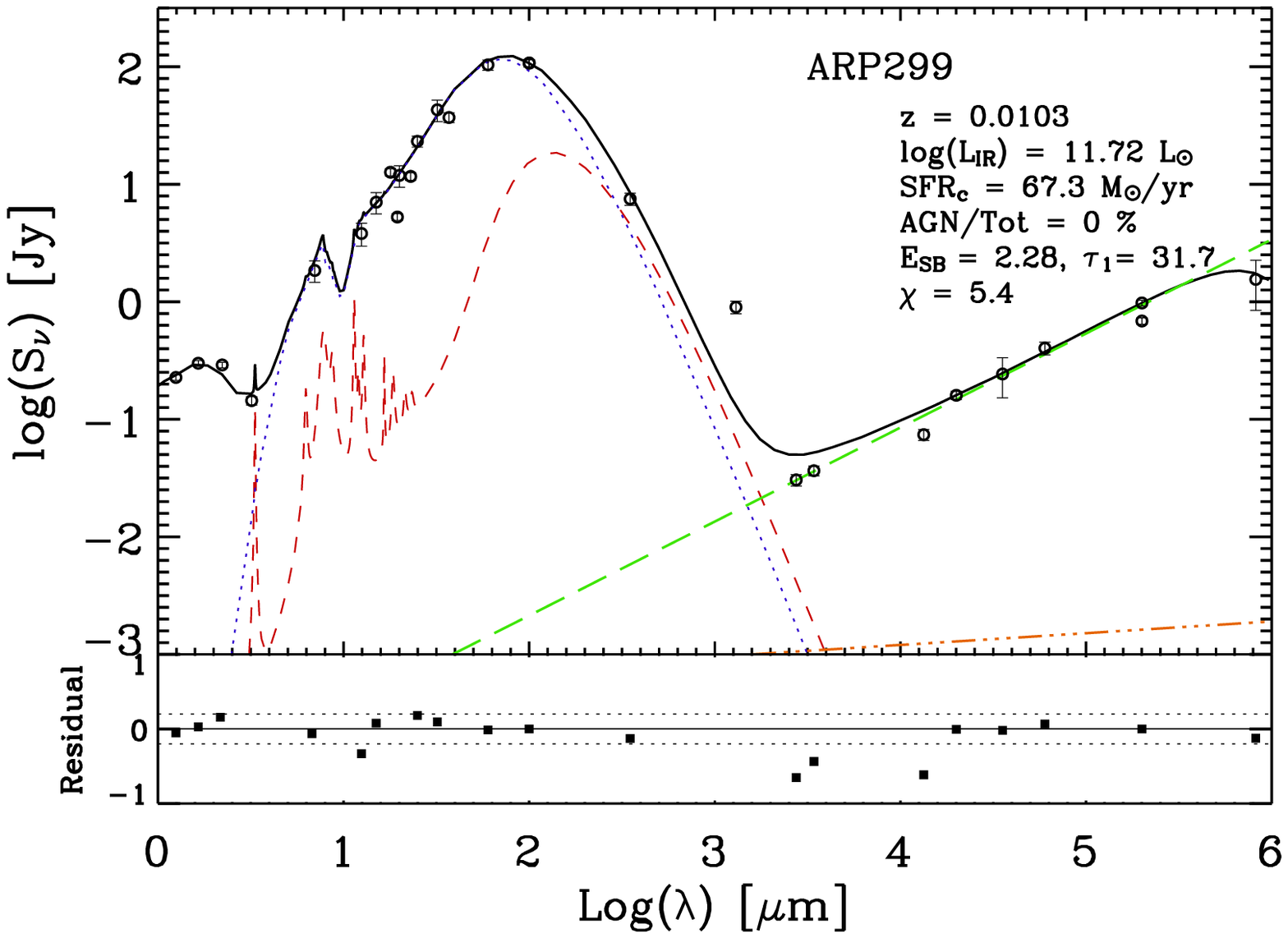}
} \caption{Continued.}
\end{figure*}

\begin{figure*}
\addtocounter{figure}{-1}

 \centerline{
\includegraphics[scale = 0.42]{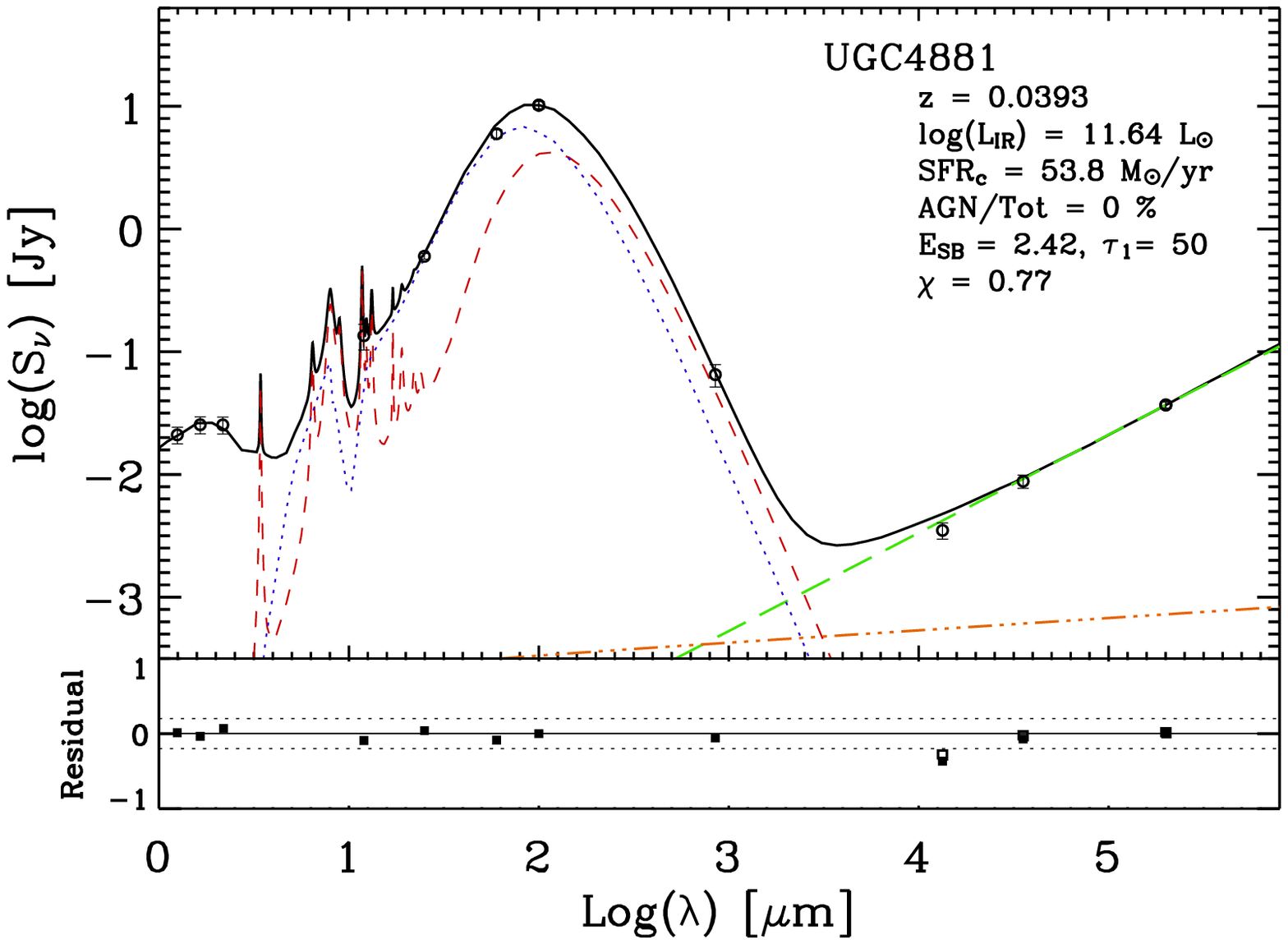}
\includegraphics[scale = 0.42]{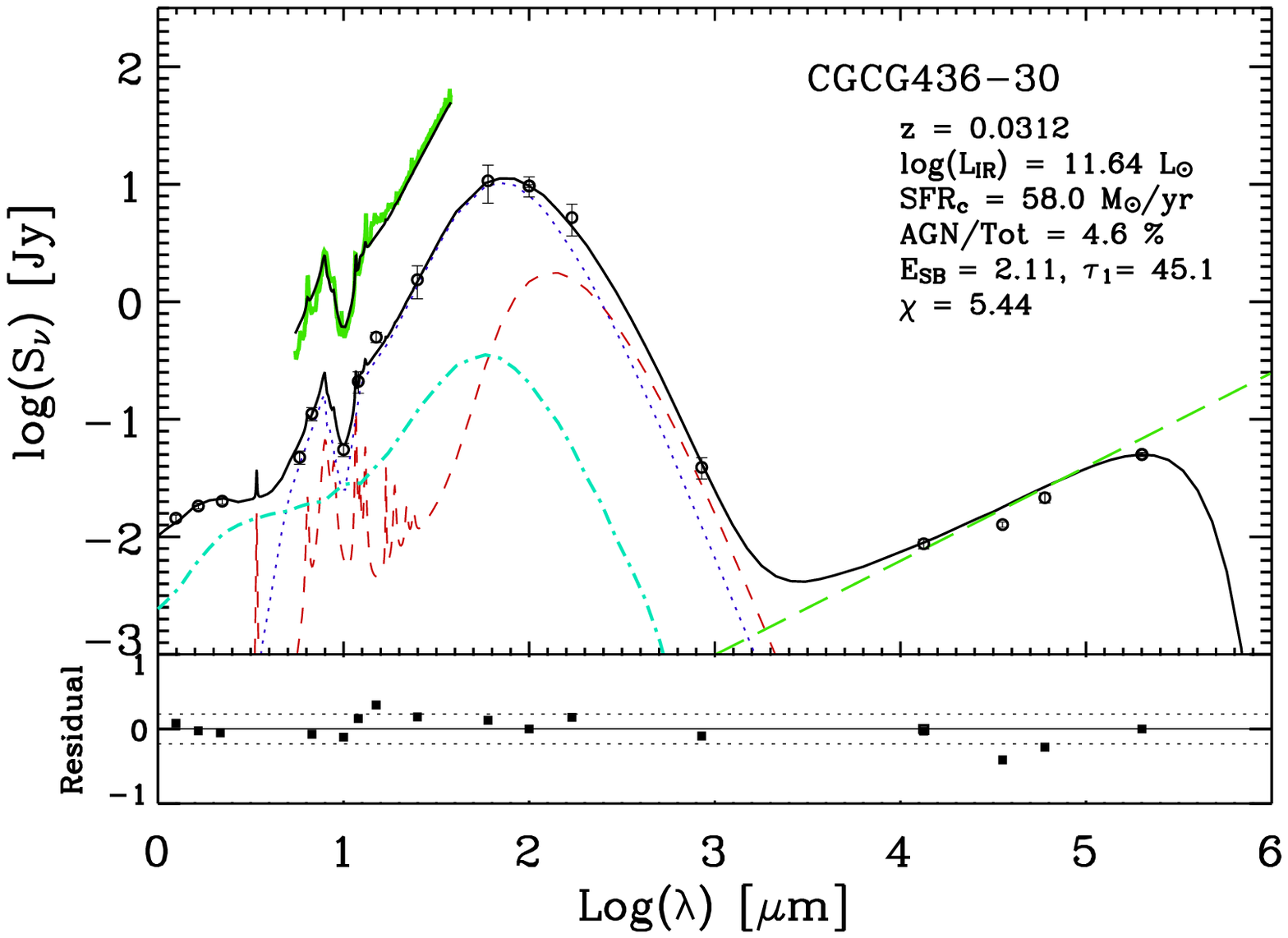}
} \centerline{
\includegraphics[scale = 0.42]{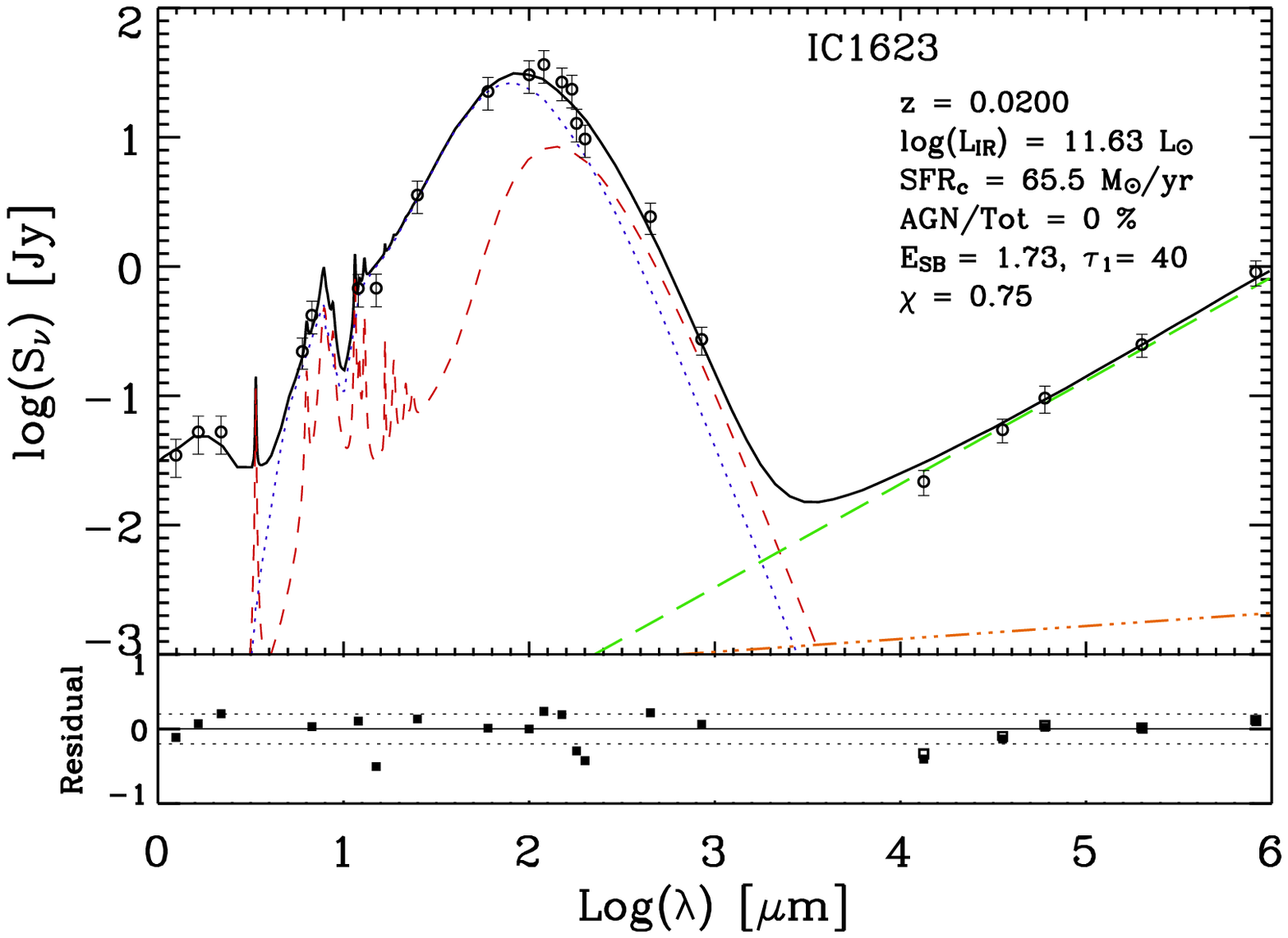}
\includegraphics[scale = 0.42]{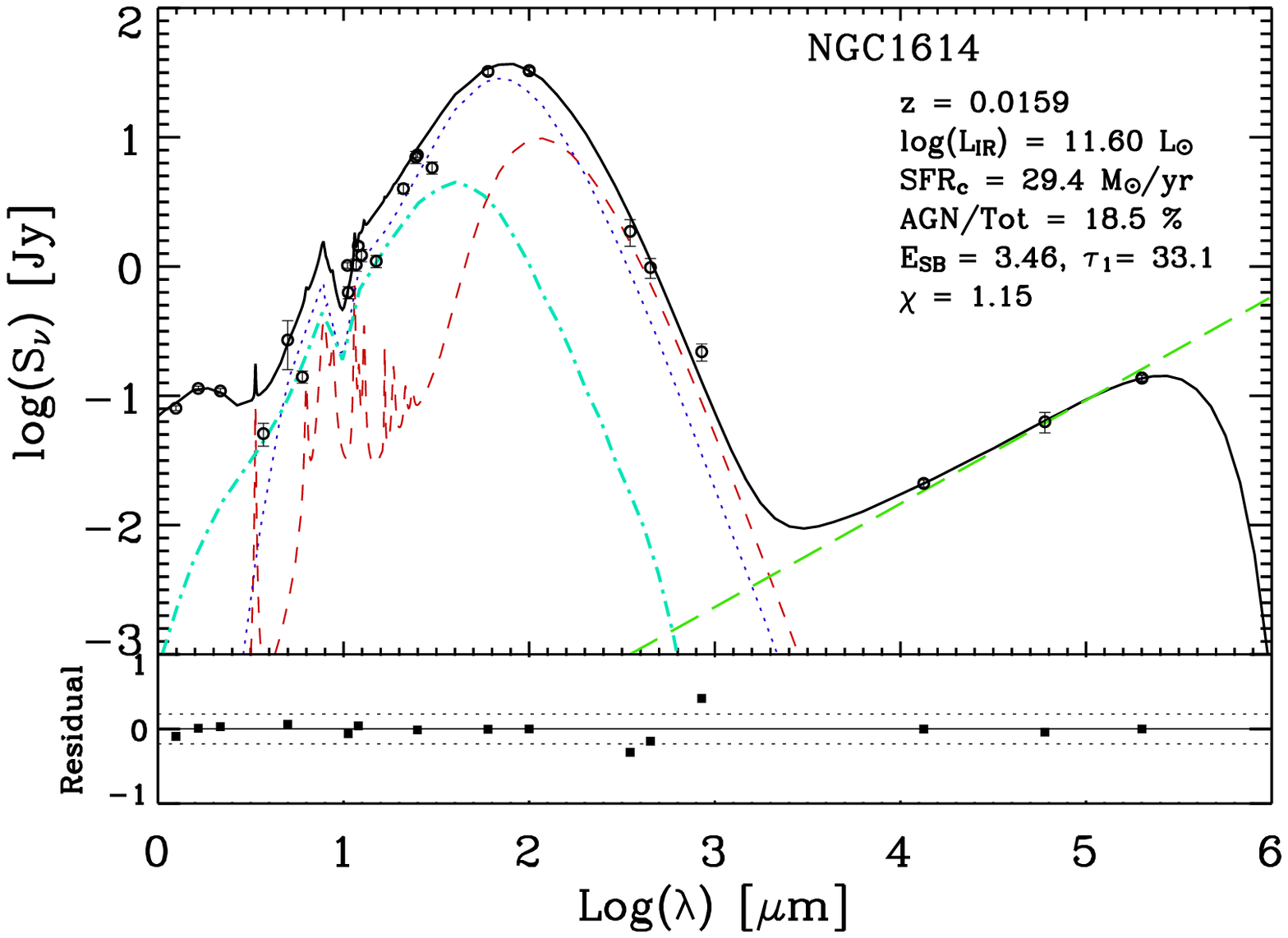}
} \centerline{
\includegraphics[scale = 0.42]{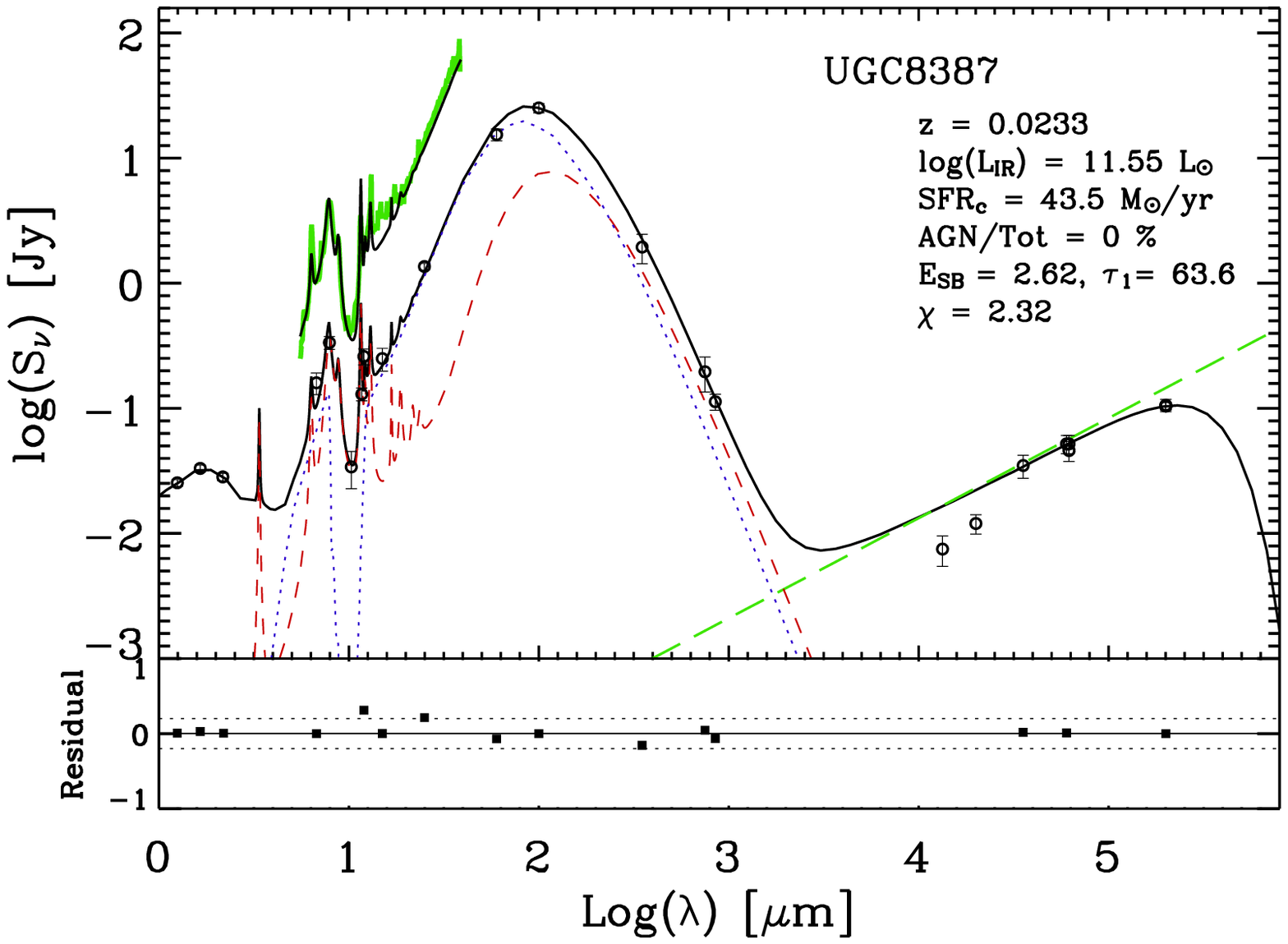}
\includegraphics[scale = 0.42]{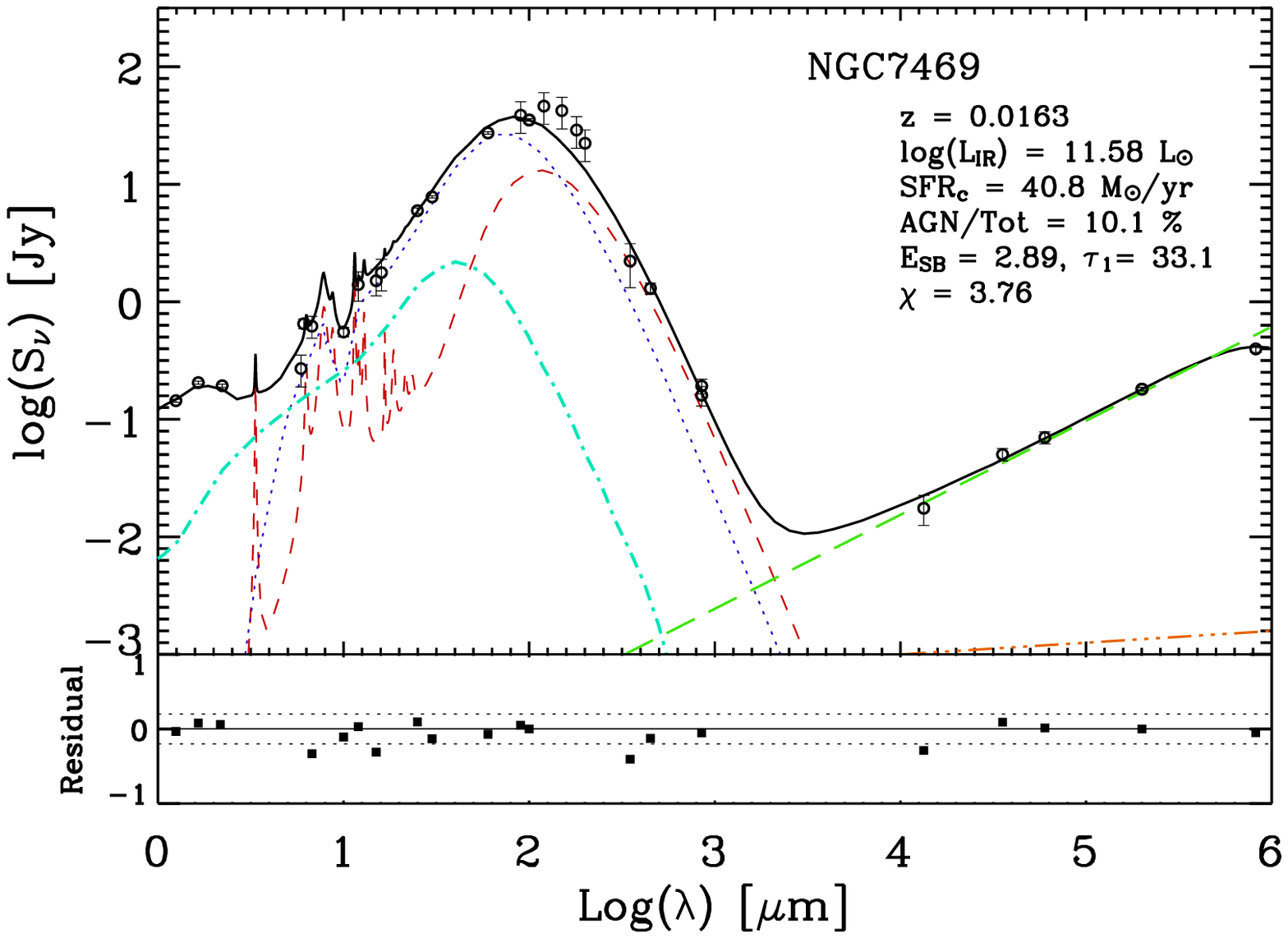}
} \centerline{
\includegraphics[scale = 0.42]{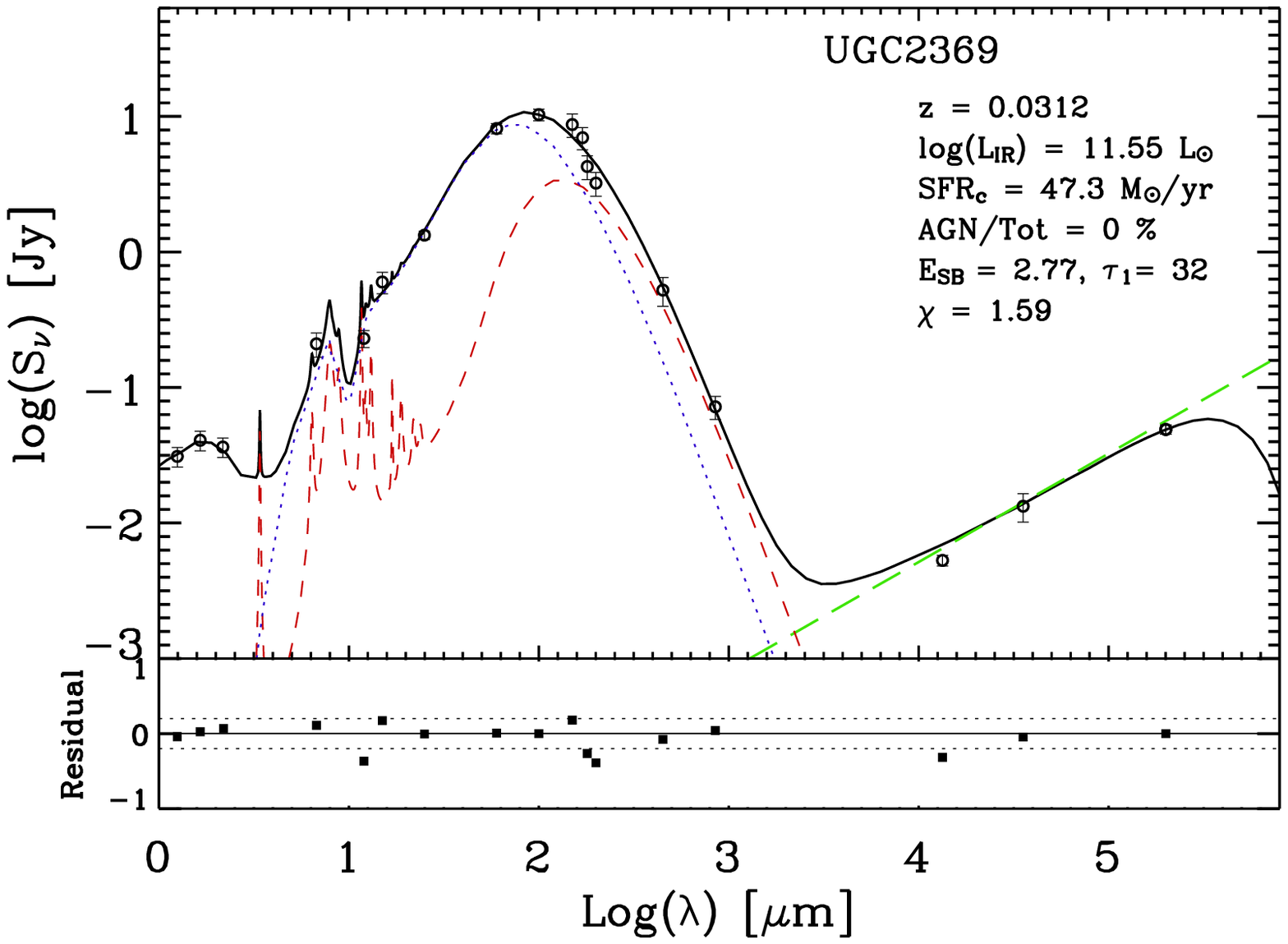}
\includegraphics[scale = 0.42]{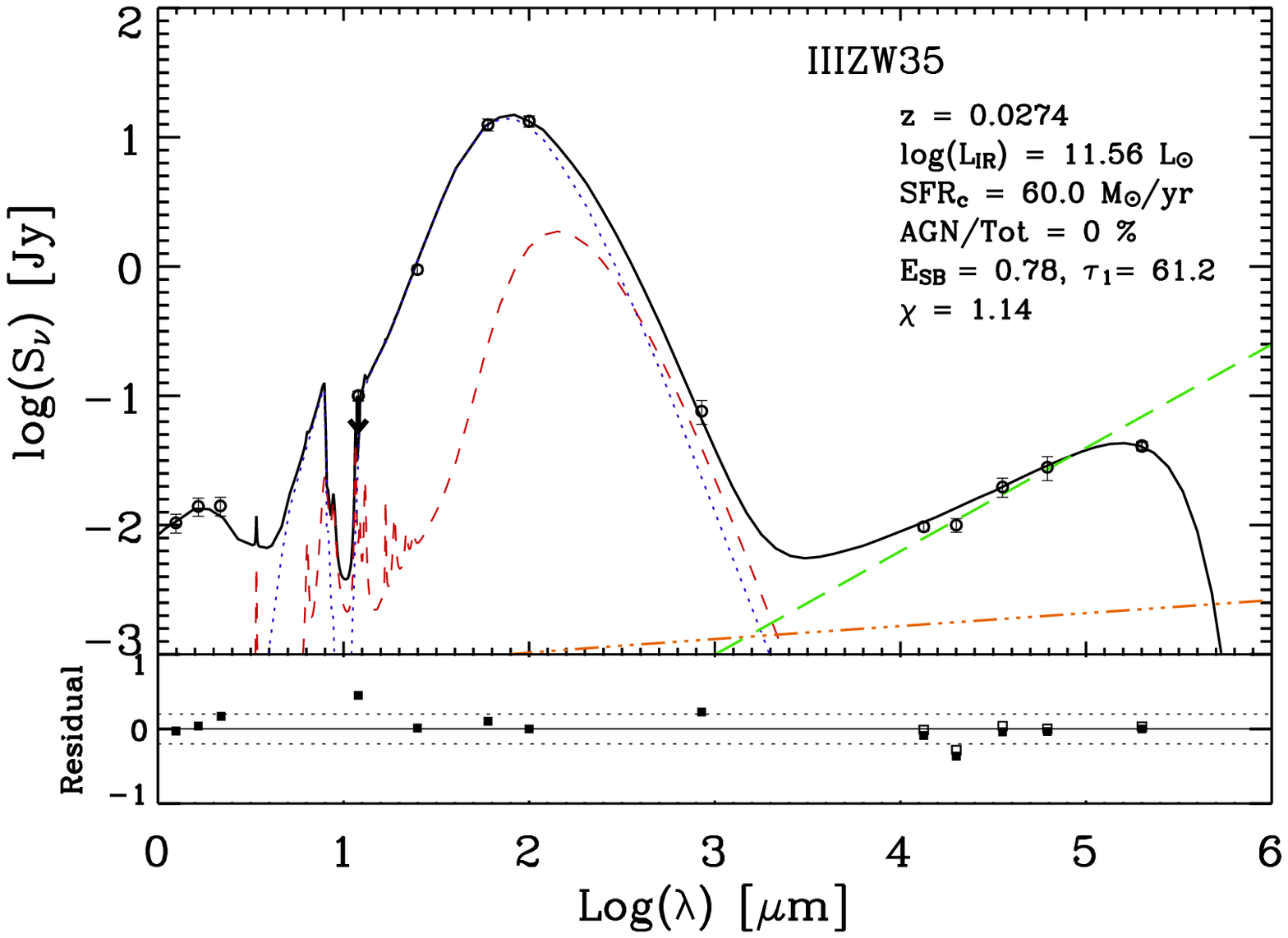}
} \caption{Continued. }
\end{figure*}

\begin{figure*}
\addtocounter{figure}{-1}

 \centerline{
\includegraphics[scale = 0.42]{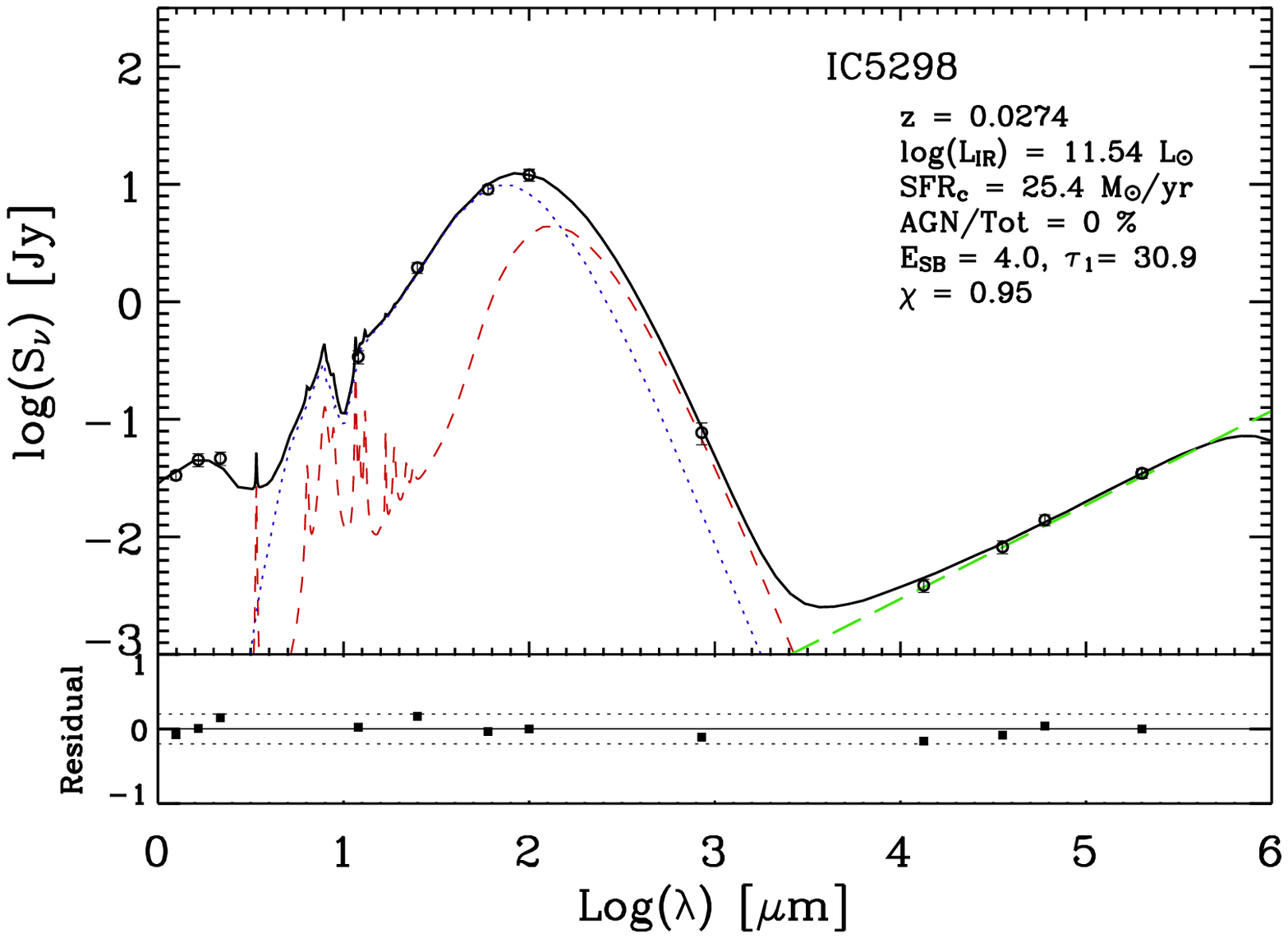}
\includegraphics[scale = 0.42]{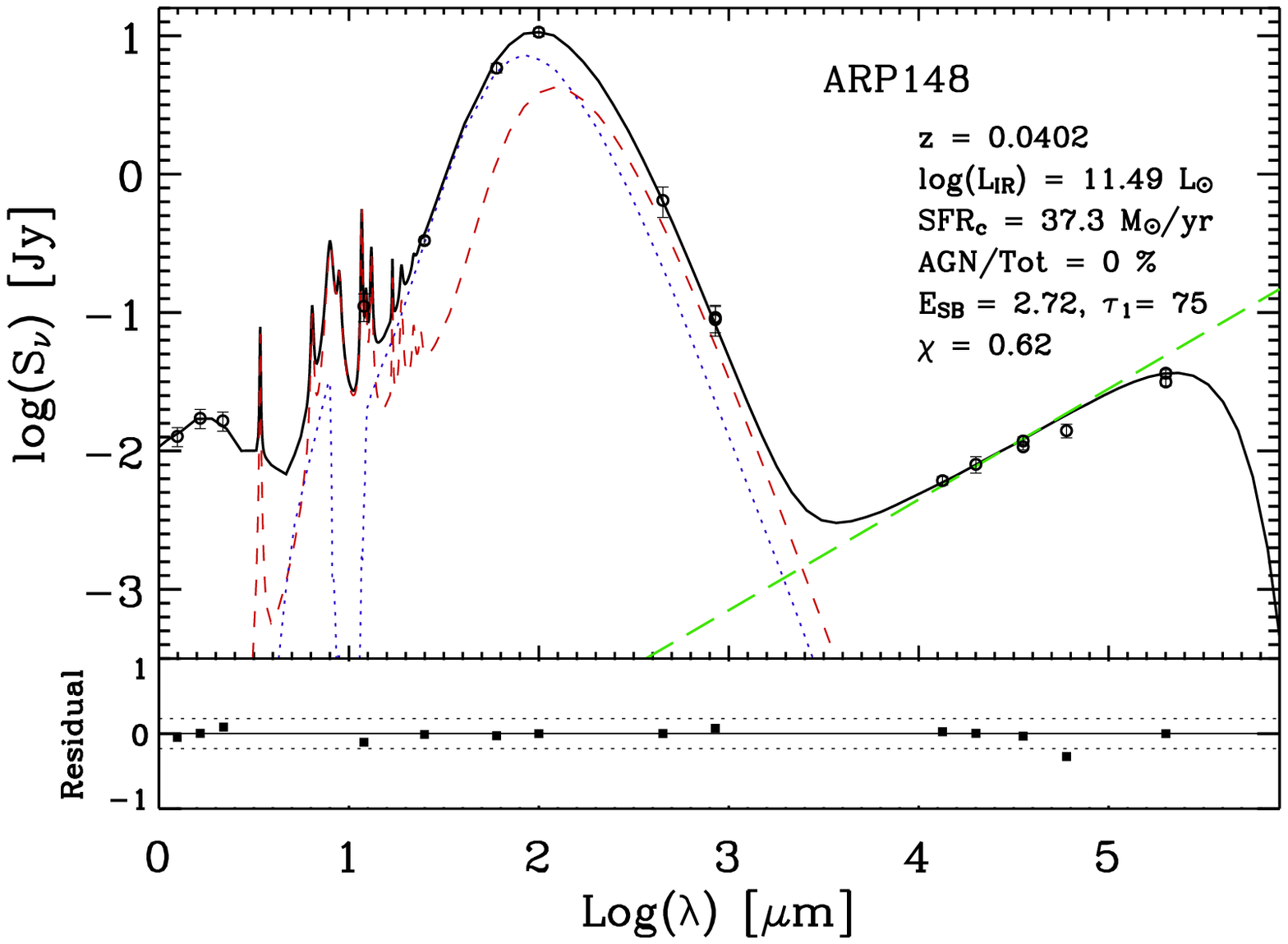}
}
 \centerline{
\includegraphics[scale = 0.42]{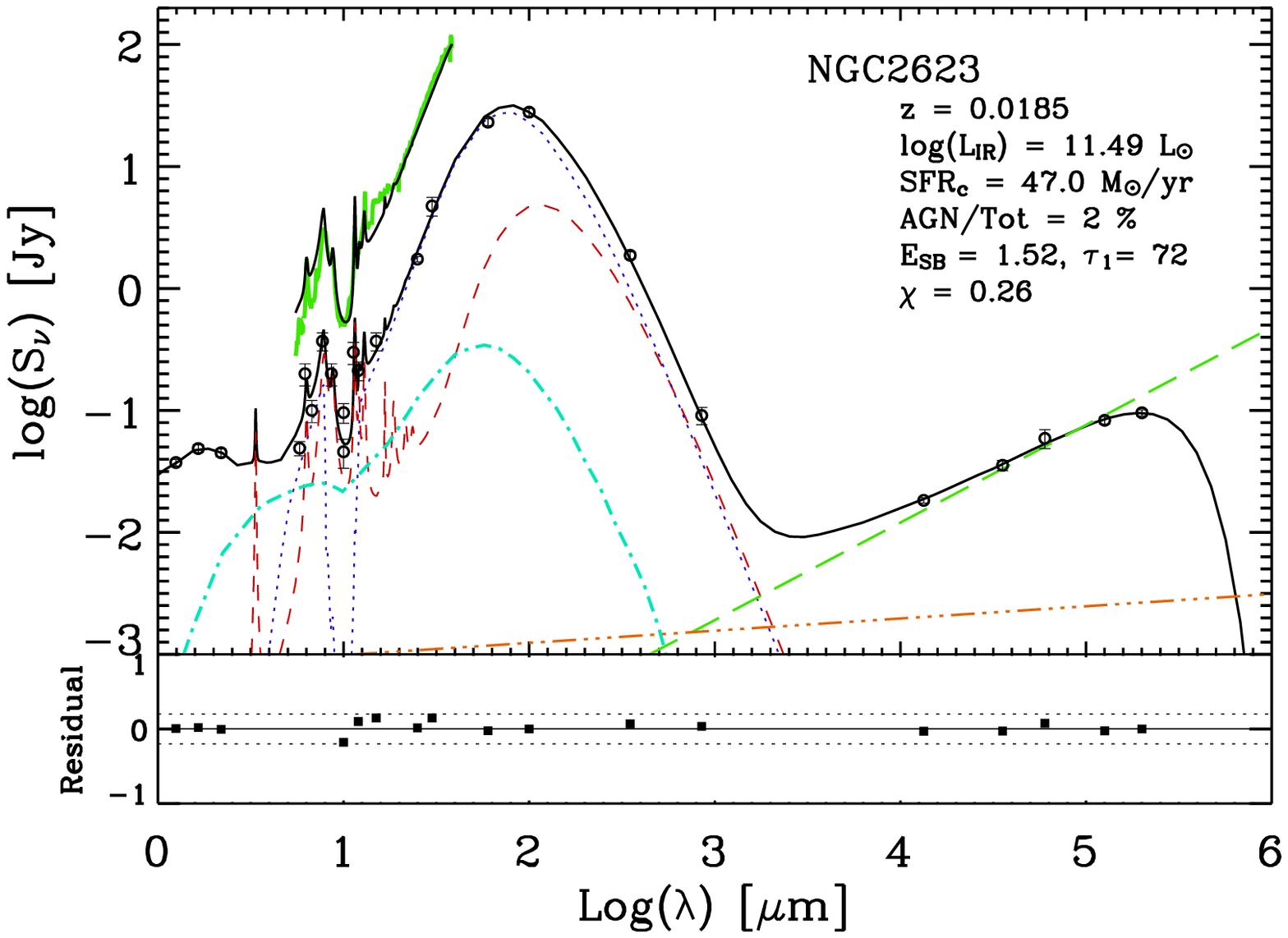}
\includegraphics[scale = 0.42]{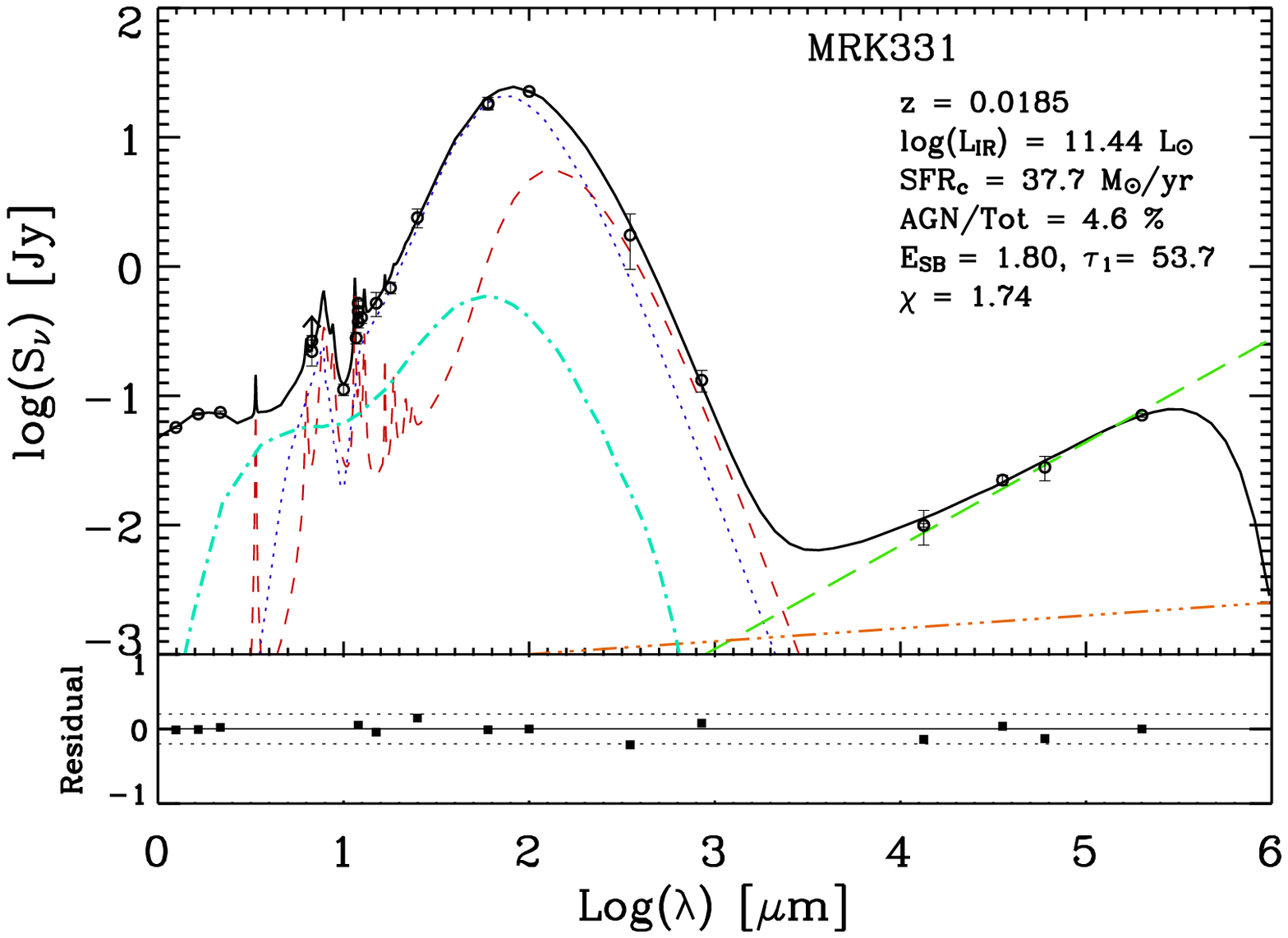}
} \centerline{
\includegraphics[scale = 0.42]{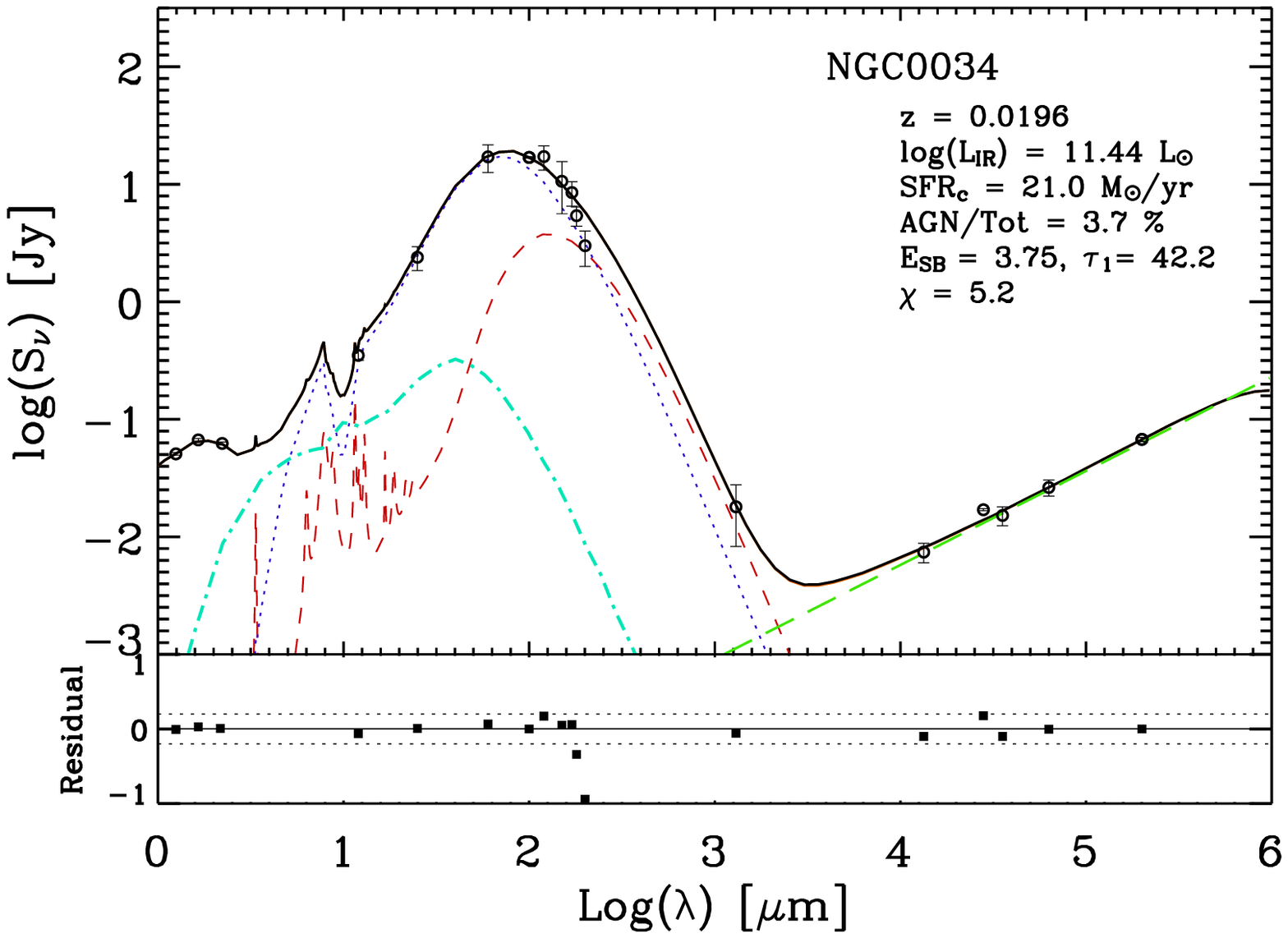}
\includegraphics[scale = 0.42]{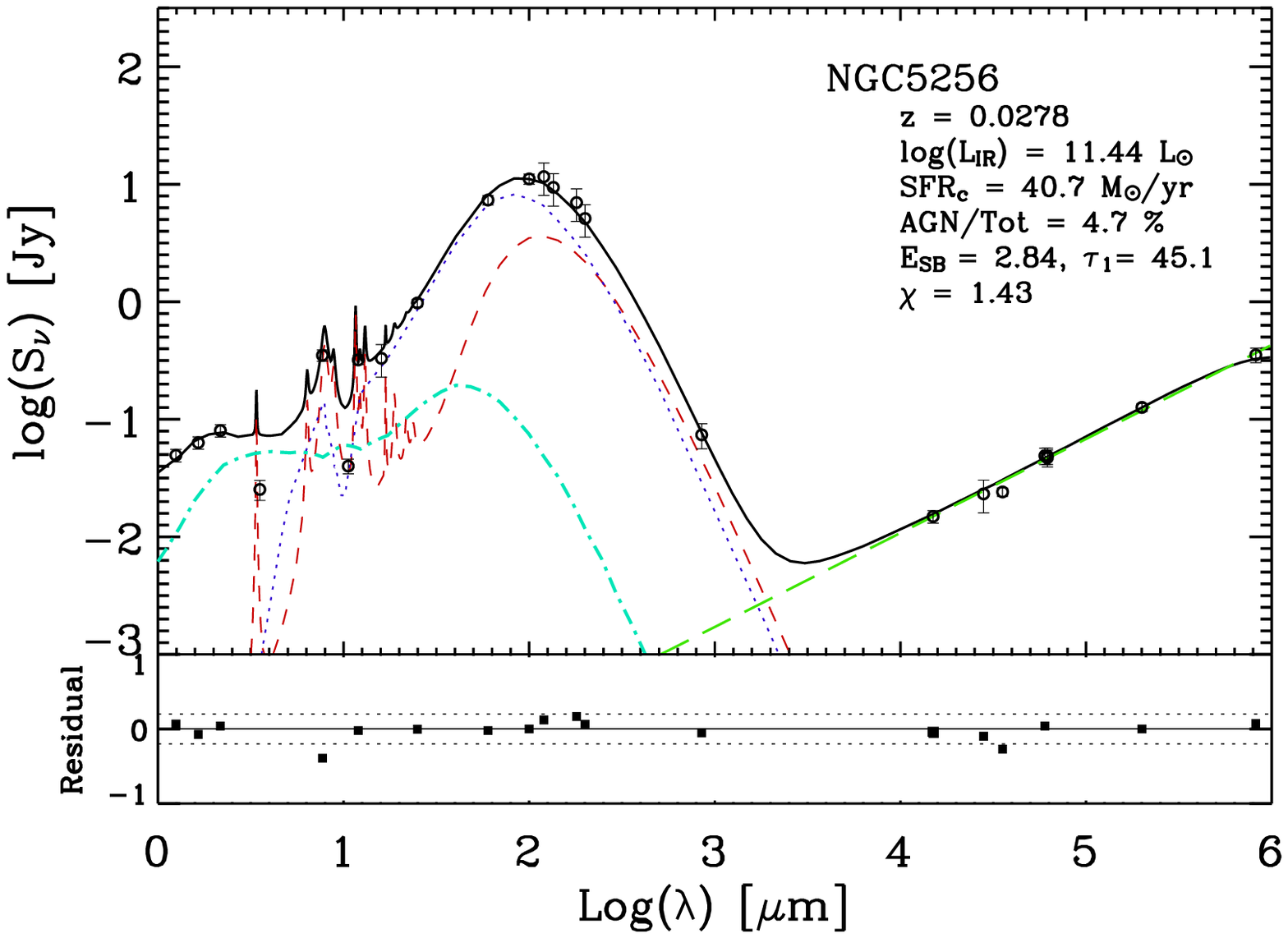}
}\centerline{
\includegraphics[scale = 0.42]{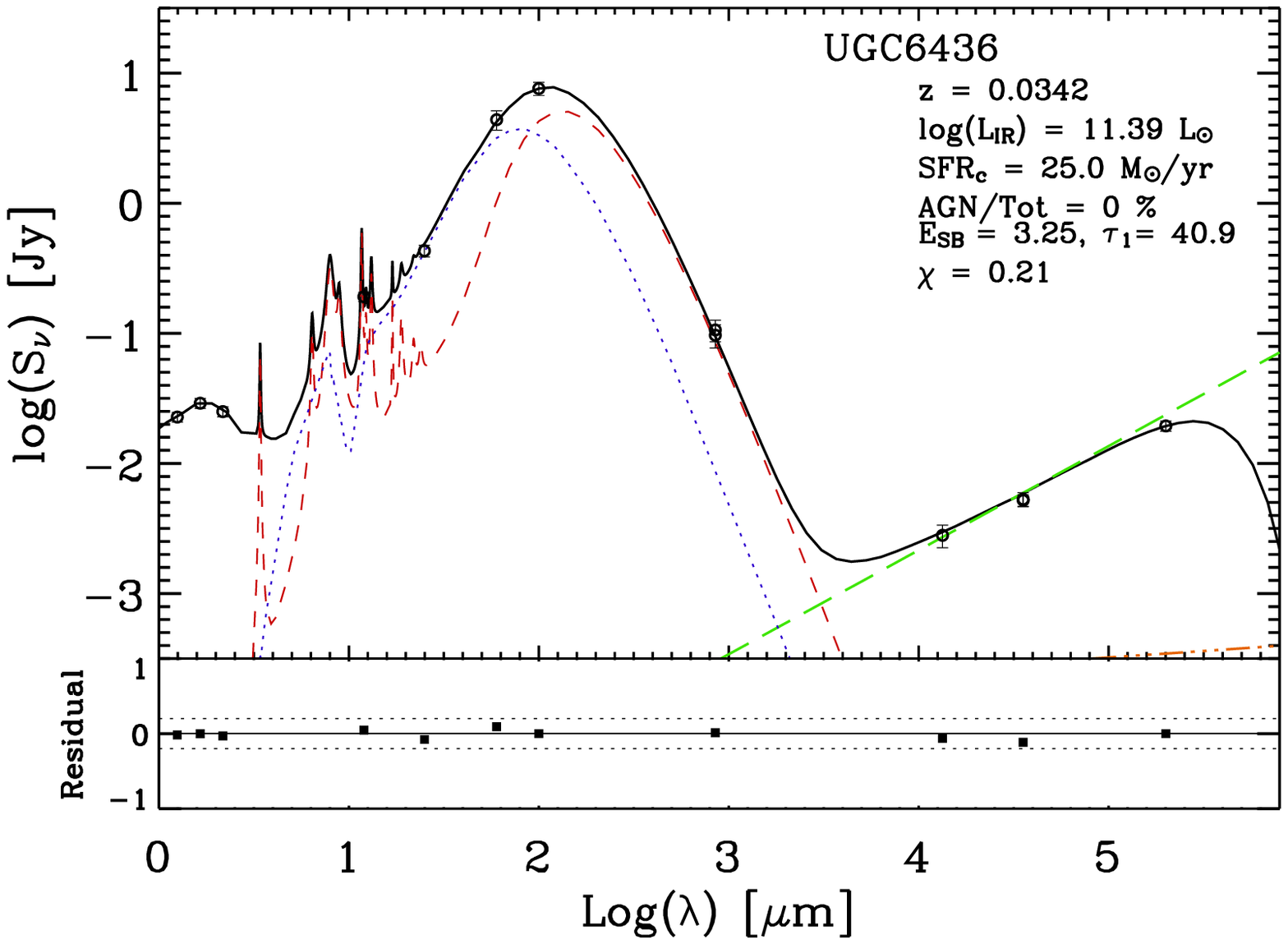}
\includegraphics[scale = 0.42]{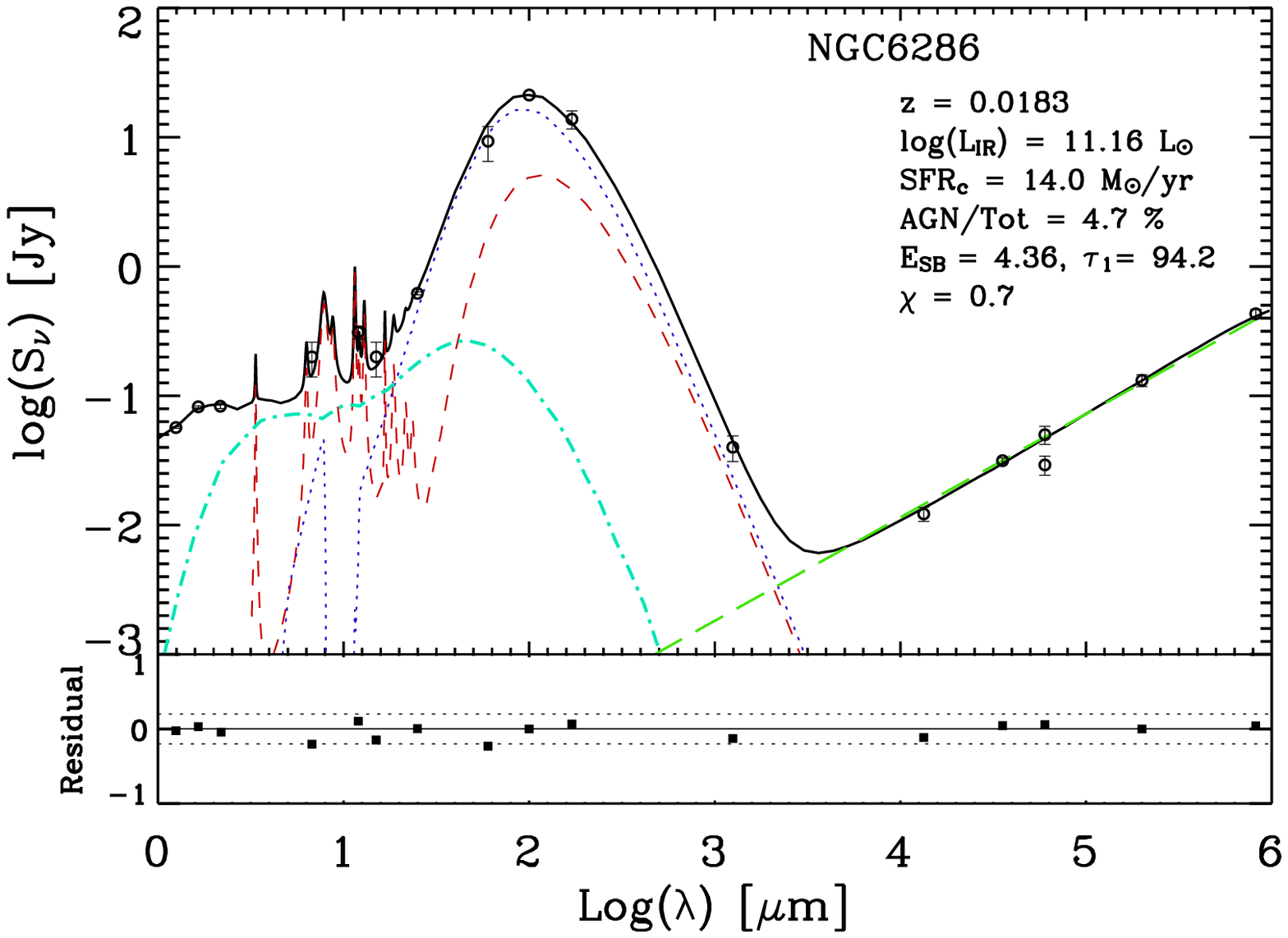}
}
  \caption{Continued.}
\end{figure*}

\begin{table}
\centering \caption{Parameters for the star formation history of the
model library. M$_{\rm{low}}$ and M$_{\rm{up}}$ are the adopted mass
limits of the Salpeter IMF; $\nu_{\rm{Sch}}$ is the efficiency of
the Schmidt-type SFR; \%$M_{\rm{b}}$ is the percentage of the mass
of gas involved in the burst at a galaxy age of 11.95 Gyr;
$t_{\rm{b}}$ indicates the e-folding time of the exponential burst
superimposed on the quiescent star formation  and age$_{\rm{b}}$ is
the time, measured from the beginning of the burst, at which the
SEDs are computed.} \label{tab:chevo}
\begin{tabular}{cccccc} \hline \hline
M$_{\rm{low}}$&M$_{\rm{up}}$&$\nu_{\rm{Sch}}$  &\%$M_{\rm{b}}$ &
$t_{\rm{b}}$ &
$\log$(age$_{\rm{b}}$) \\
$\rm{M_\odot}$&$\rm{M_\odot}$&(Gyr$^{-1}$) &$\rm{M_\odot}$&  (Myr) &
(yr)\\
\hline 0.1&100&0.5    &10      &  5 - 80 &  6.0 - 8.5 \\
\hline
\end{tabular}
\end{table}

\begin{table*}
\centering
 \caption{GRASIL parameters for the starburst SED library. $G/D$ is the
 gas to dust mass ratio; $\beta$
 is the dust emissivity index;
 $t_{\rm{esc}}$ is the escape time of newly born stars from their parent MCs;   $f_{\rm{mol}}$
 is the fraction of gas in MCs; $\tau_{1}$ is the $1 \mu$m optical depth of MCs, and $r$ is the
 core radius of the King profile for the dust-star distributions. }
  \label{tab:gra}

\begin{tabular}{cccccc} \hline
\hline

 $G/D$ & $\beta$&$t_{\rm{esc}}$ & $\tau_{1}$&$f_{\rm{mol}}$ &r\\
     &  &(Myr)
   & &&(kpc)\\
  \hline
 $100$ &1.5 - 2&5 - 80 & 10 - 180 &0.05 - 0.9 & 0.01 - 1\\

\hline
\end{tabular}
\end{table*}

\begin{table*}
\centering
 \caption{List of parameter values  for the AGN  model library.
$\Theta$ is the view angle, $r_{\rm{m}}/r_{\rm{0}}$ is the maximum
radius of the torus, $\alpha$ and $\beta$ are model parameters of
the dust density distribution;  $\Theta_h$ is is the half opening
angle of the dust free region;    $a_{\rm{max}}$ is maximum radius
of the dust grain, and $\tau_{e,9.7}$ is the equatorial optical
thickness of the dust in the torus.}
  \label{tab:agn}

\begin{tabular}{ccccccc} \hline
\hline

 $\Theta$ & $r_{\rm{m}}/r_{\rm{0}}$  & $\beta$&$\Theta_h$& $\alpha$&$a_{\rm{max}}$& $\tau_{\rm{e,9.7}}$\\
   (deg)  &  && (deg)  &
   & ($\mu$m)& \\
  \hline
 $1-90$ &30, 100, 300, 1000, 3000 & 0, 0.5, 1& 20, 30, 40 &0, 3, 6&0.3, 0.5, 1& 1, 3, 10\\

\hline
\end{tabular}
\end{table*}

\begin{table*}
\centering \caption{Best fit parameters for the starburst model, see
Section \ref{sec:results} for details. $E_{\rm{SB}}={\rm
Age_{\rm{b}}}/t_{\rm{b}}$ is the parameter which quantifies the
evolutive phase of the SB. Log(L$_{\rm{IR}}^{\rm{SB}}$) is the total
(8 - 1000 $\mu$m) IR luminosity calculated by integrating the
starburst component of the SED model. L$_{\rm{AGN}}$/L$_{\rm{Tot}}$
is the percent contribution of the AGN to the IR luminosity.
Galaxies marked with an $^*$ have warm IR colors,
$f_{25}/f_{60}>0.2$. Galaxies marked with a $^\dag$ have NIR colours
J-K$>1.7$, and those marked with a $^\S$ have available
\emph{Spitzer} IRS spectra.   The last row gives an estimate of the
errors on the parameters as described in Sect. \ref{sec:results}. }
\label{tab:fitpara}
\begin{tabular}{lccccclc|lc|c} \hline \hline
NAME
&$t_{\rm{b}}$&log($age_{\rm{b}}$)&$\beta$&$t_{\rm{esc}}$&$\tau_{\rm{1}}$&
$f_{mol}$&    r& $E_{\rm{SB}}$&Log(L$_{\rm{IR}}^{\rm{SB}}$)&L$_{\rm{AGN}}$/L$_{\rm{8-1000}}$\\
&(Myr)&(yr)&&(Myr)&&&(kpc)&&(L$_\odot$)&$\%$\\
 \hline
UGC~8058$^{*,\dag,\S}$&35&7.572&2.00&26&33.13&0.65&0.40&1.07&12.35&16\\
IR~14348-1447$^{\dag}$&10&7.170&1.95&14&49.09&0.26&0.27&1.48&12.17&8\\
IR~12112+0305&    25&6.880&2.00&7& 45.80&     0.35&      0.20&0.30&12.24&--\\
IR~05189-2524$^{*,\dag,\S}$&10&6.750&2.00&6&27.37&0.40&0.65&0.56&11.93&17\\
UGC~9913$^\S$&           35&7.340&1.80&20&64.93&     0.14&      0.55&0.62&12.10&--\\
IR~08572+3915$^{*,\S}$& 7&6.590&2.00&3&64.93&0.95&0.25&0.56&11.76&$47$\\
UGC~8696$^\S$& 20&7.650&1.95&37&56.72&0.20&0.37&2.23&11.99&$11$\\
IR~15250+3609$^\S$&    20&7.289&2.00&16& 53.66&     0.47&      0.70&0.97&11.85&$20$\\
IR~10565+2448$^\S$&  45&7.910&2.00&60&      47.43&     0.29& 0.14&1.80&11.94&--\\
UGC~5101$^\S$& 7&7.440&1.95&22&50.85&0.18&0.20&3.93&11.88&1\\
IZW~107 &      30&7.770&2.00&40&33.13&0.17&0.27&1.96&11.79&--\\
IR~01364-1042& 10&7.290&2.00&14&116.1&0.53&0.50&1.95&11.67&$11$\\
IR~10173+0828&  40& 7.200&2.00&14&42.22&0.27&0.41&0.41&11.68&--\\
Arp~299$^*$&25&7.756&2.00&43&31.75& 0.17&  0.70&2.28&11.68&--\\
UGC~4881&      40&7.986&2.00& 44&49.97&     0.22&       0.20&2.42&11.64&-- \\
CGCG~436-30$^\S$&  15&7.500&2.00&24&45.09&0.15&0.60&2.11&11.55&$5$\\
IC~1623&      25&7.635&1.80& 24&39.63&     0.17&      0.51&1.73&11.60&-- \\
NGC~1614$^*$&  20&7.840&2.00&50&33.13& 0.17&0.30&3.46&11.48&$18$ \\
UGC~8387$^\S$&   20&7.720&        1.95&33&      63.65&     0.28&      0.20&2.62&11.55&--\\
NGC~7469$^{*}$&  30&7.938&2.00&45&33.13&0.16&0.43&2.89&11.55&$10$\\
UGC~2369 & 25&7.840&2.00& 36&      31.70&     0.12&      0.67&2.77&11.55&-- \\
IIIZW~35& 45& 7.545&1.95&35&61.20&     0.18&     0.55&0.78&11.50&--\\
IC~5298$^{*}$& 20&7.900&2.00&60&30.88&0.10&0.42&4.00&11.53&--\\
Arp~148 & 20&7.735& 1.80&27&75.07&     0.20&      0.24&2.72&11.49&--\\
NGC~2623$^\S$&35&7.726&2.00& 30&71.94&0.32&  0.35&1.52&11.45&$2$ \\
Mrk~331&45&7.910&2.00&35&53.66&0.14&0.70&1.80&11.38&$5$\\
NGC~34&  20&7.875&2.00&65&42.23&0.14&0.45&3.75&11.39&$4$\\
NGC~5256&15&7.630&1.75&15&45.09&0.30&0.60&2.84&11.41&5 \\
UGC~6436&       30& 7.990&1.90&34&40.89&     0.05&     0.37&3.25&11.39&--\\
NGC~6286& 10&7.640&1.80&23&      94.25&     0.50& 0.47&4.36&11.20&$5$\\

 \hline
Error (\%)&13.0&23.0&-&36.0&14.0&24.0&-&12.0&6.0&-\\

\hline
\end{tabular}
\end{table*}

\begin{table*}
\footnotesize \centering \caption{Derived quantities from the
best-fit starburst models: $\tau_{\rm{ff}}$ is the free-free opacity
at  1.4 GHz, $C$ is the fraction, in \%, of the total energy of the
starburst reprocessed by the cirrus component. $M_{\rm{gas}}$,
$M_{\rm{D}}$, $M_{\rm{den}}$, $M_\star^{b}$, and $M_{\rm{T}}$ are
the logarithms of the final gas mass, the dust mass, the dense
molecular gas mass, the mass of stars created during the burst, and
the total mass of the galaxy calculated as $M_{\rm{gas}}+
M_{\rm{D}}+ M_\star^{tot}$. $SFR_{\rm{c}}$ is the star formation
rate averaged over the last 10 Myr, while $\langle SFR\rangle$ is
the SFR averaged over the age of the burst. T$_{\rm{th}}$ is the
percentage of thermal emission to the 22~GHz flux.  The last row
gives an estimate of the  errors on the parameters as described in
Sect. \ref{sec:results}.} \label{tab:derived}
\begin{tabular}{lrrrrrrrrrrr} \hline \hline

NAME &$\tau_{\rm{ff}}$&$C$&$M_{\rm{gas}}$&$M_{\rm{D}}$&$M_{\rm{den}}$& $M_\star^{b}$& $M_{\rm{T}}$&$\langle SFR\rangle$&$SFR_{\rm{c}}$&T$_{\rm{th}}$\\
&&$\%$&(M$_\odot$)&(M$_\odot$)&(M$_\odot$)&(M$_\odot$)&(M$_\odot$)&($\frac{M_\odot}{yr}$)&($\frac{M_\odot}{yr}$)&$\%$\\
 \hline
UGC~8058      & 0.03 & 3.0  & 10.26 & 8.26& 10.08 & 10.40& 11.41 & 681.5 & 435.8 &  20.7\\
IR~14348-1447 & 2e-3 & 12.5 & 10.44 & 8.44 &9.86 &  9.86 & 10.95 & 492.3 & 336.3 &  20.7\\
IR~12112+0305 & 0.28 & 13.8 &  10.44 & 8.44 &9.98 & 9.59 & 11.07 & 508.3 & 502.9 &  50.4 \\
IR~05189-2524 & 0.02 & 5.0 & 10.02 & 8.02 & 9.62 & 9.18& 10.51 & 260.8 & 225.8  &  66.4 \\
UGC~9913      & 0.56 &7.0   &  10.65 & 8.65 &9.79 & 9.80 & 11.10 & 288.3 & 235.1 &  33.2\\
IR~08572+3915 & 0.58 & 1.9 & 9.16  &7.16  &9.14  &8.90&10.13 &  205.5& 129.8& 78.9 &\\
UGC~8696      & 0.01 & 7.7  &  10.41 & 8.41 &9.71 & 10.25 & 11.19 & 400.8 & 137.8 &  19.3\\
IR~15250+3609 & 1.08 & 2.0  &  9.77  & 7.77 &9.44 & 9.61 & 10.67 & 209.8 & 154.5 &  29.6 \\
IR~10565+2448 & 7e-3 & 25.0 &  10.29 & 8.29 &9.75 & 10.36& 11.27 & 281.5 & 116.3 &  25.4\\
UGC~5101      & 0.13 & 20.9 &  10.55 & 8.55 &9.81 & 10.10& 11.08 & 454.8 & 71.5  &  4.8 \\
IZW~107       &0.03  & 13.0  & 10.24 & 8.24 &9.50 &  10.12& 11.06 & 225.8 & 87.2  &  22.5\\
IR~01364-1042 & 0.76 & 3.7 &  9.66  & 7.66 &9.38 & 9.55 & 10.50 & 180.7 & 91.7  &  26.0 \\
IR~10173+0828 & 0.67 & 7.0  & 10.06 & 8.06 & 9.36 & 9.28 & 10.66 & 120.3 & 109.7 &  41.7 \\
Arp~299       & 0.03 & 5.3  &  10.19  & 8.19 &9.41 & 10.07 & 11.00 & 207.4 & 67.3 &  15.6\\
UGC~4881      & 1e-3 & 21.0 &  10.05 & 8.05 &9.39 & 10.25& 11.12 & 184.0 & 53.8  &  24.7\\
CGCG~436-30   & 0.39 & 6.4 & 10.02   & 8.02 &9.20 & 9.67 & 10.65 & 146.5 & 58.0  &  23.1\\
IC~1623       & 4e-4 & 10.8 &  10.18 & 8.18 &9.41 & 9.78 & 10.82 & 139.2 & 65.5  &  19.4\\
NGC~1614      & 0.24 & 11.6  &  9.89  & 7.89 &9.12 & 10.09 & 10.95 & 179.6 & 29.4 & 18.1\\
UGC~8387      & 0.31 & 17.1 &  10.07 & 8.07 &9.52 & 9.93 & 10.84 & 160.9 & 43.5  &  14.8\\
NGC~7469 & 0.02 & 14.3 & 10.04  & 8.04 & 9.24 &10.20  & 11.06 &181.2  & 40.8  &  20.1\\
UGC~2369      & 0.13 & 11.4 &  10.28 & 8.28 &9.36 & 10.13 & 11.04 & 194.1 & 47.3  &  23.7\\
IIIZW~35      & 0.64 & 2.3  &  10.07 & 8.07 &9.33 & 9.46 & 10.64 & 81.9  & 60.0  &  29.1\\
IC~5298 & 0.03 & 12.8 & 10.28  & 8.28 &  9.28 &10.21 &11.06  & 206.6 & 25.5  &  26.8\\
Arp~148       & 0.32 & 21.6   &  10.06 & 8.06 &9.36 & 9.90 & 10.82 & 146.2 & 37.3 & 16.8\\
NGC~2623      & 0.42 & 8.7  &  9.77  & 7.77 &9.28 & 9.70 & 10.66 & 95.3  & 47.0  &  23.1\\
Mrk~331       & 0.17 & 9.9  &  10.08 & 8.08 &9.23 & 9.87 & 10.82 & 91.3  & 37.7  &  31.8 \\
NGC~34        & 0.01 & 7.8 &  9.93 & 7.93 &9.08 &  10.05 & 10.91 & 150.7 & 21.0  &  18.3\\
NGC~5256      &0.01  & 15.0 &  9.77  & 7.77 &9.25 & 9.85& 10.73 & 164.7 & 40.7  &  11.0\\
UGC~6436      & 0.19 & 27.1 &  10.36 & 8.36 &9.07 & 10.12& 11.04 & 135.3 & 25.0  &  22.2\\
NGC~6286      & 5e-3 & 13.6 &  9.65  & 7.65 &9.35 & 9.77 & 10.59 & 136.1 & 14.0  &  5.2 \\

 \hline
Errors (\%)&-&-&56.0&56.0&17.0&22.0&8.0&3.0&7.0&-\\
\hline \normalsize
\end{tabular}
\end{table*}

\begin{table*}
\centering
 \caption{AGN fitted parameters. See text for details.}
  \label{tab:agnfit}
\begin{tabular}{lccccccr|r} \hline
\hline

NAME&$\Theta$ & $r_{\rm{m}}/r_{\rm{0}}$ & $\beta$ & $\Theta_{\rm{h}}$& $\alpha$&$a_{\rm{max}}$&$\tau_{\rm{e,9.7}}$&$r_{\rm{m}}^{\rm{(a)}}$\\
   &(deg)  &    & &(deg)
   & &($\mu$m)& &(pc)\\
  \hline
UGC~8058&40&1000&0.5&20&6.0&0.3&10.0& 68.98\\
IR~14348-1447&46&3000&0.5&40&0.0&0.3&3.0&99.00\\
IR~05189-2524&41&1000&0.0&40&0.0&0.5&10.0&47.29\\
IR~08572+3915&77&1000&1.0&40&6.0&0.3&10.0&62.48\\
UGC~8696&74&1000&0.5&40&6.0&1.0&10.0&29.01\\
IR~15250+3609&90&300&1.0&40&0.0&0.3&10.0&10.38\\
UGC~5101&10&3000&1.0&40&3.0&0.3&3.0&56.44\\
IR~01364-1042&62&300&0.0&20&3.0&1.0&10.0&6.01\\
CGCG~436-30&40&1000&0.0&40&0.0&0.3&10.0&18.08\\
NGC~1614&74&1000&1.0&40&6.0&1.0&10.0&22.24\\
NGC~7469&21&1000&0.0&20&0.0&1.3&3.0&19.67\\
NGC~2623& 22&3000&0.5&20&0.0&0.3&3.0&22.44\\
Mrk~331&48&3000&0.5&40&6.0&0.3&10.0&32.05\\
NGC~34&60&1000&0.5&40&0.0&0.3&1.0&9.79\\
NGC~5256&39&3000&1.0&40&0.0&0.3&1.0&67.58\\
NGC~6286&40&3000&1.0&20&6.0&0.3&10.0&32.57\\
\hline
\end{tabular}

$^{\rm{(a)}}$ Maximum radius of the torus model. It was calculated
from the equation given in Sect. 3 in Granato \& Danese (1994) for
$r_{\rm{0}}$  and the value of the parameter $r_{\rm{m}}/r_{\rm{0}}$
given in Col. (3) of this Table.
\end{table*}
\begin{table*}
\centering
 \caption{Total (SB + AGN) luminosity from the SED models in different spectral windows and
the corresponding AGN contribution. Galaxies marked with an $^*$
have warm IR colors, $f_{25}/f_{60}>0.2$. Galaxies marked with a
$^\ddag$ have NIR colours J-K$>1.7$, and those marked with a $^\S$
have available  \emph{Spitzer} IRS spectra. }
  \label{tab:agnlum}
\begin{tabular}{lccccccc} \hline
\hline

NAME&L$_{\rm{1-5}}$/AGN &L$_{\rm{3-30}}$/AGN &L$_{\rm{8-1000}}$/AGN  &L$_{\rm{1-1000}}$/AGN&L$_{\rm{40-500}}$/AGN&L$_{\rm{3-1100}}$/AGN&L$_{\rm{bol}}$/AGN\\
   &(L$_\odot/\%$)  &(L$_\odot/\%$)     &(L$_\odot/\%$)
   &(L$_\odot/\%$) &(L$_\odot/\%$)&(L$_\odot/\%$)&(L$_\odot/\%$)\\
  \hline

UGC~8058$^{*,\ddag,\S}$&11.64/71&11.66/42&12.42/16&12.51/24& 12.18/7&12.48/21&12.51/24\\
IR~14348-1447$^{\ddag}$&10.75/56&11.48/27&12.21/8&12.23/10&12.07/6&12.23/10&12.23/10\\
IR~05189-2524$^{*,\ddag,\S}$&11.01/74&11.60/30&12.01/17&12.07/23&11.75/13& 12.05/20&12.20/23\\
IR~08572+3915$^{*,\ddag,\S}$ &10.88/91&11.64/81&12.04/47&12.09/52&11.77/30&12.09/52&12.09/52\\
UGC~8696$^\S$&10.68/14& 11.29/29&12.05/11&12.07/12&11.90/6&12.06/12&12.08/12\\
IR~15250+3609$^\S$ &10.45/30&11.37/49&11.94/20&11.97/21&11.73/8& 11.96/21&11.97/21 \\
UGC~5101$^{\ddag,\S}$&10.71/53&11.06/12&11.89/1&11.92/4&11.79/0&11.90/2&11.97/6\\
IR~01364-1042 & 10.13/16&10.85/60&11.73/11&11.74/12&11.61/3&11.74/12&11.75/12\\
CGCG~436-30$^\S$   & 10.38/42&10.92/14&11.57/5&11.61/7&11.41/3&11.59/6&11.71/10\\
NGC~1614$^*$&10.55/18&11.11/37& 11.58/18& 11.63/19&11.35/8& 11.61/20&11.65/18\\
NGC~7469$^*$&10.76/20& 11.06/25&11.55/10&11.63/12&11.36/4&11.59/12&11.70/13\\
NGC~2623$^\S$&10.25/15& 10.56/12&11.46/2&11.50/3&11.37/1&11.48/3&11.51/3\\
MRK~331&10.47/20&10.71/20&11.40/5&11.46/6&11.27/3&11.43/6&11.48/6\\
NGC~34&10.47/17&10.79/17&11.40/4&11.46/5&11.24/1&11.43/5&11.48/5\\
NGC~5256&10.79/45&10.91/25&11.43/5&11.54/12& 11.28/2& 11.47/8&11.71/16\\
NGC~6286&10.50/32&10.46/41& 11.22/5&11.31/9&11.15/1&11.26/8&11.33/9\\
 \hline
\end{tabular}
\end{table*}

\begin{table*}
\centering
 \caption{Total (SB + AGN) flux from the SED models in different IR spectral bands
 and
the corresponding AGN contribution. Galaxies marked with an $^*$
have warm IR colors, $f_{25}/f_{60}>0.2$. Galaxies marked with a
$^\ddag$ have NIR colours J-K$>1.7$, and those marked with a $^\S$
have available \emph{Spitzer} IRS spectra. For the latter galaxies
the lower row provides fluxes calculated from the \emph{Spitzer}
spectra in the IRAC 3 and IRAC 4 and MIPS 24$\mu$m bands. In the
IRAC bands, differences with models in the upper row, are mainly
attributable to effects of molecular absorption bands.}
  \label{tab:agnband}
\begin{tabular}{lccc|cccc|ccc} \hline
\hline

NAME&H&J&K&IRAC1&IRAC2&IRAC3&IRAC4&MIPS1&MIPS2&MIPS3\\
&1.23$\mu$m&1.66$\mu$m&2.16$\mu$m&3.55$\mu$m&4.5$\mu$m&5.73$\mu$m&8$\mu$m&24$\mu$m&70$\mu$m&160$\mu$m\\
&(Jy/$\%$)&(Jy/$\%$)&(Jy/$\%$)&(Jy/$\%$)&(Jy/$\%$)&(Jy/$\%$)&(Jy/$\%$)&(Jy/$\%$)&(Jy/$\%$)&(Jy/$\%$)\\
  \hline

UGC~8058$^{*,\ddag,\S}$&0.06/24&0.11/52&0.21/76&0.38/90&0.48/82&0.70/62&1.16/43&6.75/27&34.91/6&14.82/3\\
&&&&&& 0.70/62&0.85/64&6.70/27&\\

IR~14348-1447$^{\ddag}$&1.7E-3/5&2.5E-3/11&3.2E-3/34&0.01/83&0.02/79&0.04/55&0.08/31&0.57/18&6.92/6&4.22/3\\
IR~05189-2524$^{*,\ddag,\S}$&0.02/77&0.03/83&0.04/90&0.07/88&0.11/65&0.21/41&0.43/27&2.62/24&12.94/12&5.86/6\\
&&&&&&  0.15/57   &   0.28/41     &  2.78/23&\\
IR~08572+3915$^{*,\ddag,\S}$&1.7E-3/26&2.8E-3/44&3.7E-3/63&0.02/95&0.07/97&0.17/95&0.26/92&1.67/72&6.78/27&1.83/18\\
&&&&&& 0.19/85   & 0.34/71 & 1.54/78&\\
UGC~8696$^\S$& 0.02/2&0.03/4&0.03/11&0.02/34&0.03/38&0.08/27&0.20/22& 2.40/31&22.82/6&12.07/2\\
&&&&&& 0.07/27 &0.16/27  &2.10/35&\\
IR~15250+3609$^\S$&4.3E-3/0&5.5E-3/0&4.9E-3/0&3.7E-3/17&0.01/59&0.05/64&0.12/55&1.27/47&7.23/6&2.70/2\\
&&&&&& 0.04/80&0.10/66&1.25/48&\\
UGC~5101$^{\ddag,\S}$&0.02/38&0.03/51&0.03/63&0.03/64&0.03/58&0.07/26&0.20/8& 0.96/4&15.51/0&13.52/0\\
&&&&&&  0.09/20&0.14/11&0.93/4&\\
IR~01364-1042&4.0E-3/0&5.2E-3/1&4.9E-3/5&4.6E-3/31&4.7E-3/62&0.01/65&0.04/48&0.50/60&7.68/2&3.26/0\\
CGCG~436-30$^\S$&0.01/26&0.02/34&0.02/48&0.02/64&0.03/53&0.07/27&0.15/15&1.31/8&10.69/3&5.36/1\\
&&&&&& 0.04/47 &0.12/19 &1.31/8&\\
NGC~1614$^*$& 0.09/3&0.11/6&0.11/13&0.11/37&0.18/41&0.44/35&1.01/33&7.44/37&35.28/7& 18.72/2\\
NGC~7469$^*$&0.15/6&0.19/10&0.18/19&0.19/42& 0.23/47&0.45/32&1.09/19& 5.40/23&34.43/3&22.17/1\\
NGC~2623$^\S$& 0.04/2&0.05/5&0.05/13&0.04/37& 0.04/48&0.09/27&0.21/12& 1.73/7&29.29/1&15.04/1\\
&&&&&& 0.04/61 &0.14/18&1.70/7&\\
MRK~331& 0.06/1&0.07/5&0.07/18&0.08/50& 0.09/57& 0.16/35&0.35/17&2.05/11&22.51/2&14.04/1\\
NGC~34&0.05/1&0.07/3&0.06/12&0.06/47& 0.08/49& 0.15/32&0.29/22&2.40/8&18.13/1& 9.74/0\\
NGC~5256&0.05/24&0.07/34&0.08/49&0.08/64& 0.08/70& 0.12/43&0.34/16&0.98/12&9.76/1&7.66/0\\
NGC~6286&0.06/5&0.08/13&0.09/32&0.10/61&0.09/77&0.13/58&0.33/23&0.61/29&16.93/1&15.48/0\\

\hline
\end{tabular}
\end{table*}

\subsection{AGN models:}

A complete description of the AGN models and the numerical method
developed to compute the emitted SED can be found in Granato \&
Danese (1994) and Granato et al. (1997). Only the basic features are
summarized here.

Several papers  compared the  observed IR SED of AGN with
predictions from radiative transfer models within dusty tori, in
order to investigate the validity of the unified model and constrain
possible geometries for the obscuring structure (e.g. Pier et al.
1993, Granato et al. 1994, Efstathiou et al. 1995, Granato et al.
1997, Alonso-Herrero et al. 2001, and Nenkova et al. 2002).

The Granato \& Danese (1994) code solves, through an iterative
numerical scheme, the transfer equation of the radiation that
originates from the central optical-UV source, in a general
axisymmetric dust distribution. The dust is composed of a mixture of
grains, in thermal equilibrium with the radiation field, and extends
out to a maximum radius $r_{\rm{m}}$.

In the absence of precise physical ideas concerning the structure of
the obscuring torus, several geometries -- and associated
free-parameters -- are plausible and indeed have been investigated
in the papers quoted above (flared discs, tapered discs, cylinders,
with and without substantial clumping etc).

A problem which is usually overlooked in this kind of study, is that
dust optical properties around AGN are likely to be, to some extent,
peculiar.  Indeed, Maiolino et al. (2001a), Maiolino et al. (2001b)
presented evidence for 'anomalous' properties of dust in AGN. Here
the term 'anomalous' is relative to the standard properties of dust
grains thought to be responsible for the average Milky Way
extinction law and cirrus emission. It is not surprising that the
properties of dust grains might be very different  in the dense and
extreme environment of an AGN. In particular, dust distribution
biased in favor of large grains, ranging up to 1-10 $\mu$m, much
larger than the usual cut at $\sim 0.3 \mu$m of Mathis (1977) type
models (e.g. Draine 1984, Silva et al. 1998), have been invoked.

Thus, we have adopted several different geometries, but also
investigated models in which the size distribution of grains extends
to radii, $a_{max}$, larger than the standard value. Of course, the
uncertainty on optical properties of dust further limits the already
moderate success of this approach in constraining the geometry (see
below).

The SED fitting is performed through comparison with libraries of
models. Each library consists of several hundreds of models
belonging to a given "geometry class", in which typically 4-5
parameters are varied assigning  3-4 different values to them over a
quite wide range. In particular we have considered  anisotropic
flared discs (for definitions, see Efstathiou et al. 1995). The
geometry consists of a structure whose height above the equatorial
plane $h$ increases linearly with the distance in the equatorial
plane $r$ (Fig.\,1a in Efstathiou et al. 1995). Therefore the
dust-free region is exactly conical, with half-opening angle
$\Theta_h$. We also introduce (at variance with respect to
Efstathiou 1995) a dependence of the dust density $\rho$, and
therefore of the radial optical depth, on the polar angle $\Theta$.
The general form used for this dependence and that on the spherical
coordinate $r$ is:
\begin{equation}
\rho(r,\Theta) \propto r^{-\beta}  \exp(-\alpha\, \cos^2 \Theta)
\end{equation}
where $\alpha$ and $\beta$ are model parameters.

The two additional parameters used to fully characterize the models
 are the ratio between the outer and inner (i.e.
sublimation) radii $r_{\rm{m}}/r_{\rm{0}}$, and the equatorial
optical depth to the nucleus at $9.7 \mu$m,  $\tau_{e,9.7}$

The parameter values used in  the libraries are:
$$r_{\rm{m}}/r_{\rm{0}}=30, 100, 300, 1000, 3000$$
$$\beta=0, 0.5, 1$$
$$\Theta_{h}=20,30,40 \, \mbox{\degr}$$
$$\alpha=0, 3, 6$$
$$a_{max}=0.3, 0.5, 1.0 \, \mu\mbox{m}$$
$$\tau_{e,9.7}=1, 3, 10$$

All possible combinations of parameter values are considered in the
libraries. For each corresponding model we compute the SED for
viewing angles from 0\degr~to 90\degr~in steps of 10 degrees.

The  impacts of changing various parameters on the predicted SED
have been explored in Granato et al. (1994) and Granato et al.
(1997). Here we recall only that, roughly speaking,
$r_{\rm{m}}/r_{\rm{0}}$ is related to the width of the IR bump,
whilst $\tau_{e,9.7}$ controls mainly the NIR slope of the SEDs, as
observed from obscured directions, as well as its anisotropy.

Indeed, the global result of the various work is that the nuclear
SED alone does not allow a strong constraints on the precise
geometry of the torus. Several individual analysis have been too
optimistic in this respect, due essentially to an insufficient
exploration of the parameter space.  Some constraints can be put on
the torus extension and on the optical thickness, and good fits are
obtained with flared disc models.

However, we point out that in this paper our aim is just to asses
the possible contribution of the warm AGN component to the total SED
of ULIRGs, without any pretence of discriminating the precise torus
characteristics.
\begin{figure}[h]
\includegraphics[scale = 0.48]{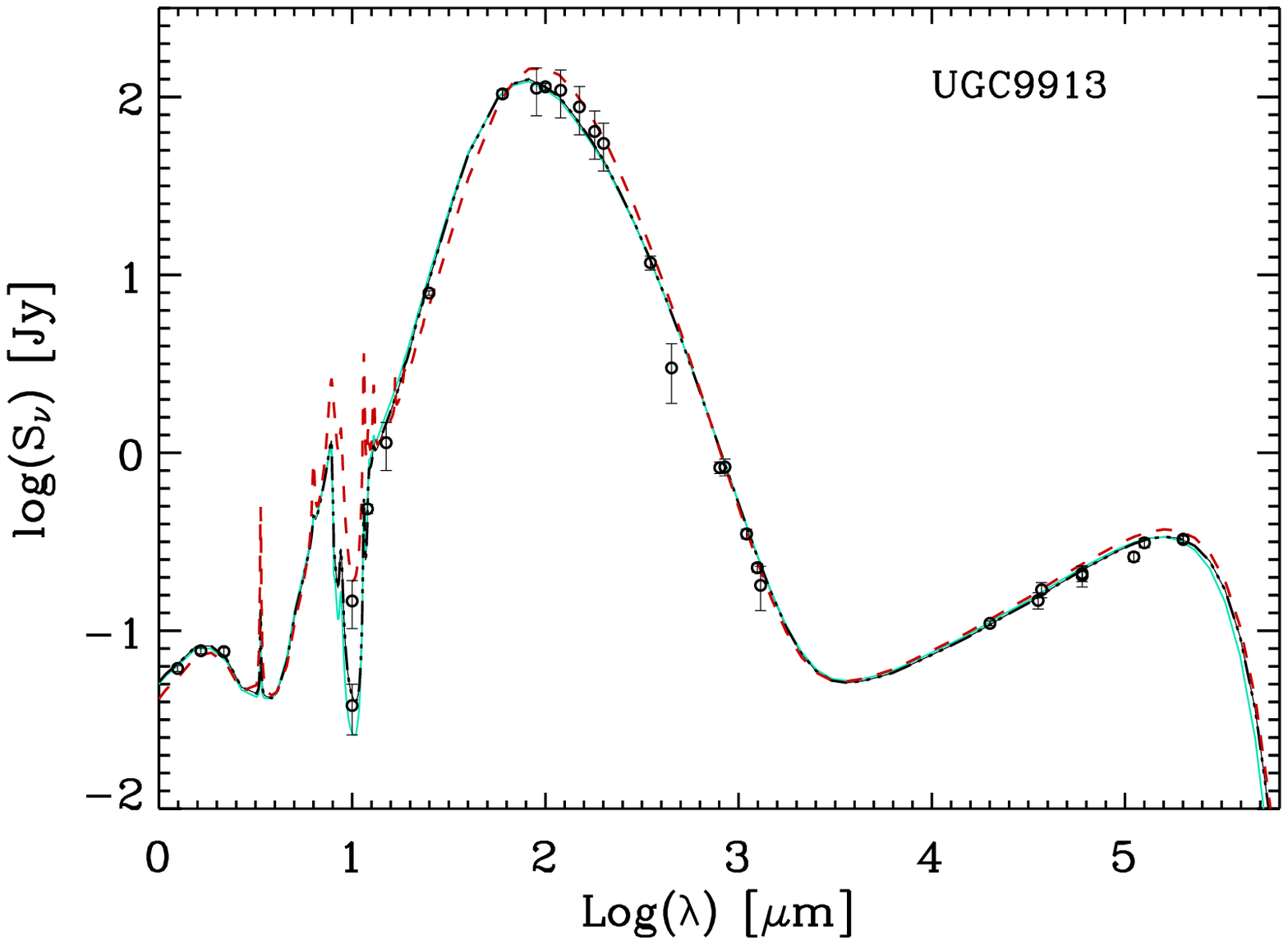}

\caption{Fits to the UGC~9913 observed and simulated SEDs performed
to estimate the errors in the different model parameters and derived
quantities. The open circles represent the observational data. The
three dots-dashed line is the best fit to the data while the thin
solid line and the dotted line represent the best fits to two
simulated data-sets drawn randomly from the observed values $\pm$ a
full standard deviation.} \label{Fig:error}
\end{figure}

\begin{figure}[h]
\includegraphics[scale = 0.48]{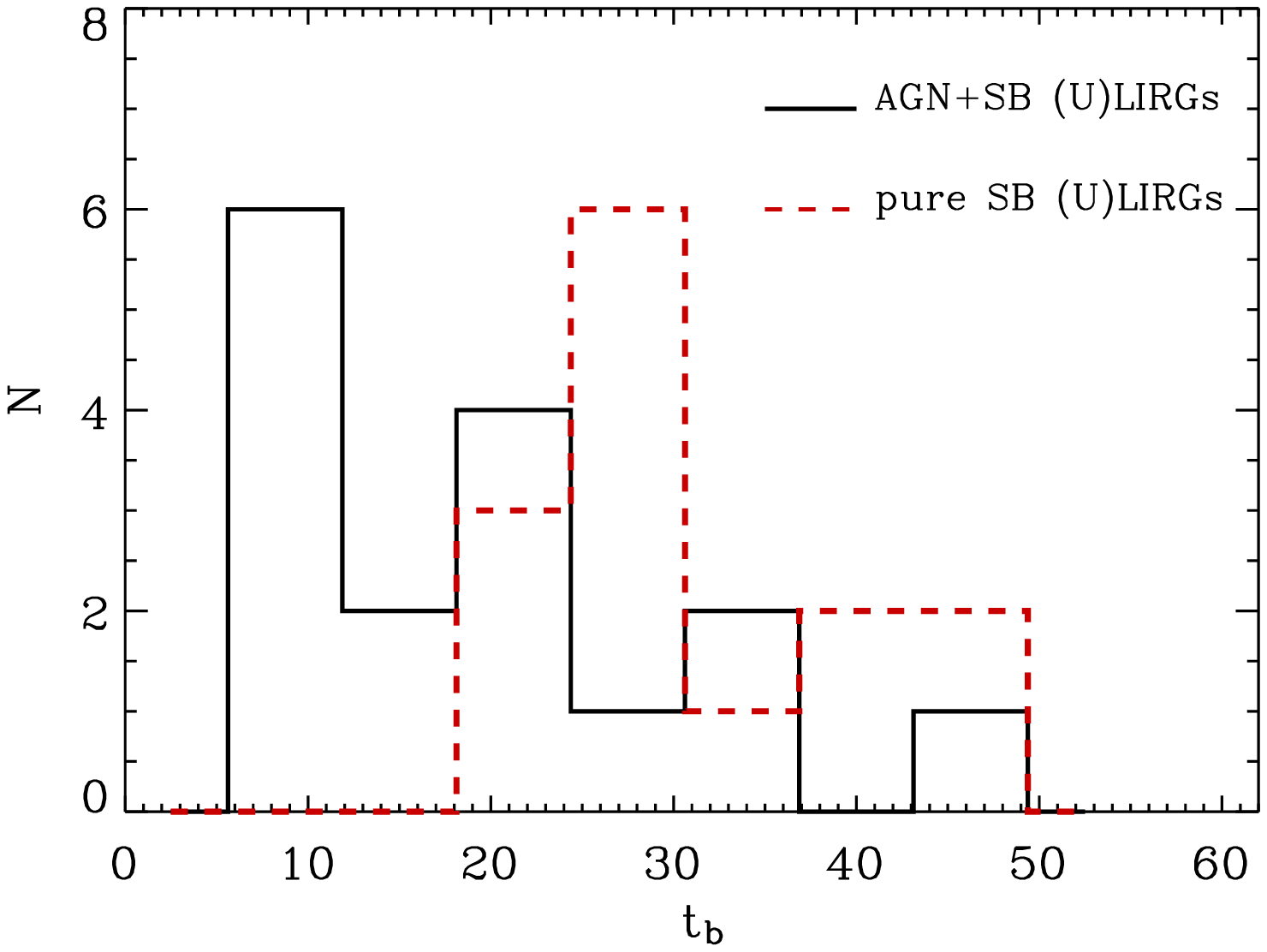}

\caption{Distributions of the e-folding times of the burst for pure
SB (U)LIRGs (dashed line) and for mixed (AGN+SB) (U)LIRGs (solid
line).} \label{Fig:tb}
\end{figure}

\begin{figure*}
\includegraphics[scale = 0.48]{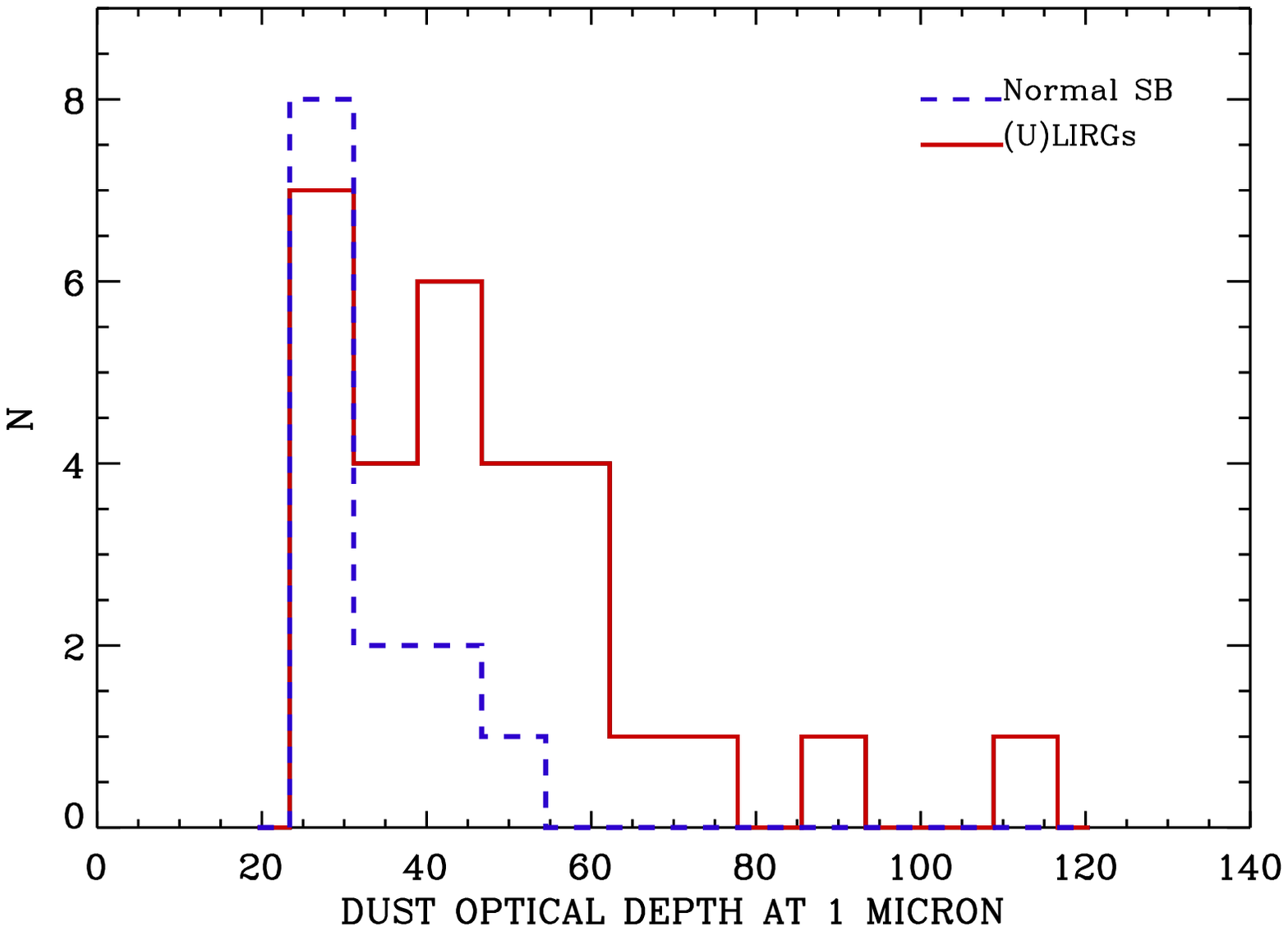}
\includegraphics[scale = 0.48]{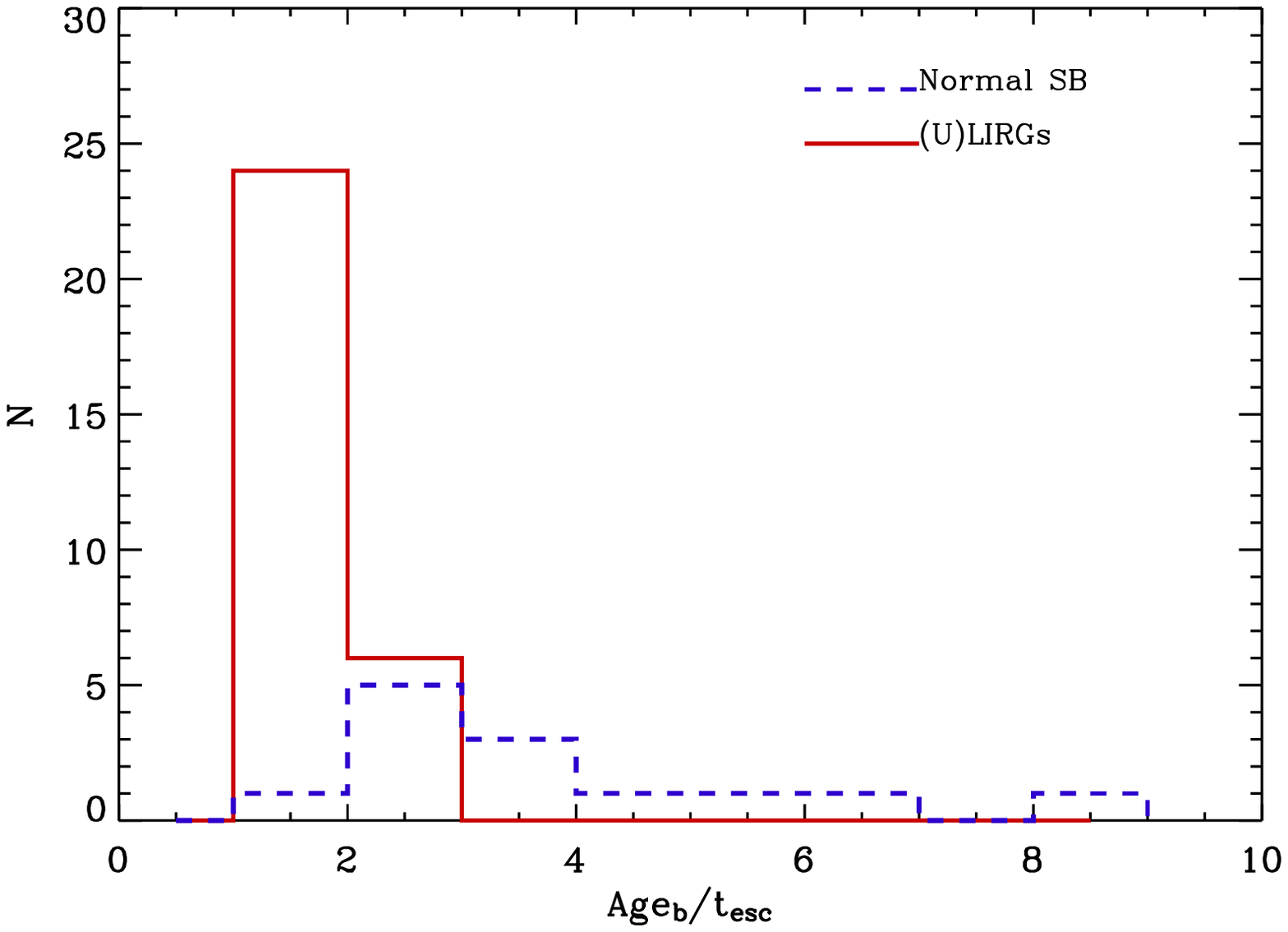}
\caption{\textit{Left panel}: histogram of the distributions of
$\tau_1$ for our sample of (U)LIRGs (solid line), and a sample of
normal starburst galaxies (dashed line). \textit{Right panel:}
histogram of the distributions of the ratio
$Age_{\rm{b}}/t_{\rm{esc}}$ for the (U)LIRGs (solid line), and the
starburst galaxies (dashed line). This ratio gives the degree of
obscuration of the newly born stars. See text for details.}
\label{Fig:osc}
\end{figure*}

\begin{figure}
\includegraphics[scale = 0.5]{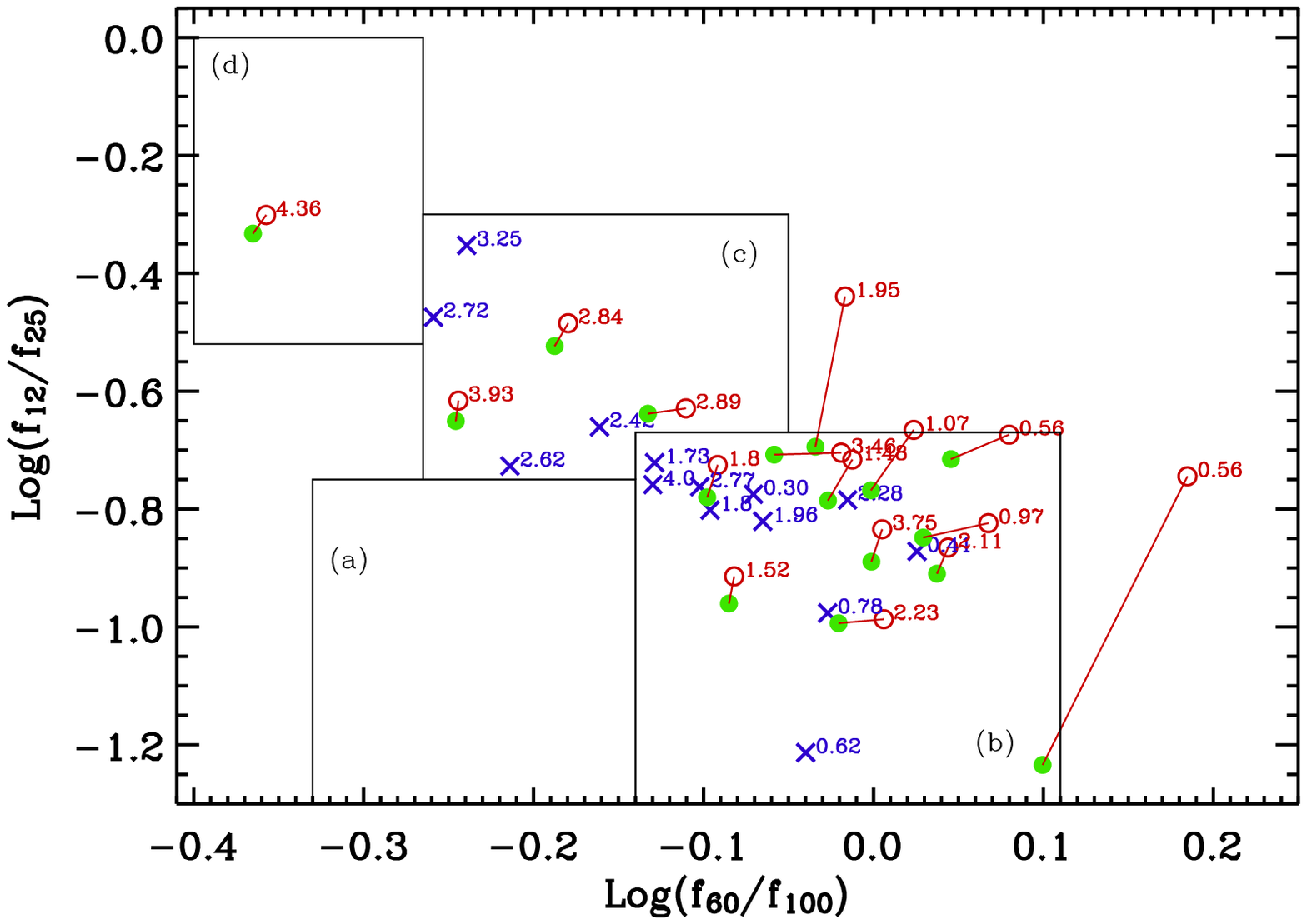}

\caption{IRAS colour-colour diagram for our sample (U)LIRGs. Crosses
denote the pure starburst galaxies, while open circles denote the
galaxies with AGN. The filled circles denote the  mixed (SB+AGN)
galaxies but with the IR colours corrected from the AGN
contributions. The numbers correspond to the values of the parameter
E$_{\rm{SB}}$. The four closed regions highlight the four
evolutionary phases of our starburst model: (a) early phase, (b)
peak phase, (c) the old starburst phase, and (d) the post-starburst
phase. } \label{Fig:phase}
\end{figure}

\section{Results}
\label{sec:results}
\subsection{Model fits}

The good spectral coverage from the NIR to the radio of our sample
imposes strong constraints, not only on the global properties of the
galaxies, i.e. SFR, but also on other important physical parameters.
In the absence of an  AGN, the fit to the total IR  and radio
luminosities  is
 related to the star formation rate, while the detailed shape of
the  SED is affected mainly by the optical depth of the dust
($\tau_1$), the escape time ($t_{\rm{esc}}$), the age of the burst
(age$_{\rm{b}}$), the e-folding time of the burst (t$_{\rm{b}}$),
the fraction of gas in molecular clouds (f$_{\rm{mc}}$) and the
sub-mm dust spectral index ($\beta$).

The optical depth mainly affects the MIR spectral range by varying
the contribution of the MC emission in the MIR. $t_{\rm{esc}}$
affects the NIR region, the PAH emission, and FIR colour
($f_{60}/f_{100}$) by changing the contribution of un-extinguished
young red supergiants to the NIR, by the heating of the cirrus from
un-extinguished young stars, and by the heating of MCs from
extinguished young stars, respectively. Age$_b$ affects the radio
spectral index, and in combination with t$_{\rm{b}}$, it has also
been used to parameterize the evolutive phase of the starburst as
$E_{\rm{SB}}={\rm Age_{\rm{b}}}/t_{\rm{b}}$. Higher values of
$E_{\rm{SB}}$, indicating more evolved starbursts, are associated
with higher NIR fluxes and cooler FIR colour (see Vega et al. 2005
for a thorough discussion). f$_{mc}$ affects the FIR colour and the
sub-millimetre region, while $\beta$ mainly affects the
sub-millimetric region.

Summarizing, earlier evolutionary phases, longer escape times and
lower dust masses in MCs combine to heat the molecular clouds
producing warmer FIR colour. The sub-millimetre spectral region
provides information on the dust emissivity index, and also on the
total dust mass, because in this region emission from the MCs and
from the diffuse components are comparable. The form of the radio
spectrum is defined mainly by the age of the starburst, in the sense
that younger objects have flatter radio slopes.

The presence of an AGN may also effect the shape of the SED by
adding an extra warm component. Typical evidence of AGN in our
observed SEDs is the presence of a near power-law shape in the NIR
to FIR spectral region.

In Fig. \ref{Fig:sbsed} we show the full SED fits for the galaxies
of our sample. The corresponding best-fit parameters of the
starburst components are shown in table \ref{tab:fitpara}. The last
three columns of the table  display the value of the evolutionary
phase of the starburst, $E_{\rm{SB}}$, the logarithm of the IR
(8-1000 $\mu$m) luminosity integrated over the SB component of the
SED, and the fractional contribution of the AGN to the 8-1000 $\mu$m
flux indicated as AGN/Tot.

Some important physical quantities derived from the best fit models
are listed in Table \ref{tab:derived}. In Col. (2) we show the
required free-free opacity at 1.4 GHz, $\tau_{\rm{ff}}$,  to fit the
1.4 GHz data, i.e. $f_{\rm{1.49 GHz}}^{\rm{obs}}= f_{\rm{1.49
GHz}}^{\rm{mod}}\times e^{-\tau_{\rm{ff}}}$. Column (3) lists the
$\%$ fraction of the total energy of the starburst reprocessed by
the cirrus component, $C$; Cols. (4), (5), (6), (7) and (8), show
the logarithms of the current total mass of gas (i.e. MC+cirrus),
$M_{\rm{gas}}$; the total mass of dust, $M_{\rm{D}}$, calculated as
$M_{\rm{D}} = M_{\rm{gas}}/(G/D)$; the mass involved in the MC
component, $M_{\rm{den}}$, calculated from the parameter
$f_{\rm{mol}}$ as $M_{\rm{den}}= f_{\rm{mol}}\times M_{\rm{gas}}$,
the mass converted in stars during the burst, $M^{\rm{b}}_\star$,
and of the final baryonic galaxy mass, $M_{\rm{T}}$, calculated as
$M_{\rm{gas}}+M_{\rm{D}}+M^{\rm{tot}}_\star$, respectively. By
construction, the SFR in our starburst models is a decreasing
function of time, with a characteristic e-folding time $t_{\rm{b}}$
of the order of a few tens of Myr. Therefore, the SFR may change
considerably during the age of the burst. We show in Cols. (9) and
(10) respectively, the SFR averaged over the entire burst, $\langle
SFR\rangle$, and the SFR in the last 10 Myr, i.e. the current star
forming rate $SFR_{\rm{c}}$, as generally probed by line emission
diagnostics. Notice that non-thermal emission probes typical
lifetimes of the SNII production, $\sim$ 50 Myr. Finally, the
percentage of model thermal emission to the 22 GHz flux is given in
the  Col. (11).

The last rows in Tables \ref{tab:fitpara} and \ref{tab:derived} list
the estimate of the  errors on the parameters. To estimate the
errors we have assumed that the SED of UGC~9913 is representative of
the starburst component of our sample of (U)LIRGs. We then
calculated several simulated SEDs for UGC~9913, by randomly varying
the observational data by $\pm$ the corresponding observational
error and, for each new SED, a new fit was performed. The fractional
errors on the parameters were then estimated as the ratio between
their range of values  and their mean value:

\begin{equation}
error^{\rm{par}}= \frac{Value_{\rm{max}}^{\rm{par}}-
Value_{\rm{min}}^{\rm{par}}}{\left\langle Value^{\rm{par}} \right
\rangle}.
\end{equation}

Figure \ref{Fig:error} shows, as an example, three fits
to the UGC~9913 obtained in this way.
The largest values of the errors are derived for the
total mass of gas and dust ($M_{\rm{gas}},M_{\rm{D}}$) because the
emission in these objects is dominated by the molecular component
while the cirrus component is not well constrained. The
errors for the other parameters are typically $< 25 \%$.

In Table \ref{tab:agnfit}, we list the values of the 7 parameters
for the best-fit AGN models for the 16 galaxies with emission
excesses that could not be fitted by the SB model alone. As already
anticipated, one of the galaxies, IRAS~01364-1042, required an AGN
contribution of 11.5 \%, while the  number of passbands was eleven.
However, the scope of this work is not the physical characterization
of the AGN component but the contribution of the AGN to the
luminosity expressed as an excess over the SB luminosity. In Table
\ref{tab:agnlum}, we show the contribution of the AGN emission to
the luminosity in different IR spectral ranges. In particular, the
spectral ranges  between 8 - 1000 $\rm \mu$m, 3 - 1100 $\rm \mu$m,
and 40 - 500 $\rm \mu$m, are those used in different calibrations
for the total IR luminosity (Takeuchi et al. 2005). In Table
\ref{tab:agnband} we show the flux and the corresponding AGN
contribution in  2MASS and \emph{Spitzer} bands. These fluxes were
calculated on the SED by using the appropriate filter responses. For
the galaxies with \emph{Spitzer} spectra the lower row provides
fluxes calculated on the observed spectra in the IRAC 3 and IRAC 4
and MIPS 24$\mu$m bands. In the IRAC bands, differences with models
in the upper row, are mainly attributable to effects of molecular
absorption bands (Lahuis et al. 2007). The maximum AGN contribution
is always confined in the NIR and MIR spectral range. Actually, it
was already known both from theoretical studies (e.g. Granato et al.
1997) and direct observations (e.g. Alonso-Herrero et al. 2001) that
the AGN radiation field can heat the dust up to 1000-1500 K,
resulting in a dust thermal emission peaking in the NIR to MIR
range. Moreover, all galaxies in our sample with J-K$ \geq 1.7$ (see
Table \ref{tab:agnlum}) cannot be modelled with pure SB and have AGN
contributions to the 1 - 5 $\rm \mu$m region higher than  50 $\%$.
However, the presence of J-K colours as blue as those of typical of
SB galaxies, cannot be used to discard the presence of an important
AGN contribution to the IR luminosity. This is, for instance, the
case of IRAS~15250+3609, with a J-K$ = 1.16$, typical of a stellar
population, but whose SED fitting required a AGN contribution $\sim
20 \%$ to the total IR luminosity. Another typical AGN diagnostic in
the NIR-MIR range  is the $25/60$ $\mu$m ratio (De Grijp et al.
1985, Sanders et al. 1988). Thus, "warm" galaxies, with $25/60 \geq
0.2$ are likely dominated by the AGN, while "cold" galaxies ($25/60
< 0.2$) are probably dominated by star formation. In Table
\ref{tab:geninfo}, we mark with a  $^*$  "warm" galaxies of our
sample. However, as a result, we can see in Table \ref{tab:agnlum}
that not all the warm galaxies needed an AGN contribution, e.g.
Arp~299 ($25/60 = 0.23$) or IC~5298 ($25/60 = 0.22$), and on the
contrary, some cold galaxies,  IRAS~15250+3609 ($25/60 = 0.18$), and
UGC~8696 ($25/60 = 0.10$), needed AGN contributions to the total IR
luminosity higher than $10 \%$.

The FIR to sub-mm spectral range is the least sensitive to the
presence of an AGN. In the 40 - 500 $\mu$m range, only two galaxies
needed AGN contributions higher than  $10\%$, IRAS~05189-2524, and
the completely AGN dominated, IRAS~08572+3915. Similar results have
also been found by Schweitzer et al. (2006) for a sample of 27
Palomar-Green QSOs.

In summary, of the 30 galaxies of our sample, 16 needed an AGN
contribution in order to fit the SEDs. Only 1 galaxy,
IRAS~08572+3915, seems to need a dominant AGN contribution ($\sim 47
\%$) to the total IR luminosity. Seven of these galaxies require an
AGN contribution to the IR luminosity (8-1000 $\mu$m) lower than the
estimated error in the IR luminosity ($\sim 6 \%$, see last row of
Col. 10 in Table \ref{tab:fitpara}). The NIR to MIR spectral range
was well constrained only for the 11 galaxies with a MIR spectrum.
Of those, 8 needed an AGN contribution in order to explain their
SED+ MIR spectrum, and only UGC~9913, IRAS~10565+2448, and UGC~8387
could be well modelled with pure starburst models.

\begin{figure*}
\includegraphics[scale = 0.5]{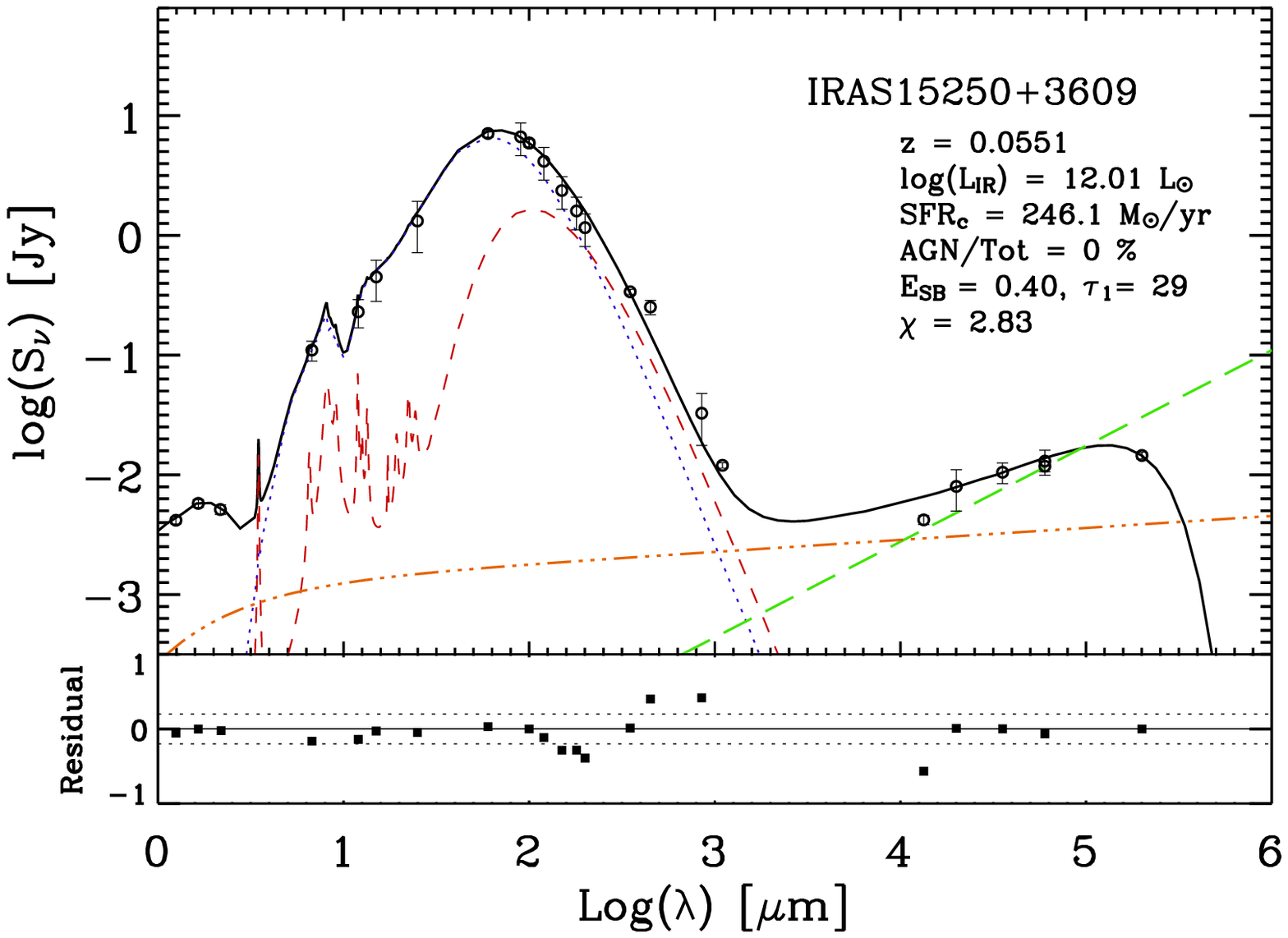}
\includegraphics[scale = 0.5]{fig1h.ps}
\caption{\textit{Left panel}: best fit to the IRAS~15250+3609 SED
without the inclusion of the \emph{Spitzer} spectrum. \textit{Right
panel:} best fit to the IRAS~15250+3609 SED with the inclusion of
the \emph{Spitzer} spectrum. The meaning of lines and symbols are
the same as in Figure \ref{Fig:sbsed}.} \label{Fig:spi}
\end{figure*}

\begin{figure}
\includegraphics[scale = 0.5]{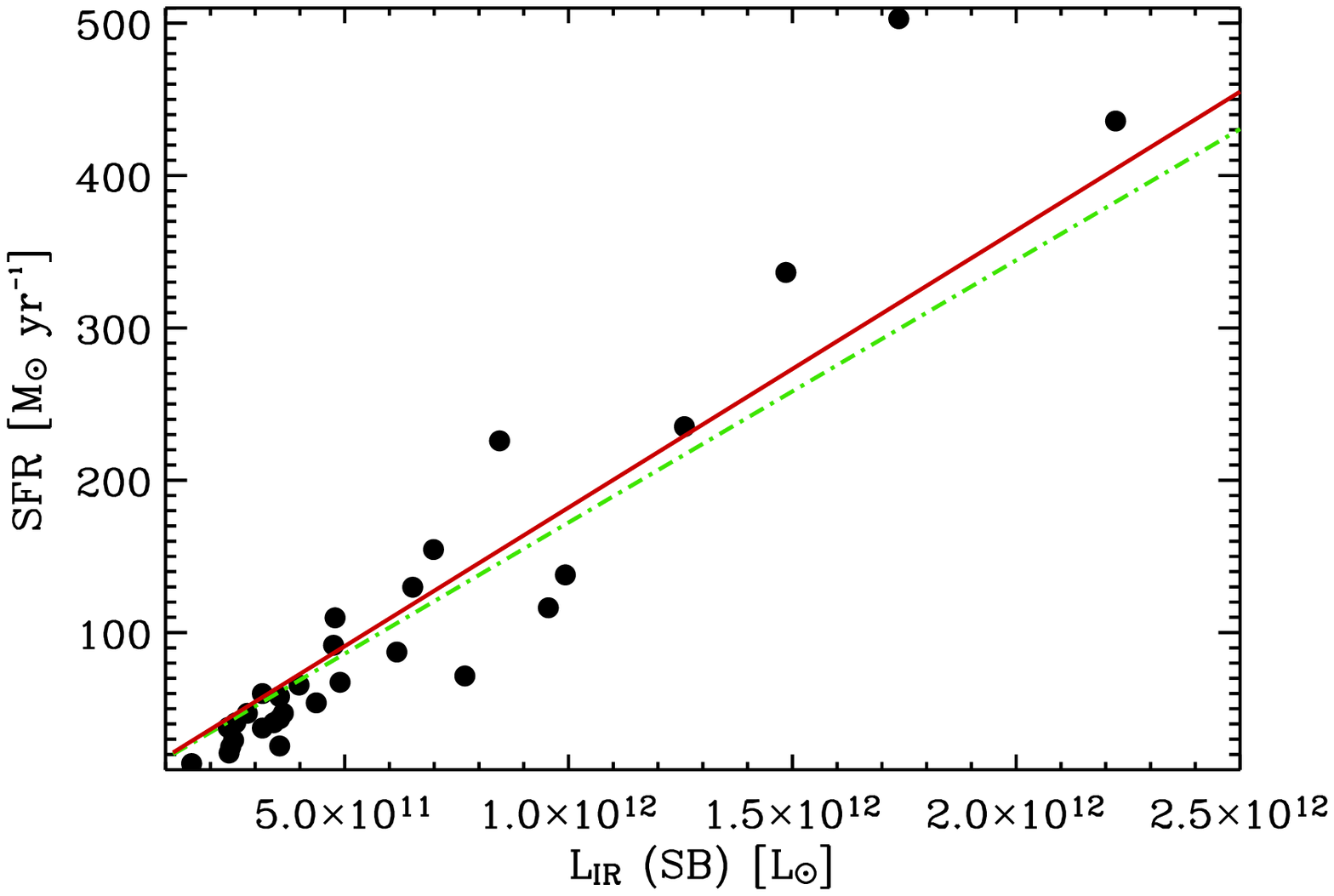}
 \caption{ Correlation between AGN corrected IR luminosity and the
 derived SFR averaged over the last 10~Myr for our sample of (U)LIRGs.
The circles denote the individual galaxies, the solid
 line represents the best fit to the correlation, and the dashed line
 represents Kennicutt's relation.
\label{Fig:sfrlir}}
\end{figure}

\section{Discussion}
\label{sec:discussion}

One of the main results emerging from our investigation is that in
all but one galaxy, IRAS~08572+3915, the starburst dominates the
bolometric ($\sim$ IR) emission of the source, with a fractional
contribution that is always larger than 80\%. About half of the
sample can actually be fitted by a pure starburst.

We thus begin the discussion by looking into the main features of
the SB components of (U)LIRGs and their possible differences with
respect to normal starburst galaxies.

\subsection{The Starburst Component}
All the pure-starburst (U)LIRGs in our sample have e-folding times
ranging between 20 to 50 Myrs, while some shorter e-folding times
are obtained for the mixed (AGN + SB) sub-sample.  In
Fig.\ref{Fig:tb}, we show the distribution of $t_{\rm{b}}$ for pure
SB and mixed (U)LIRGs. A K-S test for the two distributions
indicates a high probability that the two distributions are
different. This would indicate that (U)LIRGs with an AGN are
characterized by a more rapidly declining SFR, than pure starburst
(U)LIRGs. An exciting interpretation of this finding is that the SFR
itself suffers negative feedback from the AGN.

On the contrary, no significant differences are found in the other
parameters, i.e. dust opacities, escape times, etc.

The ranges of ages varies between $3.9 \times 10^6$ yr for
IRAS~08572+3915 to $9.8 \times 10^7$ yr for UGC~6436, with a small
trend for older bursts in pure SB objects.

More interesting is the comparison of the MC optical depth,
$\tau_1$, and the degree of obscuration,  measured by the quantity
$age_{\rm{b}}/t_{\rm{esc}}$, between (U)LIRGs and normal starburst
galaxies. Both parameters indicate that (U)LIRGs constitute a
population of very dust enshrouded objects with respect to normal
starburst galaxies. In Fig. \ref{Fig:osc}, we show the distributions
of $\tau_1$ (left panel) and $age_{\rm{b}}/t_{\rm{esc}}$ (right
panel) for our sample of (U)LIRGs (solid line) and the sample of
normal starburst from Vega et al. (2005) (dashed line). Both
histograms indicate that (U)LIRGs and starburst are two very
distinct populations and that, consequently, the infrared properties
of (U)LIRGs cannot be obtained from those of starburst galaxies just
by re-scaling them to higher star formation rates.

The values of the SB evolutionary parameter, $E_{\rm{SB}}$,  show
that many LIRGs and ULIRGs (15/30) are found in the peak phase of
the starburst phenomenon ($0.20\leqslant E_{\rm{SB}}<2$). None of
galaxies is in the early phase ($E_{\rm{SB}}<0.2$); 14 are in an
older phase ($2<E_{\rm{SB}}<4$), and only one, the least IR luminous
NGC~6286, is found in the post-starburst phase. The starburst phase
defined by the IR colours (an observational quantity) is in perfect
agreement with that defined by the $E_{\rm{SB}}$ parameter (a
theoretical quantity), see Fig. \ref{Fig:phase}. In the same figure
we notice that when the estimated AGN contribution is subtracted
from the fluxes in the IR bands, (U)LIRGs with AGN fall on the same
trend as pure starburst (U)LIRGs (open circles connected to solid
circles), indicating that our determination of the AGN contributions
is robust.

Another important quantity derived from our fits is the mass of
dust. Dunne et al. (2000) have published dust mass measurements for
their SLUGS sample, which includes some of our objects. In general,
there is a very good agreement between their determinations from 850
$\mu$m observations and ours. However, for three galaxies,
IRAS~08572+3915, IRAS~15250+3909, and NGC~1614, their estimates are
considerable higher than ours, reaching one order of magnitude in
the case of IRAS~08572+3915. The likely origin of the discrepancy is
that Dunne et al. (2000) determinations are based on the assumption
that the dust emitting in the sub-mm range is cold (T$\sim 20$ K),
while GRASIL accounts for a distribution of dust temperatures within
the ISM. In the above three galaxies we can see, from the fits, that
the sub-mm spectral range is still dominated by a warm MC component
which thus require a significantly lower dust mass. We believe that
our estimate is more correct, since it is based on the fit to the
whole MIR-FIR-sub-mm spectral range.

The total baryonic mass derived for our sample of (U)LIRGs is also
in good agreement with dynamical mass estimates (e.g. Genzel et al.
2001, Colina, Arribas \& Monreal-Ibero 2005, Dasyra et al. 2006).
Moreover, our estimates of the total gas mass are in agreement with
those derived from available observations in HI and in CO, see for
instance Cols. 7 and 8 in Table 4 by Dunne et al. (2000), provided
that a reduced conversion factor (X-factor) between CO luminosity
and mass of molecular gas is used (Solomon et al. 1997).

Finally, in Fig. \ref{Fig:sfrlir} we plot the current SFR
$SFR_{\rm{c}}$  versus the AGN corrected IR luminosity of the
best-fit models, $L_{\rm{IR}}^{\rm{SB}}$. The solid line represents
the best linear fit to the data, i.e. $ SFR = 1.82 \times 10^{-10}
L_{\rm{IR}}$ ($M_\odot$ L$\odot^{-1}$). By comparison we also plot
the SFR/$L_{\rm{IR}}$ calibration of Kennicutt (1998) (dashed line).
The two calibrations coincide within the accuracy of the models
(Kennicutt 1998).

\subsection{\emph{Spitzer} spectra}
\label{sec:spitzer}
 In our starburst models the MIR spectral domain is largely dominated by the MC component.
 However, in (U)LIRGs this spectral region  may be significantly
 affected by the presence of an AGN.
 The interpretation of
 the MIR broad band fluxes may be affected
 by a degeneracy between the optical depth of the MCs and
 the contribution of the putative AGN.
 Indeed observed warm MIR colours may be almost equally well reproduced
 by increasing the MC optical depth and, at the same time,
 increasing the AGN contribution
 (see e.g. IRAS~08572+3915 in Fig. 1 of Prouton et al. 2004.)
 However an increase of the MC optical depth has the
 effect of strengthening   the $9.7 \mu$m
 silicate absorption feature. Thus, a possible way
 to break this degeneracy is to complement
 the broad band fluxes with the 5 - 35 $\mu$m  \emph{Spitzer}
 IRS spectrum which carries a great deal of additional
 information.

We thus searched for \emph{Spitzer} IRS spectra of our ULIRG sample
in the SSC archive and found them for 11 galaxies. For these
galaxies the fitting procedure includes also the \emph{Spitzer}
spectra which are shown in Fig. \ref{Fig:sbsed}, together with the
corresponding models, upwardly displaced by one dex, as  solid thick
and solid thin lines, respectively.

The largest effect, after accounting for the analysis of the
\emph{Spitzer} spectra, was found for IRAS~15250+3609. As an
illustration purpose,  Fig. \ref{Fig:spi} shows how the inclusion of
the IRS spectrum changes our spectral classification of
IRAS~15250+3609. The left panel shows the best fit to the  broad
band SED alone. The 20 continuum passbands are very well fitted
by a pure SB model with e-folding time of 25 Myr, age of $10^7$ yr,
$\tau_1$ of 29 and a large current star forming rate, $SFR_{\rm{c}}
\sim 246$ $M_\odot/yr$. However, the IRS spectrum of IRAS~15250+3609
shows the presence of a deep silicate absorption feature at $9.7
\mu$m (right panel). With a starburst alone, it is not possible to
reproduce the large optical depth required by this silicate
absorption feature and the emission in the MIR-FIR continuum, at
once. Indeed, the fit shown in the right panel includes an obscured
AGN (torus inclination $\sim 90^\circ$) contributing $\sim 20\%$ of
the IR (8-1000 $\mu$m) luminosity, and about $50 \%$ to the MIR
luminosity. We notice that, though the contribution of the AGN is
only 20\% of the IR luminosity, other quantities have changed by a
significant factor. The optical depth of MC has almost doubled, and
now it fits better in the range of (U)LIRGs. The current SFR has
decreased from $\sim 246$ M$_\odot/yr$  to $\sim 154$ M$_\odot/yr$
while the model still fits very well the radio data. This happens
because the evolutionary phase becomes more advanced
(E$_{\rm{SB}}\sim 1$ instead of 0.4) and what actually fuels the
radio flux is the the SNII rate, which probes a time interval five
times larger than the 10 Myr adopted for the current SFR. It is
worth noticing that the common belief (e.g. Elvis et al. 1994) is
that the presence of the AGN eventually dilutes the $9.7 \mu$m
feature. IRAS~15250+3609 actually shows the opposite case, since the
$9.7 \mu$m  feature is intrinsically less pronounced in the less
opaque MCs of the pure starburst model.

In the other 10 galaxies, the inclusion of the IRS spectrum did not
change our classification, although it allowed a more accurate
estimate of the SB parameters and, of the AGN contribution.

In summary, of the subsample with IRS spectra, three galaxies,
UGC~9913, IRAS~10565+2448 and UGC~8387, were still well fitted by
pure SB models. Seven galaxies, UGC~8058,  UGC~8696, UGC~5101,
NGC~2623 CGCG~436-30, IRAS~05189-2524 and IRAS~08572+3915 were
fitted with almost the same SB and AGN components. In
IRAS~15250+3609 the  IRS spectra revealed the presence of an almost
entirely obscured AGN contributing 20\% of the IR luminosity. This
is where  the combined analysis of the IRS spectra and panchromatic
SED exploits its maximum capability.

\subsection{Star formation versus AGN}

Sixteen galaxies (53 \%) of our sample of 30 (U)LIRGs require an AGN
contribution in order to fit their SEDs. However, only 9 (30\%) need
a contribution to the IR luminosity which is higher than 10 $\%$. In
most  cases it is lower than 20 \%. Only one galaxy,
IRAS~08572+3915, seems to be AGN dominated ($\gtrsim 50 \%$, see the
discussion below). In 2/30 (6\%) objects, the AGN contribution
(8-1000 $\mu$m) is less than 2\%. In these objects the AGN
contribution is more important at NIR wavelengths, where it may
reach 15\% or more, as for UGC~5101 ($\sim 50\%$), while the AGN
contribution to the IR (8-1000) luminosity is $\sim 1 \%$. However,
it is worth remarking that UGC 5101 shows an evident power law NIR
SED and [NeV]15$\mu$m emission in the \emph{Spitzer} IRS spectrum
 (Armus et al. 2004). In the remaining 5/30 (17\%) the AGN contribution (8-1000)
is of the order of 5\%.

For those objects that have been observed with \emph{Spitzer}, we
find that our estimates of the AGN contribution agree very well with
those derived by Armus et al. (2007) using  \emph{the line ratios
Ne[V]/Ne[II] or O[IV]/Ne[II]}, although many of their values are
quoted as upper limits. We find a different result for two galaxies,
IRAS~14348-1447 and IRAS~15250+3609. The first  was classified as a
pure SB by Armus et al. (2007), based on the strength of the PAH
features. However, we require an AGN contribution of about 10 $\%$
to the IR luminosity, in order to match its red NIR colour (J-K$ =
1.7$). In the case of IRAS~15250+3609, Armus et al. find that the
power source remains buried even in the MIR, while our analysis of
the NIR to radio SED (including the IRS spectrum) requires a buried
AGN contributing about 20 \% to the IR luminosity.

Armus et al. (2007) have estimated the AGN fractions also with other
methods, using either the equivalent width of the $6.2\;\rm \mu m$
PAH feature, or the mid-infrared spectral slope, or the ratio
between the total PAH flux and the IR flux. These methods provides
AGN fractions that are in general very much higher than those
obtained by analyzing the high ionization line ratios which are
similar to our values (see appendix for comparisons for individual
objects). Therefore, our work supports the notion that \emph{neither
the mid-infrared slope nor the PAH equivalent widths or relative
fluxes provide a good estimate of the AGN contribution} (see
conclusions in Vega et al. 2005).

Summarizing the above results, we find that the contribution of the
AGN is well determined in 10/30 objects (33\%), while in 5/30
objects it is minimal (5\%). In the remaining 15/30 objects (50\%)
the SED can be well fitted by almost pure starbursts because the
total contribution of the AGN is $< 1\%$, less than the accuracy of
the model in all bands. Even in the objects where the presence of
the AGN is well determined, only in one case, IRAS~08572+3915, does
the AGN dominate the bolometric luminosity. In general, the optical
spectral types of the galaxies (Col. 9 in Table \ref{tab:geninfo})
agree very well with the results from the SED analysis. All galaxies
classified as "\emph{HII}" were well modelled with pure SB models.
While those classified as "\emph{Sy}" needed some AGN contributions.
The situation is less clear for galaxies classified as Liners. Some
of them (4) were modelled with pure SB models, while 5 of them,
IRAS~14348-1447, IRAS~15250+3609, IRAS~01364-1042, CGCG~436-30 and
NGC~6286, required of the presence of a buried AGN.

Since only 11/30 objects have published  \emph{Spitzer} IRS data
that can be used to add further constraints to our fitting
procedure, we cannot definitely rule out that the AGN contribution
be larger and, in particular, that our conclusions on some pure
starbursts need revision. It is, however, less likely that the
inclusion of \emph{Spitzer} spectra changes our conclusion of the
overall dominant contribution of the SB in (U)LIRGs (see Sect.
\ref{sec:spitzer}).

\subsection{SED fits at radio frequencies}
\label{sec:radiofits} In paper I we found a systematic deficit of
the 22 GHz emission in our galaxies when compared with that expected
from the radio spectral slope calculated at lower frequencies
(Clemens et al. 2007 ). Actually, Fig. \ref{Fig:sbsed} shows that
the 22 GHz data fall below the best fit model in 19/30 of the
galaxies. In 16 out of 19 objets the differences between models and
22 GHz data are small, most of them within the $20-30 \%$, as can be
seen from the residual plots of the same figure. For 9 objects the
22 GHz deficit is within the observational error. For the other 7 we
notice that the 22 GHz flux, though below the total predicted radio
emission, actually falls above the predicted synchrotron emission
(long dashed line in the plots). Thus a possible explanation is that
the deficit at 22 GHz may be due a deficit of thermal emission
caused by absorption of ionizing photons by dust within the HII
regions (Vald\'es et al. 2005). In the above 7 objects we estimate
that the fractional absorption is between 20\% to 40\%, as can be
seen from the open squares in the residual plots of Fig.
\ref{Fig:sbsed}. Finally, three galaxies, UGC~8387, IZW~107 and
Arp~199, show a deficit in the high frequency radio emission that
can not be explained by observational errors (unless we have
severely underestimated it) or by dust absorption. We are currently
investigating the eventual origin of this discrepancy (Clemens et
al. in preparation).

Two galaxies  of the sample, UGC~8058 and UGC~8696, show variability
at radio frequencies (Fig. \ref{Fig:sbsed}, Paper I) on a timescale
of a few years (Condon et al. 1991). In the case of UGC~8058 the
higher radio fluxes in Fig. \ref{Fig:sbsed} are from observations
taken on the same day. They are higher by about 0.4 dex and have a
slope which is strikingly similar to that at the lower radio fluxes.
The latter data result from observations taken over a period of
about 15 Years. The variability may be an indication that the AGN
actually powers sporadic emission which dominates at radio
frequencies (Lonsdale et al. 2003). On the other hand it is likely
that the lower fluxes are the genuine emission from the starburst,
because, in the other alternative and given the amplitude of
variability, they should show a much larger scatter around the
synchrotron relation. In the fitting procedure we have thus assumed
that the lower level radio emission is representative of the
starburst. Both  galaxies have published IRS spectra which, together
with the broad band SED, are well fitted by a composite model, SB
plus AGN.

\begin{figure*}
\centerline{
\includegraphics[scale = 0.8]{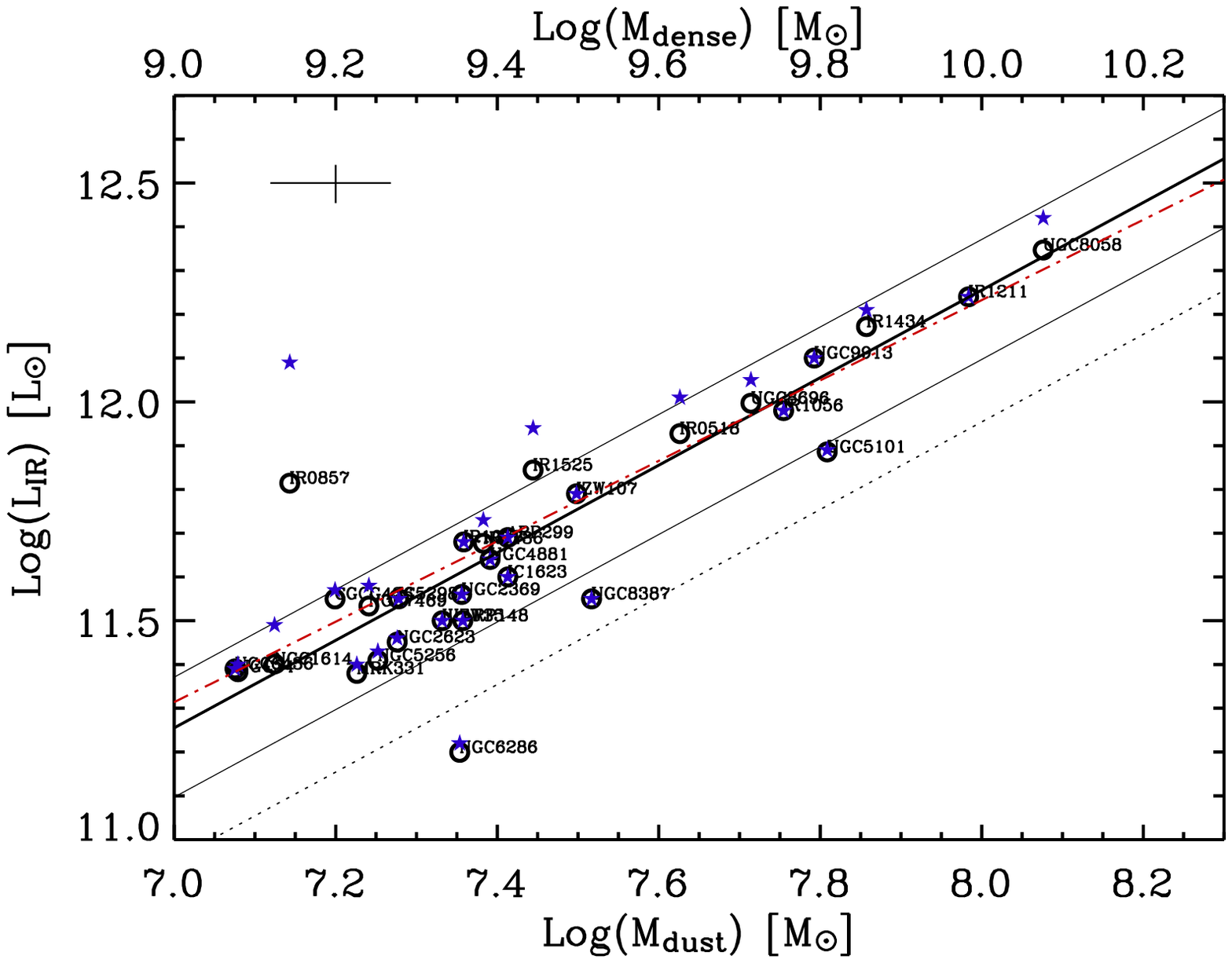}}
\caption{Correlation between the dust  mass in the dense component,
derived from our best fit models, and the IR luminosity for our
sample of 30 (U)LIRGs. In the upper axis, we display the equivalent
molecular mass for a dust-to-gas ratio of 100. Filled stars denote
the total IR luminosity, calculated from the 4 IRAS bands (Sanders
\& Mirabel 1996), while the black open circles denote the IR
luminosity corrected for the AGN contribution. The latter was
obtained by integrating the luminosity of the starburst component of
the best-fit model. The thick solid line is a linear fit to this AGN
corrected IR luminosity, i.e. the IR luminosity due to the starburst
emission component, while the dot-dashed line corresponds to the fit
to the total IR luminosity. See text for details. The dotted line
corresponds to the fit between the IR luminosity and the dense
molecular mass as traced by HCN emission obtained by Gao \& Solomon
(2004a). The cross to the upper left denotes the typical errors of
the
 IR luminosities and masses derived with our models as explained in
Sect. \ref{sec:results}.} \label{Fig:GAO}
\end{figure*}

\begin{figure*}[t]
\centerline{
\includegraphics[scale = 0.7]{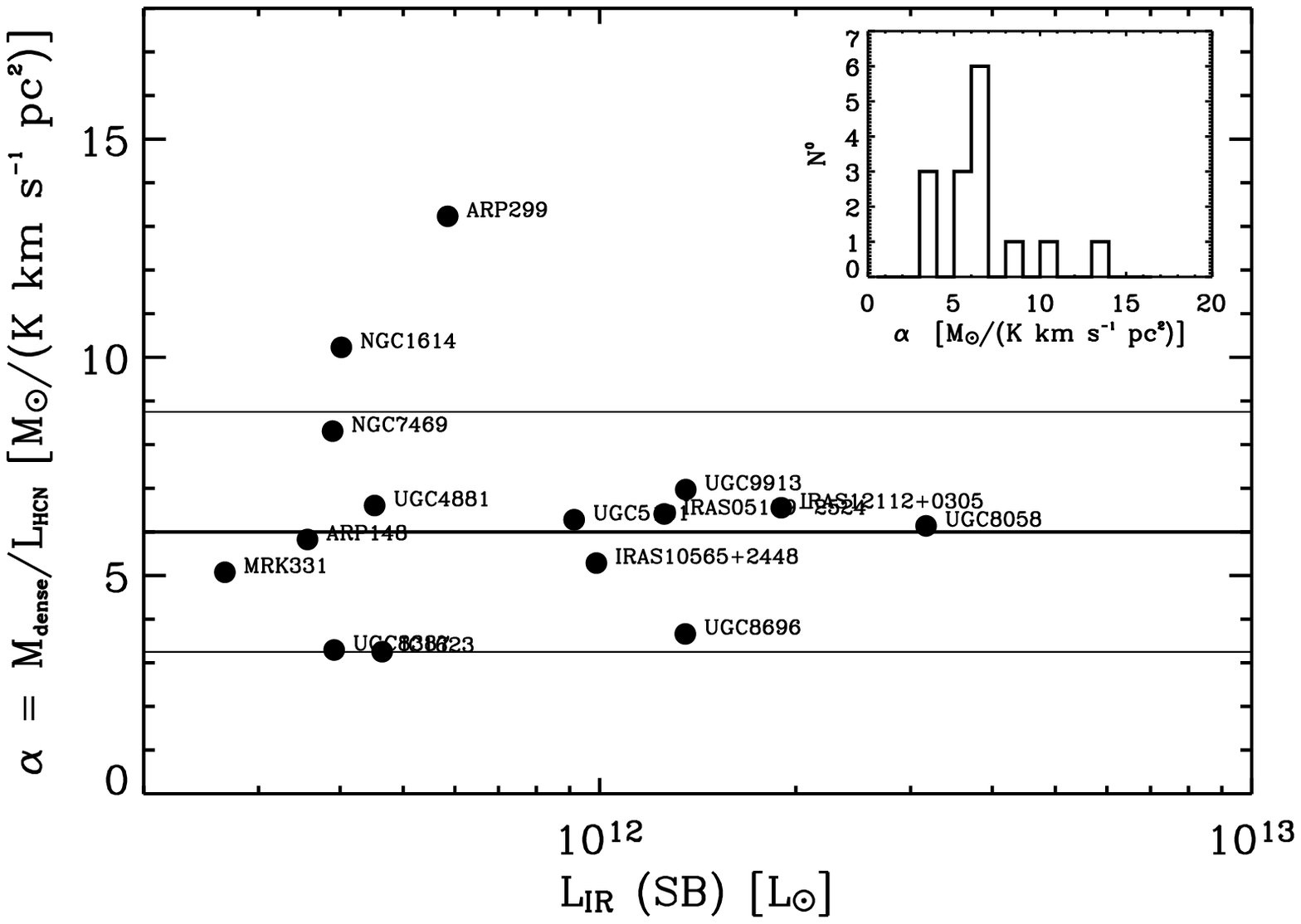}
} \caption{Values of $\alpha_{\rm{HCN}}$ for the 15 galaxies in
common between the Gao \& Solomon sample and our sample of (U)LIRGs
as a function of the starburst IR luminosity. The filled circles are
the data, the thick solid line represents the median value of
$\alpha_{\rm{HCN}}$, the thin solid lines represent the derived
$\sigma$ of the distribution in $\alpha_{\rm{HCN}}$. The histogram
of the distribution is shown to the upper right.}
\label{Fig:GAOmass}
\end{figure*}

\subsection{Molecular masses in very actively star forming galaxies}
Our SED fitting technique allows a fairly accurate determination of
the main physical parameters of the power sources in compact
(U)LIRGs, and in particular, a full characterization of the
starburst component. One of the quantities that we derive directly
from the fit is the mass of the obscuring dust present in the dense
molecular clouds. This can be converted into a mass of dense
molecular gas associated with star formation, after assuming a
suitable gas/dust mass ratio. This method of determining the mass of
dense molecular gas in galaxies with high star-formation rates, may
constitute a viable alternative to the more traditional methods
based on the molecular tracers of high density regions (e.g. HCN,
see Gao \& Solomon 2004a and references therein).

In order to obtain molecular gas masses from the dust masses, we
have adopted a conservative value of G/D=100 which is typical of
nearby star forming galaxies. This value seems adequate also for our
sample because there is evidence that the metallicity in these
objects is around solar (Rupke, Veilleux \& Baker 2007).
Furthermore,  our chemical evolution models indicate that the
metallicity of the gas changes by less than 30\% during the whole
duration of the burst. This allows us to assume a constant dust to
gas ratio for all the objects, independent of their evolutionary
status.

In Fig. \ref{Fig:GAO} we plot  the infrared luminosity (8 - 1000
$\mu$m) against the values of the dust masses of our objects. In the
upper horizontal axis we show the corresponding molecular masses,
obtained by assuming G/D=100. The existence of the tight linear
correlation between the infrared luminosity and the dust mass of the
dense star forming component, reproduced in Fig. \ref{Fig:GAO}, is a
natural consequence of the fact that the infrared emission in our
objects is dominated by the MC component. The correlation
coefficient is R$^2 = 0.874$, and the non-parametric Kendall tau
coefficient is $\tau_{\rm{K}}=0.74$, indicating a very low
probability of non-correlation. Assuming G/D=100 we find that the
best-fit (logarithmic) relation for the galaxies of our sample is
\begin{equation}
\log(L_{\rm{IR}}^{\rm{SB}})= (1.01 \pm 0.08) \log(M_{\rm{den}}) +
(2.20 \pm 0.79)
\end{equation}
or, in linear form,
\begin{equation}\label{equ:gao2}
L_{\rm{IR}}^{\rm{SB}}/M_{\rm{den}}= 183 (\pm 11) \ \rm L_\odot
M_\odot^{-1}.
\end{equation}
This relation is plotted in the figure as
a thick solid line, while the thin solid lines correspond to the
one-sigma deviation around the mean relation.

It is also remarkable that objects with a significant contribution
from an AGN lie above this relation. However, when the contribution
of the AGN to the IR luminosity is subtracted, these objects nicely
fit on the relation. The unique exception is IRAS~08572+3915 which
seems to require an IR contribution from the AGN even higher than
the value of 47\% obtained from the SED fit.

All these findings support the existence of a fairly robust correlation
between the infrared luminosity and dust/gas mass of the star forming component,
which could possibly be used to determine the fractional contribution of the AGN
in luminous IR objects where the gas mass is known by other methods.

As a check of our finding, we have compared our  results with that
obtained by Gao and Solomon (2004a) for their sample of 64 star
forming galaxies.  For these objects, they derived a tight
correlation between the HCN luminosity and the IR luminosity, that
was interpreted  as evidence that both originate  in the same
region. They obtained the mass of dense molecular gas from the
observed L$_{\rm{HCN}}$ by adopting a conversion factor
$M_{\rm{dense}}/L_{\rm{HCN}} $
\begin{equation}
\label{equ:alpha}
\alpha_{\rm{HCN}} =
2.1 \sqrt{n(H_2)}/T_{\rm{b}}=10 \ \rm M_\odot (K\cdot km\cdot
s^{-1}\cdot pc^2)^{-1},
\end{equation}
assuming a typical density of $n(H_2) \sim 3\times 10^4$ cm$^{-3}$
and brightness temperature T$_{\rm{b}}=35$ K. Their relation between
IR luminosity and dense mass is shown in Fig \ref{Fig:GAO} as a
dotted line:
\begin{equation}\label{equ:gao3}
L_{\rm{IR}}^{\rm{SB}}/M_{\rm{den}}= 90 \ \rm L_\odot\cdot
M_\odot^{-1},
\end{equation}
Their relation falls below our relation by about a factor of two. In
order to understand the origin of this discrepancy we searched the
Gao \& Solomon sample for common objects and derived the conversion
factors, $\alpha_{\rm{HCN}}$, between our estimated gas masses and
the HCN luminosities quoted by them.

Table \ref{tab:gao1} lists the values of the IR and HCN luminosities
of the 15 common galaxies\footnote{Actually, there are 14 galaxies
in common between the two samples. We also included in the study
IRAS~12112+0305 from the Graci\'a-Carpio et al. (2006) sample} while
Fig. \ref{Fig:GAOmass} shows the derived $\alpha_{\rm{HCN}}$ against
the IR luminosity. The median value of $\alpha_{\rm{HCN}}$ for the
common objects is 6.0 M$_\odot$(K km s$^{-1}$pc$^{2})^{-1}$ with
87$\%$ of the galaxies having $\alpha_{\rm{HCN}}=6.\pm 2.75$. The
exceptions are Arp~299, and NGC~1614 that are also outliers in the
Gao \& Solomon correlation. These galaxies have values of
$L_{\rm{IR}}/L_{\rm{HCN}} \sim 3000$, while the mean value of
$L_{\rm{IR}}/L_{\rm{HCN}}$ in  their sample is 900 - 1200, implying
 significantly  weaker HCN emission.

Our median value of $\alpha_{\rm{HCN}}$ is about half that derived
by Gao \& Solomon (2004a) for their full sample. However the
$\alpha_{\rm{HCN}}$ factor derived by by Gao \& Solomon (2004a) is
certainly biased towards normal galaxies, that constitute the
majority of their sample, and may not be representative of the more
extreme conditions of the molecular gas in (U)LIRGs. That a lower
$\alpha_{\rm{HCN}}$ factor would be more appropriate for (U)LIRGs
was also argued by Gao \& Solomon (2004a), who noticed that this
factor is a sensitive function of the brightness temperature,
T$_{\rm{b}}$, that can be higher in (U)LIRGs. Indeed, that
conversion factors between molecular gas mass and luminosity may be
lower in more extreme star forming conditions has already been found
for CO emission (Solomon et al. 1997, Downes \& Solomon 1998).
Moreover brightness  temperatures reaching hundreds of kelvins have
been found in massive Galactic star forming cores (Boonman et al.
2001).

Making the same approximation as Gao \& Solomon,  that the HCN
emission originates in gravitationally bound cloud cores, and
assuming the same core density of n(H$_2)\sim 3\times 10^4$
cm$^{-3}$, we estimate from equation \ref{equ:alpha} that our
$\alpha_{\rm{HCN}}$ would require a T$_{\rm{b}}\sim 65_{-21}^{+28}$
K.

Actually, an accurate determination of the conversion factor between
HCN and the mass of the very dense molecular gas needs more
extensive studies, including observations of HCN lines at higher
frequencies. In this respect, we anticipate that preliminary results
of the combined analysis of the SED and the high excitation HCN
transitions in Arp~220 (Vega et al. in preparation) indicate that
these high brightness temperatures are likely, but also that other
factors (abundances, densities, velocity gradients) affect the
derived value of $\alpha_{\rm{HCN}}$. Greve et al. (2006), in their
very extensive study of the molecular emission in Arp~220,  derived
a likely range of dense molecular masses of [0.6 - 2.0] $\times
10^{10}$ M$_\odot$. Their derived lower limit corresponds to our
derived value. We notice also that the upper limit of Greve et al.
exceeds the dynamical mass estimated by Downes \& Solomon (1998).

\begin{figure*}
\centerline{\includegraphics[scale = 0.43]{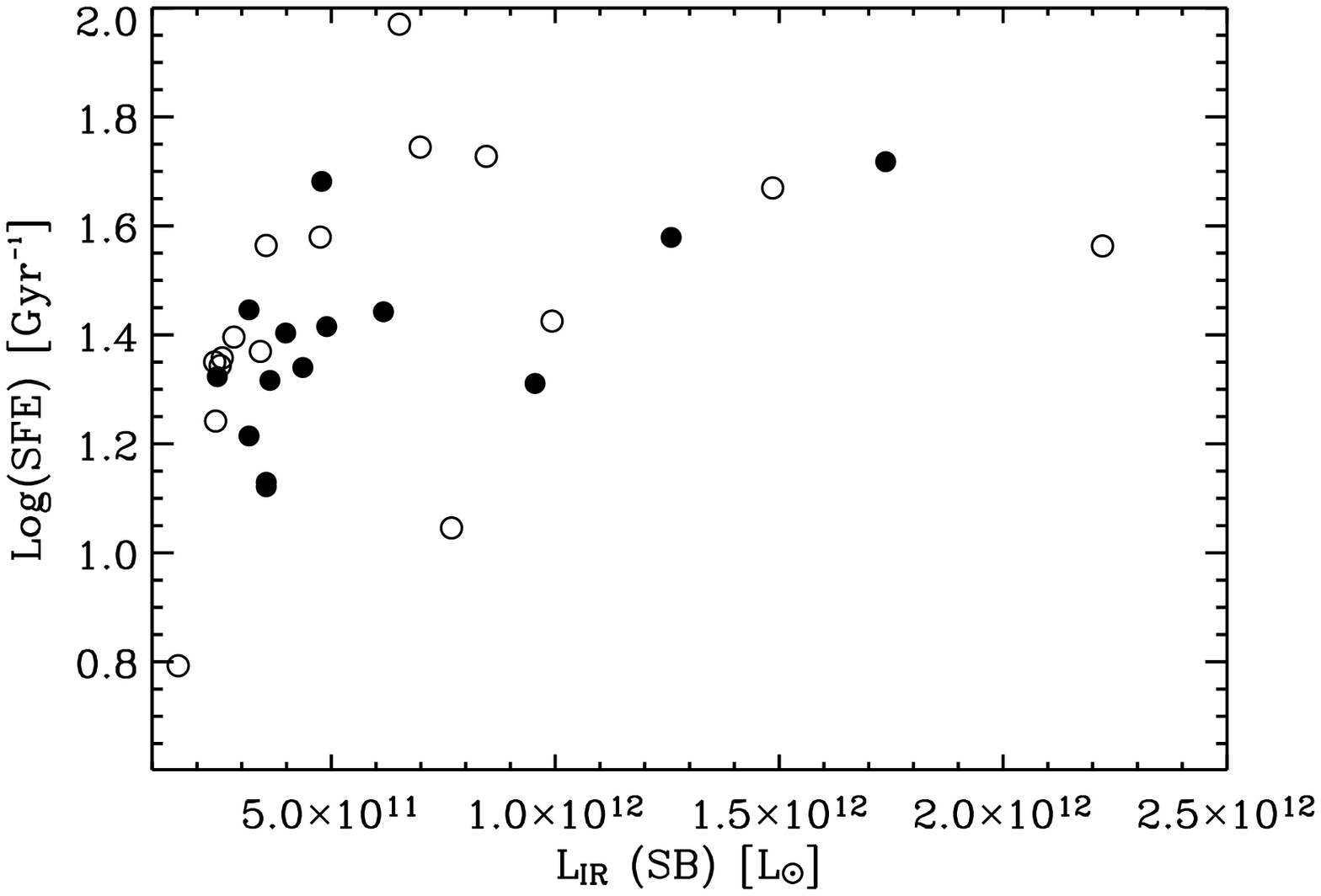}
\includegraphics[scale = 0.43]{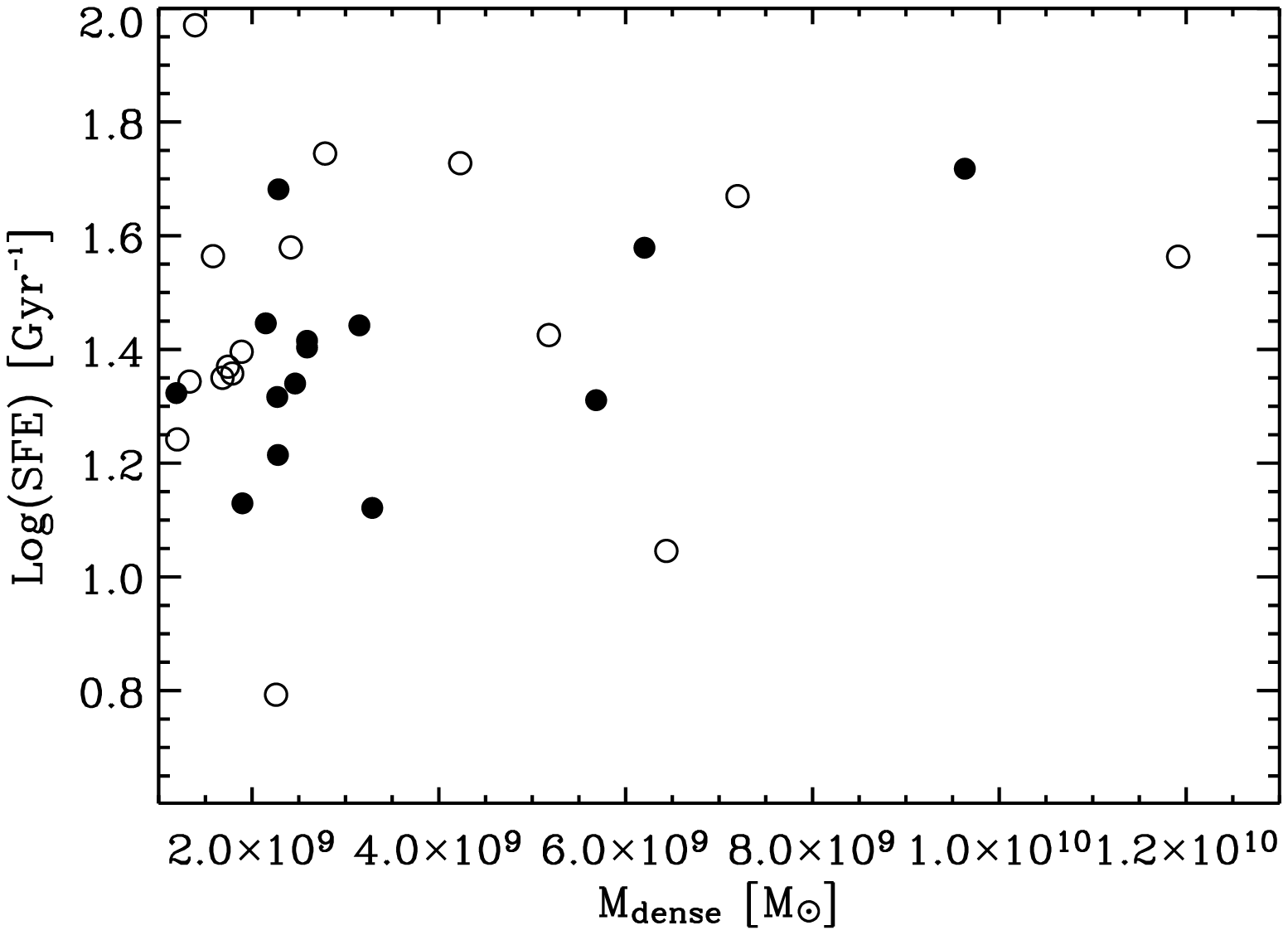}}
\centerline{\includegraphics[scale = 0.43]{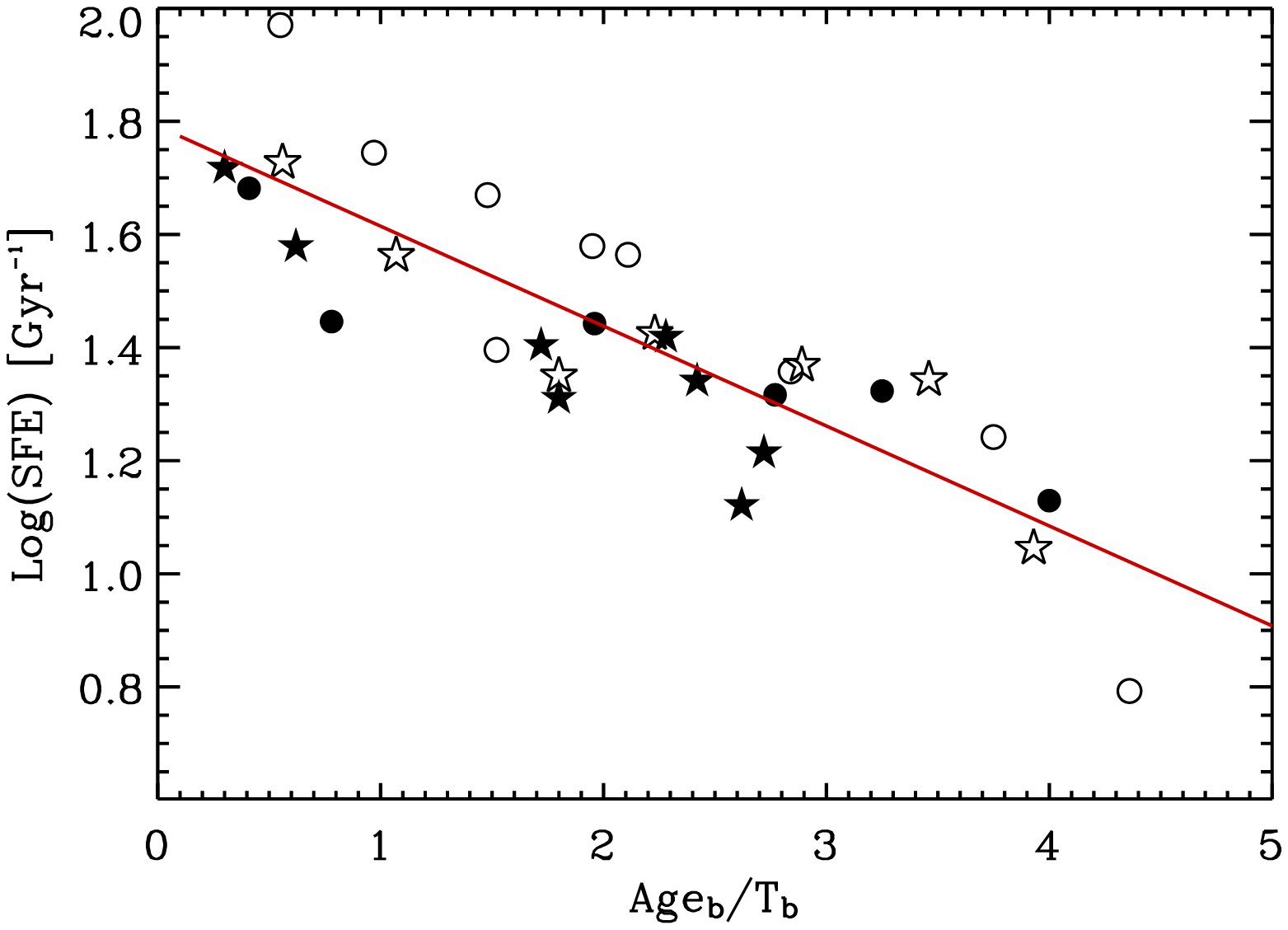}
\includegraphics[scale = 0.43]{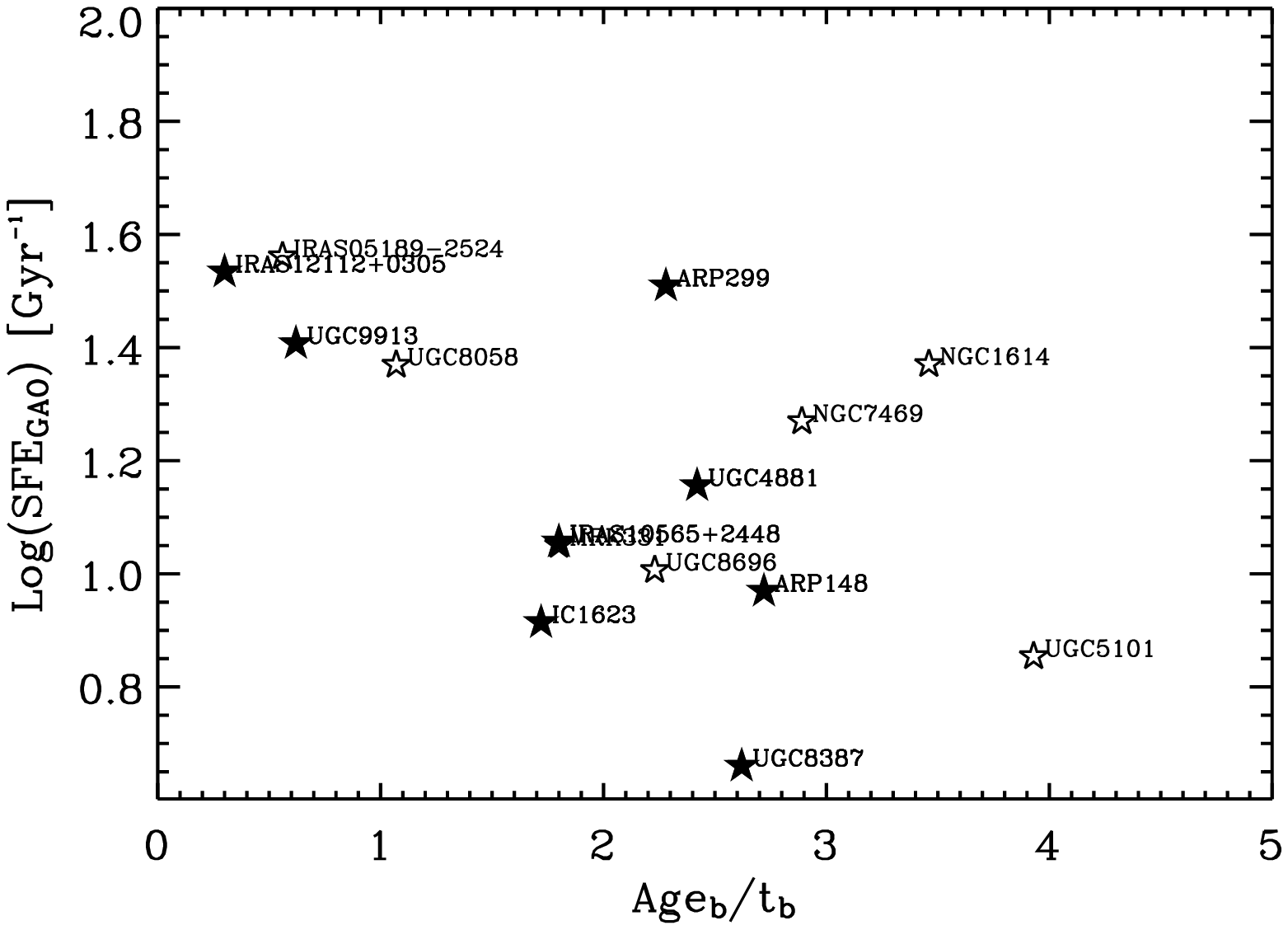}}
\caption{ Correlations of the SFE with different quantities:
\textit{Upper left:} Correlation of AGN corrected IR luminosity from
the best-fit models and the derived SFE for our sample of (U)LIRGs.
\textit{Upper right:} SFE vs. the derived dense gas masses of the
models. \textit{Lower left:} SFE as derived from our models versus
the evolutive phase of the starburst. Galaxies in common with the
Gao \& Solomon sample are marked as stars. Filled symbols correspond
to galaxies fitted with pure SB models while open symbols correspond
to those fitted with mixed (AGN + SB) models. \textit{Lower right:}
As adjacent panel but for the subset of galaxies in common with the
Gao \& Solomon sample and calculated using their derived molecular
gas mass and our estimate of the recent SFR. Filled stars correspond
to galaxies fitted with pure SB models while open stars correspond
to those fitted with mixed models.\label{Fig:sfe_fase}}
\end{figure*}

We finally notice that IRAS~08572+3915 falls significantly above the
correlation, suggesting that we could have overestimated the SB
contribution.
 However, a further decrease of the SB component is
incompatible with the radio SED, dominated by thermal emission, so
that a fraction of it should be attributed to the AGN.
Alternatively, the starburst in IRAS~08572+3915 could be
characterized by a metallicity significantly lower than the average
in the sample. In fact, while the typical metallicity of (U)LIRGs is
1.1 solar (Rupke, Veilleux \& Baker 2007) that of IRAS~08572+3915 is
only 0.8 solar (as obtained from the N2 calibrator, Denicol\'o,
Terlevich \& Terlevich 2002 and optical emission lines Veilleux et
al. 1995). In this case, the thermal radio emission is enhanced
(with respect to FIR emission) by about 10\%, but, and more
importantly, its dense gas mass will be 40\% higher than that
derived from the assumed dust/gas ratio. This would push the galaxy
on the border of the $L_{\rm{IR}}-M_{\rm{gas}}$ correlation.

\subsection{The star forming efficiency, SFE}

With the ratio between the current star formation rate and the mass
of dense gas, we obtain a quantity which measures the reciprocal of
the SF time scale and which is usually called star formation
efficiency (SFE, Gao et al. 2004a). This ratio is plotted in the
different panels of Fig. \ref{Fig:sfe_fase}. There is only a weak
correlation between the SFE and the IR luminosity of the SB or the
mass of dense gas, while an anti-correlation becomes fairly evident
(Kendall coefficient, $\tau_{\rm{K}}=-0.85$) when the SFE is plotted
against the evolutionary phase of the SB (bottom left panel). The
efficiency of the star formation is larger in the early-peak phases
of the starburst (E$_{\rm{SB}} \lesssim 1.$) and then tails off to
values characteristic of more normal galaxies (Gao \& Solomon 2004)
as  the starburst reaches the later phases.
\begin{equation}\label{equ:sfe}
\log(SFE)=-(0.18 \pm 0.02) \cdot E_{\rm{SB}}+(1.79 \pm 0.05)
\end{equation}
Stars in the bottom panels mark objects in common with Gao \&
Solomon (2004b). The right panel shows that this anti-correlation
exists even adopting the gas masses derived by them, although with a
larger scatter. This shows that the existence of this
anti-correlation is not a spurious result of the method we have used
to determine the gas mass. On the contrary, we are confident that
our method provides more reliable SFR and gas mass estimates,
because it  reduces the scatter in spite of increasing the number of
objects. We notice also that Gao \& Solomon (2004b) claim that the
star formation rate per unit mass of dense gas is constant for all
galaxies, including ULIRGs, and that the star formation law in terms
of dense molecular content  is well represented by a  power law with
exponent $\sim$1.0.

Such as rapid decrease of the star formation efficiency as the
starburst evolves, strongly suggests that feedback plays a major
role in the evolution of these extreme objects. To check whether
there are significant differences between pure and mixed ULIRGs, we
have plotted the former objects with filled symbols and the latter
objects with open symbols. In the lower left panel of Fig.
\ref{Fig:sfe_fase} there appears to be a separation between the two
classes of ULIRGs, with the efficiency in early mixed objects being
on the average $\sim$ 40\% larger than in early pure starbursts.
Given the small number of objects, this is far from being a definite
conclusion. However, if confirmed, it will shed important new  light
on the AGN-Starburst connection. At present,  we cannot say whether
a higher efficiency in mixed (U)LIRGs is  due to an enhancement of
the SFR due to positive AGN feedback, or  a lower mass of  dense
gas, due to negative AGN feedback.

While $E_{\rm{SB}}$ (=${\rm Age_{\rm{b}}}/t_{\rm{b}}$) is a quantity
entirely derived from our best fit models it would be desirable to
put the above finding on observational grounds, to check its
independence from the modelling process. As already mentioned, Vega
et al. (2005) found a tight correlation between $E_{\rm{SB}}$ and
the FIR colour $f_{60}/f_{100}$. This correlations goes in the sense
that as the SB evolves, i.e. $E_{\rm{SB}}$ increases, the infrared
colour gets cooler, i.e. $f_{60}/f_{100}$ decreases.

\begin{figure}
\includegraphics[scale = 0.5]{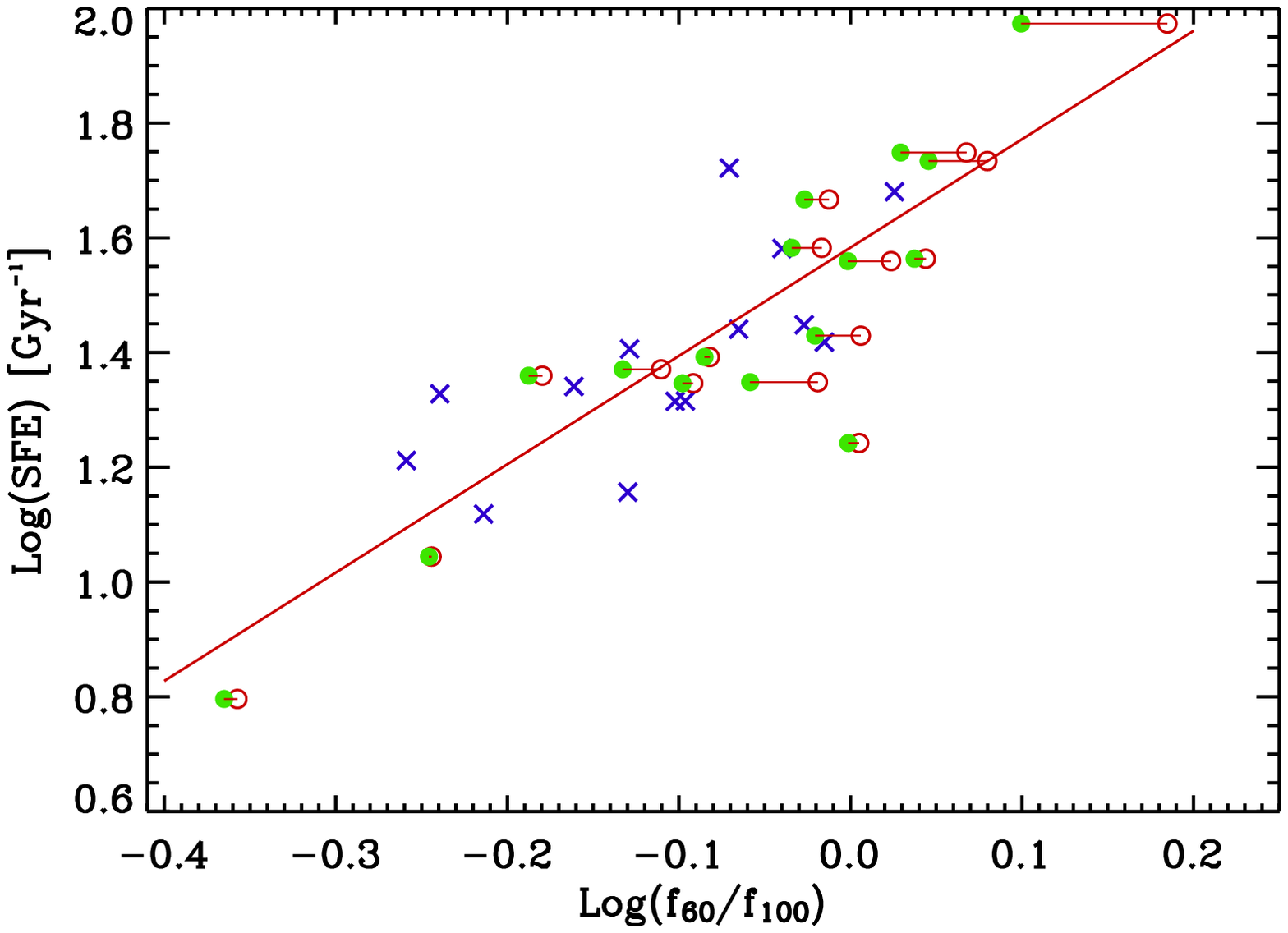}
 \caption{Correlation between the far-infrared color
 ($f_{60}/f_{100}$) and the derived SFE for our sample of (U)LIRGs.
 The meaning of the symbols are the same as in Fig. \ref{Fig:phase}.
 The thick solid line is the best fit to this correlation.
\label{Fig:sfe_col}}
\end{figure}

Indeed, by using the FIR colour instead of our indicator of the
evolutive phase of the starburst, we obtain the correlation shown in
Fig. \ref{Fig:sfe_col}. A Kendall test indicates a very strong
correlation between the SFE and the IR colours corrected for the AGN
contribution, $\tau_{\rm{K}}=0.78$. This correlation remains strong
($\tau_{\rm{K}}=0.77$) if the observed IR colors, i.e. without AGN
contribution correction, are used.
\begin{equation}
\log(SFE)= (1.74 \pm 0.20)\cdot \log(f_{60}/f_{100})+(1.55 \pm 0.03)
\end{equation}

Instead of SFE we could have used the ratio LIR/L$_{HCN}$ and
obtained a correlation analogous to that shown in Figure 9a of Gao
\& Solomon (2004). Notice, that considering only ULIRGs in this
Figure, a definite trend of decreasing efficiency at cooler
$f_{60}/f_{100}$ colour, with slope $\sim 1.72$, is seen. Therefore,
this strong correlation between the observed colour  and the
observed SFE indicates that this variation of the SFE is real. The
additional information on the evolutive phase of the starburst and
on the AGN contamination, allowed by a realistic modelling
procedure, eliminates a large part of the dispersion in the star
formation efficiency found by Gao \& Solomon (2004).

\begin{table*}
\centering \caption{Molecular data for the galaxies in common
between the Gao and Solomon (2004a) sample and our sample of
(U)LIRGs. Column (1) lists the name of the galaxy; (2) the
luminosity distance in Mpc; (3) the total L$_{\rm{IR}}$, in units of
$10^{11} L_\odot$, given by Gao \& Solomon (2004a) but corrected for
our different luminosity distance; (4) the L$_{\rm{IR}}$ due to the
starburst component from our fitted model; (5) the L$_{\rm{HCN}}$
from Gao \& Solomon, in units of $10^{8}$K km s$^{-1}$pc$^{2}$, and
corrected by using our calculated luminosity distance; and (6) the
$\alpha_{\rm{HCN}}$ needed to convert our molecular mass (Col. 4 in
table~\ref{tab:derived}) to the observed L$_{\rm{HCN}}$. }
\label{tab:gao1}
\begin{tabular}{lcccccccc} \hline
\hline
 NAME&$D_{\rm{L}}$&L$_{\rm{IR}}$ & L$_{\rm{IR}}^{\rm{SB}}$&L$_{\rm{HCN}}$&$\alpha_{\rm{HCN}}$ \\
 &(Mpc)&($10^{11}L_\odot$) & ($10^{11}L_\odot$)&($10^{8}$K km s$^{-1}$pc$^{2}$)&(M$_\odot$/K km s$^{-1}$pc$^{2}$) \\
 \hline
UGC~8058&173.987&31.678&22.243&19.414&6.190\\
IR~12113+0305&309.289&19.000&19.000&14.1&6.772\\
IR~05189-2524&175.661&12.565&5.943&6.596&6.320\\
UGC~9913&73.465&13.560&13.560&8.898&6.929\\
UGC~8696&155.376&13.537&10.204&13.57&3.780\\
IR~10565+2448&177.947&9.889&9.889&10.754&5.229\\
UGC~5101&162.193&9.143&7.705&10.250&6.298\\
Arp~299&41.500&5.849&5.849&1.956&13.140\\
UGC~4881&161.812&4.520&4.520&3.758&6.531\\
IC~1623&81.452&4.641&4.641&8.448&3.100\\
NGC~1614&64.489&4.019&3.507&1.301&10.133\\
UGC~8387&95.008&3.918&3.918&9.979&3.318\\
NGC~7469&66.000&3.802&3.548&2.094&8.298\\
Arp~148&141.647&3.566&3.566&3.908&5.802\\
Mrk~331&74.931&2.663&2.419&3.317&5.120\\
\hline
\end{tabular}
\end{table*}

\subsection{Implications for the evolution of LIRGs and ULIRGs}

 All but one of the 30 compact (U)LIRGs we have analyzed are
predominantly powered by a starburst, contributing more than 80\% of
the bolometric light. Sixteen galaxies show evidence for the
presence of an AGN, but only nine  have an AGN contribution that
exceeds 10\% of the IR luminosity, although it may be larger at
NIR-MIR wavelengths.

Since there is much evidence that IRAS bright galaxies are
associated with gas-rich mergers, this could suggest that strong AGN
activity is only triggered in certain types of mergers. Our sample
is biased toward IRAS bright objects, and the mergers involved
should be those in which disruption and/or infall of gas is
expected, such as discussed by Sopp, Alexander \& Riley (1990).
However, modelling of such encounters does not indicate that the
dynamical parameters (prograde, retrograde orbit etc) influence the
fueling of an active nucleus, at least for Seyfert galaxies (Keel,
1996).

The alternative explanation is that strong AGN activity is common to
most such mergers, but lasts for an order of magnitude less time
than the starburst. This would correspond to AGN lifetimes of a few
million years. The latter figure is in very good agreement with AGN
lifetimes derived from the comparison of the ratio between SNIIb/c
and SNII rates in normal spirals and in Seyfert II galaxies
(Bressan, Della Valle \& Marziani 2002).

\section{Conclusions}
\label{sec:conclusion}

Based on homogenized literature data and on our own radio
observations (Prouton et al. 2004, Clemens et al. 2007) we have
constructed well-sampled,  broad band, NIR to radio spectral energy
distributions of 30 luminous and ultra-luminous infrared galaxies,
selected from the sample of compact (U)LIRGs of Condon et al.
(1991). For one third of the sample, the broad band SED has also
been complemented by the published  low resolution \emph{Spitzer}
IRS spectrum. The library of SEDs, available in electronic form (see
Tables B.2 and B.3), constitutes a real work bench for models aimed
at interpreting such powerful and obscured sources. It is tempting
to say that the template SEDs may be used for the analysis of huge
star forming objects detected at high redshifts, if it were not that
one of the strongest warnings emerging from this work is that it is
dangerous to simply re-scale observed SEDs to highest luminosities.
On the contrary, the aim of the present investigation was to
determine physical parameters and environmental conditions that
accompany very high star formation rates, that can be used to model
high redshift galaxies.

To this end, we have analyzed the observed SEDs with a
chemo-spectro-photometric code that includes one of the most
advanced and realistic treatments of the complex interaction between
star formation, starlight and dust reprocessing (GRASIL). In
addition, emission from an AGN, based on the same homogeneous
physics, was added in those cases where the pure starburst model
failed to fit the data.

Our results can be summarized as follows:

\begin{itemize}
\item 16/30 sources show evidence for the presence of an AGN.
However only in 9/30 sources does
 the contribution of the AGN to the total infrared luminosity
  exceed 10\%.
  Only one galaxy, IRAS~08572+3915, seem to be  powered by an AGN.
 We therefore find that the fraction of starburst dominated objects (i.e.
 pure SB galaxies and mixed galaxies  with AGN contributions lower than $10\%$)
 in our sample, is 70\%. This value is in good agreement than
 that found by  Farrah et al. (2003).

\item In a previous paper we noticed that the results of fitting
(U)LIRGs SEDs could be affected by a degeneracy between optical
depth of the starburst MCs and the presence of the AGN, since both
may affect the MIR emission. By sampling the silicate absorption
feature at $9.7 \mu$m, low resolution \emph{Spitzer} IRS spectra can
actually break this degeneracy and provide a more accurate
quantification of the AGN/SB parameters.

The largest revision after including the \emph{Spitzer} IRS
spectrum, has been found for IRAS~15250+3609, whose broad band SED
could be fitted by a pure starburst, whereas reproducing also the
$9.7 \mu$m feature, revealed a dust obscured AGN contributing  20\%
of the IR luminosity. The classification of the other three pure
starbursts remained unchanged. In those cases where the presence of
the AGN were already suspected from other diagnostics, the inclusion
of \emph{Spitzer} IRS observations, confirmed this presence and
increased the robustness of the derived parameters.

\item Our measured AGN fractions are consistent with
those reported by Armus et al. (2007) based on
the mid-infrared line ratios  Ne[V]/Ne[II] and O[IV]/Ne[II].
These lines may thus provide a fairly accurate measure of the AGN
contribution to the bolometric luminosity of infrared luminous
galaxies. The PAH equivalent widths and mid-infrared spectral slope
on the other hand, overestimate (often by large factors) the AGN
fraction.

For two galaxies, our findings are markedly different from those of
Armus et al., namely IRAS~14348-1447 and IRAS~15250+3609. For the
former, we find an AGN contribution to the IR luminosity $\sim 10
\%$, while they  find no evidence of an AGN from the MIR analysis.
Indeed, this AGN contribution is required by its red NIR colour,
J-K$ = 1.71$ which can not be explained with stellar populations
alone. The power source of IRAS~15250+3609 could not be determined
by Armus et al., while our analysis indicates that it is a mixed
object with a very obscured AGN, that contributes $\sim 20 \%$ of
the IR luminosity.

\item If an active nucleus forms part of a an evolutionary phase of ULIRGs
then we expect this phase to be energetically important for an order
of magnitude less time than the starburst. From our modelling, this
would correspond to a few Myr. However, there is no evidence in our
sample that the AGN phenomenon is correlated with a particular phase
of the starburst, although all 4 AGNs in the early phase of the SB
make a  large IR contribution, while only 1 of the 12 associated
with the later SB phases, has similar IR power.

\item There is evidence that the star formation in objects with an
AGN decreases more rapidly than in pure starbursts. In fact, while
about 75\% of SBs with AGN  have t$_b\leq$20 Myr, all pure SBs  have
t$_b\geq$20 (and only 3 with t$_b$=20). A possible explanation is
that the surrounding star forming regions have suffered from  AGN
feedback, either positively as a sudden enhancement of the SFR, or
negatively by a more rapid decrease of the efficiency.

\item Since the MIR-FIR spectral region in our sample (U)LIRGs is dominated
by MC emission, we can determine quite accurately the mass in dust
associated with the dense star forming regions. We can convert this
to the molecular gas mass, using a typical dust/gas ratio. We find a
relation between the infrared luminosity and molecular gas mass with
the same slope as that  derived by Gao \& Solomon (2004a) from
observations of HCN emission. However, we derive a constant of
conversion between HCN flux and mass of dense molecular gas a factor
of 2 smaller than that  assumed by these authors. Such a low
conversion factor for the (U)LIRG population
 is suggested by Gao \&
Solomon (2004a) themselves and by other investigations (e.g.
Graci\'a-Carpio et al. 2006).

Notable outliers from this correlation are those few objects with an
energetically important AGN. After correcting their IR luminosity
for the AGN contribution, they also lie on the correlation. This is
an independent check of the robustness of our method. If confirmed
by the study of a larger sample of (U)LIRGs this correlation can be
used to evaluate the AGN fraction in distant ULIRGs with known IR
and HCN luminosities (e.g. such those observed by Gao et al. 2007).

\item Using the ratio between the current SFR and the mass
of dense molecular gas we obtain an estimate of the reciprocal of
the consumption time that we call {\sl efficiency} of star
formation. While there seems not to exist clear correlations between
SFE and IR luminosity of the starburst or mass of dense molecular
gas, an evident anti-correlation appears when SFE is plotted against
the starburst phase. The SFE decreases by about one order of
magnitude as the SB evolves from its early phase to the more
advanced phases.

This anti-correlation is not due to a circularity in our fitting
method, because it is also evident if we adopt the gas masses
obtained by Gao \& Solomon (2004a), although with a larger
dispersion. Indeed, most of the dispersion in the star formation
efficiency derived observationally from HCN emission (Gao \&
Solomon, 2004b), can be attributed to this evident fall in
efficiency during the lifetime of a starburst.

\end{itemize}

\vspace{2cm}

\noindent

\begin{acknowledgements}
O.V. and M.C. acknowledge the  the support of the I.N.A.F. research
fellowships. O.V.  also acknowledges the support of the INAOE and
the Mexican CONACYT projects 36547 E and 39714 F.  A.B., G.L.G. and
L.S. acknowledge the warm hospitality of INAOE and the partial
funding by the European Community by means of the Maria Curie
contract MRTN-CT- 2004-503929, MAGPOP.  We also thank the anonymous
referee for a careful and helpful report. This publication makes use
of data products from the Two Micron All Sky Survey, which is a
joint project of the University of Massachusetts and the Infrared
Processing and Analysis Center/California Institute of Technology,
funded by the National Aeronautics and Space Administration and the
National Science Foundation. This work is also  based on data taken
with the \emph{Spitzer} Space Telescope, which is operated by the
JPL,Caltech under a contract with NASA.
\end{acknowledgements}

\begin{appendix}

\section{Individual objects}
\label{api:individual}
 Below we  compare our results with the
relevant literature concerning the power sources for each of our
sample objects. The AGN contributions described refer to the
wavelength range $8 - 1000\;\rm \mu m$.

\smallskip
\noindent
{\bf UGC~08058 (Mrk~231)}\\
The identification of this source as the nearest broad absorption
line quasar is supported by a wide range of observational data.
Compact radio emission from its nucleus (Lonsdale et al. 2003), its
X-ray luminosity (Maloney \& Reynolds 2000), short term variability
at X-ray wavelengths (Gallagher et al. 2002) and low $3.3\;\rm \mu
m$ PAH feature equivalent width (Imanishi \& Dudley 2000) all point
to the presence of an AGN.

However, a central starburst is also probably present and accounts
for a significant fraction of the soft X-ray luminosity (Braito et
al. 2004). Davies, Tacconi \& Genzel (2004) estimate that
star-formation in the nuclear disk accounts for 25-40\% of the
bolometric luminosity. Taylor et al. (1999) use high- resolution
multi-wavelength radio observations to deduce that the central
starburst component forms stars at a rate of $220 \rm
M_{\odot}\,yr^{-1}$.

Condon, Frayer \& Broderick (1991) classify the source as
AGN-dominated based on the radio morphology and infrared-radio flux
ratio, as do Lutz et al. (1999) using mid-infrared spectroscopy.
Farrah et al. (2003) require a 30\% contribution from an AGN for
their SED fit. An AGN fraction of $<10\%$ is found by Armus et al.
(2007) using the line ratios  Ne[V]/Ne[II] and O[IV]/Ne[II], whereas
values of close to 100\% are obtained from the PAH equivalent widths
or the mid-infrared spectral slope.

Our best fit to the SED plus the IRS spectrum requires an active
nucleus accounting for 16\% of the infrared luminosity. Notice
however, that our fit is based on the assumption that the lower
radio fluxes are due to the SB.

\smallskip

\noindent
{\bf IRAS~14348-1447}\\
The source lies at a position in the mid-infrared diagnostic diagram
of Genzel et al. (1998) that suggests an AGN contribution $\lesssim
25\%$, and both Lutz et al. (1999) and Charmandaris et al. (2002)
classify this source as starburst dominated, based on mid-infrared
ISO data.

Risaliti et al. (2006) find a high equivalent width for the
$3.3\;\rm \mu m$ PAH feature and flat near-infrared continuum in
both nuclei, suggesting a starburst origin, although Imanishi,
Dudley \& Maloney (2006) find weak evidence for the presence of and
AGN in the NE nucleus using similar data. Franceschini et al. (2003)
find no evidence for the presence of an AGN at hard X-ray
wavelengths.

Takagi, Arimoto \& Hanami (2003), model the uv to far-infrared SED
with a starburst model with current star formation rate of $327\;\rm
M_{\odot}\,yr^{-1}$ but do not consider an AGN component.

Farrah et al. (2003) find that this source has an AGN accounting for
17\% of the IR luminosity. Armus et al. (2007) estimate the AGN
fraction using \emph{Spitzer} data to be between $<1\%$ ($\rm
Ne[V]/Ne[II]$ line ratio) and 65\% (mid-IR slope). Imanishi et al.
(2007) do not find AGN signatures in the \emph{Spitzer}  spectrum
but find strong evidence of the AGN in the 3-4 $\mu$m spectral
range.

Our models do not fit the SED unless an AGN component accounting for
$\sim 8\%$ of the IR luminosity is included.

\smallskip

\noindent
{\bf IRAS~12112+0305}\\
Colina et al. (2000) find that the star formation as traced by
H$\alpha$ emission is dominated by that in the two nuclei. Risaliti
et al. (2006) and Imanishi, Dudley \& Maloney (2006) find a high
equivalent width for the $3.3\;\rm \mu m$ PAH feature, no $3.4\;\rm
\mu m$ absorption and a flat near-infrared spectrum in both nuclei,
giving no evidence for the presence of an AGN. Hard X-ray data
support this conclusion (Franceschini et al. 2003). Veilleux,
Sanders \& Kim (1999) also find no spectroscopic evidence for a BLR
at near-infrared wavelengths. In the diagnostic diagram of Genzel et
al. (1998) based on the strength of the $7.7\;\rm \mu m$ PAH feature
and the ratio of high and low excitation mid-infrared emission lines
the object lies at a position implying an AGN contribution of $<
25\%$. Lutz et al. (1999) classify the IR spectrum as a starburst.

Armus et al. (2007) find that the AGN fraction estimated from the
line ratio $\rm Ne[V]/Ne[II]$ is $<1\%$ whereas if the SED is fitted
from $1-1000\;\rm \mu m$ the estimate is 65\%.

We find a good fit to the SED without the need for an AGN component,
consistent with the SED fits of Farrah et al. (2003).

Takagi, Arimoto \& Hanami (2003), model the uv-far-infrared SED as a
starburst and find a current star formation rate of $244\;\rm
M_{\odot}\,yr^{-1}$ which is about half that of our model, but their
Salpeter IMF is truncated at $60\;\rm M_{\odot}$ rather than
$100\;\rm M_{\odot}$.

\smallskip
\noindent
{\bf IRAS~05189-2524}\\
Imanishi \& Dudley (2000) find that the small equivalent width of
the $3.3\;\rm \mu m$ PAH feature shows that $> 90\%$ of the $3-4
\;\rm \mu m$ flux comes from an AGN. They conclude that the
bolometric luminosity output is dominated by an AGN power source.
The mid-infrared imaging of Soifer et al. (2000) leads them to a
similar conclusion. Dudley (1999), on the other hand, finds that the
$8 - 13\;\rm \mu m$ spectrum is indicative of a starburst by the
presence of the $11.3\;\rm \mu m$ PAH feature. He argues that the
discrepancy between the mid-IR classification and the optical
classification as an AGN can be reconciled if the AGN is responsible
for the near-IR emission but contributes little at mid- and
far-infrared wavelengths. The X-ray and near-infrared spectroscopy
of Severgnini et al. (2001) lead to the conclusion that the AGN
accounts for $\sim 10\%$ of the IR luminosity.

Armus et al. (2007) have used \emph{Spitzer} IRS spectra to classify
the source based on various mid-IR line ratios and the mid-IR
continuum slope, and find AGN fractions from 33 to 95\%.

We  fit the SED plus the IRS spectrum with the inclusion of an AGN
component that accounts for 17\% of the infrared luminosity. This is
lower than the AGN fraction of 1/3 found by Farrah et al. (2003).

\smallskip
\noindent
{\bf UGC~9913 (Arp~220, IRAS~15327+2340)}\\
Smith et al. (1998) resolved the compact radio emission into
separate supernova remnants and subsequent, high sensitivity VLBI
monitoring of both nuclei has now succeeded in obtaining a direct
estimate of the supernova rate (Lonsdale et al. 2006) of $4 \pm
2\;\rm yr^{-1}$. The implied star formation rate is sufficient to
supply the bolometric luminosity of the system.

Shier, Rieke \& Rieke (1996) concluded that a large fraction of the
luminosity must come from an AGN unless the IMF for recent star
formation is very biased towards high mass stars. They note,
however, that very high extinction may be confusing such
conclusions.

Condon, Frayer \& Broderick (1991) classify the source as
starburst-dominated from the radio morphology, infrared spectral
index and infrared-radio flux ratio, and Soifer et al. (2002)
classify both the eastern and western nuclei as starburst dominated
based on the strength of the $11.3\;\rm \mu m$ PAH feature. The
position in the mid-infrared diagnostic diagram of Genzel et al.
(1998) implies at most a negligible contribution from an AGN. Lutz
et al. (1999) also classify the source as a starburst using
mid-infrared spectroscopy.

Risaliti et al. (2000) find no X-ray evidence for the presence of an
AGN.

Takagi, Arimoto \& Hanami (2003), model the uv to far-infrared SED
with a starburst model with current star formation rate of $260\;\rm
M_{\odot}\,yr^{-1}$, and Farrah et al. (2003) find that less than
1\% of the IR-luminosity of this source comes from an AGN.

Armus et al.(2007) use \emph{Spitzer} IRS data to estimate AGN
fractions from $2-90\%$. Imanishi et al. (2007), by using different
MIR diagnostics, conclude that an AGN is possibly powering  the MIR
spectrum of this galaxy. However, we find a good fit to the SED plus
\emph{IRS} spectrum without the inclusion of an AGN.

\smallskip
\noindent
{\bf IRAS~08572+3915}\\
This source consists of two nuclei separated by 5 \arcsec. It is the
north-west component which dominates in both the infrared and radio
(Sanders et al. 1988; Condon et al. 1991). Surace \& Sanders (2000)
found evidence for stars no older than $100$ Myr and despite the
warm IRAS colour, Veilleux et al. (1999) found no clear evidence for
the presence of an AGN from NIR spectroscopy. Farrah et al. (2003)
find an AGN contribution of  34\% to the IR luminosity. The
inclusion of radio data allows us to provide a  more accurate
quantification of the starburst parameters. Our derived age is $\sim
4$ Myr versus Farrah et al.'s upper limit of 57 Myr. Imanishi et al.
(2007) find strong evidence of AGN in the high resolution
\emph{Spitzer} spectrum, and conclude that this galaxy is AGN
dominated. Prouton et al. (2004) fail to obtain a fit to the NIR to
radio SED with a starburst model alone. They need a contribution of
between 37 to 47 \% (for the cases of low and high dust optical
depth respectively, i.e. $\tau_1= 20$ and $\tau_1= 45$) in order to
fit the SED. The inclusion of  the \emph{Spitzer} spectrum in the
fit, allows us to provide a more precise determination of the MC
optical depth, and therefore, of the AGN contribution. Our SED plus
IRS spectrum fit required the highest AGN contribution of our sample
of 47\%, with a $\tau_1 =65$. Notice however, that our fit to the
MIR spectral range is the poorest of our sample of (U)LIRGs, and
seems to require a higher AGN contribution (see the discussion in
Sect. \ref{sec:radiofits}).

\smallskip

\noindent
{\bf UGC~8696 (Mrk~273)}\\
The mid-infrared spectrum of Dudley (1999) suggests the presence of
an AGN that dominates the far-infrared luminosity, and Lutz et al.
(1999) also classify this source as AGN-dominated based on ISO data.
However, the sub-arcsecond $2.2\;\rm \mu m$ imaging of Eales et al.
(1990) shows that the central source is resolved. Imanishi \& Dudley
(2000) find that $\sim 50\%$ of the $3-4\;\rm \mu m$ flux originates
from an AGN. Indeed, in the mid-infrared diagnostic diagram of
Genzel et al. (1998) this source lies at a position implying an AGN
contribution of $\sim 50\%$. Soifer et al. (2002), in fact,  find
that the the NE nucleus is starburst dominated whereas the SW
nucleus ($1^{\prime\prime}$ away) is AGN dominated using mid-IR
spectroscopy.

Condon, Frayer \& Broderick (1991) classify the source as
starburst-dominated from the radio morphology and infrared-radio
flux ratio. Risaliti et al. (2000) find that there is no evidence
for the presence of an AGN in hard X-rays despite it's optical
classification as a Seyfert 2.

Farrah et al. (2003) find  a  11\% AGN contribution to the
IR-luminosity whereas Armus et al. (2007) find contributions of
between 17 and 33\% based on the Ne[V]/Ne[II]
 and O[IV]/Ne[II] line ratios respectively (from the PAH equivalent width or mid-IR
spectral slope they find $\sim 70\%$). Our fit to the SED plus the
IRS spectrum includes an AGN component that contributes 11\% of the
far-infrared luminosity.

Takagi, Arimoto \& Hanami (2003), model the uv to far-infrared SED
with a starburst model with current star formation rate of $207\;\rm
M_{\odot}\,yr^{-1}$, higher than our value of $138\;\rm
M_{\odot}\,yr^{-1}$.

\smallskip

\noindent
{\bf IRAS~15250+3609}\\
Lutz et al. (1999) use mid-infrared spectroscopy to classify the
source as starburst powered and Takagi, Arimoto \& Hanami (2003),
model the uv to far-infrared SED with a starburst model with current
star formation rate of $185\;\rm M_{\odot}\,yr^{-1}$. Imanishi et
al. (2006) do not find any evidence of an AGN in the $3-4\;\rm \mu
m$ range despite strong absorption features seen at mid-infrared
wavelengths (Spoon et al. 2002). Klaas et al. (2001) classify the
infrared to millimetre spectrum as the result of starburst activity.
Armus et al. (2007) were not able to clarify the power mechanism of
this galaxy from their analysis of the IRS spectrum.

We fit the SED plus the IRS spectrum, with  an  AGN component of
about 20\%. This is \emph{inconsistent} with Farrah et al. (2003)
who find that the source requires an AGN supplying 60\% of the IR
luminosity.

\smallskip

\noindent
{\bf IRAS~10565+2448}\\
Dudley (1999) detects strong PAH features in the mid-infrared
spectrum which is consistent with the optical classification of the
spectrum as HII and Goldader et al. (1995) detect a strong
Br$\gamma$ line. Downes \& Solomon (1998) find a compact nuclear
ring of molecular gas containing an estimated $4\times 10^9\;\rm
M_{odot}$ of molecular ISM.

Farrah et al. (2003) require an AGN component accounting for 5\% of
the infrared luminosity for an acceptable SED fit. We find a good
fit to the SED and the IRS spectrum without any AGN component at
all.

\smallskip

\noindent
{\bf UGC~5101}\\
A $12.3\; \rm mJy$ radio source lies $1^{\prime}$ to the north-east.
IRAS fluxes reduced by 8\%.

Much observational evidence exists for the existence of an AGN in
this source. Imanishi \& Maloney (2003) and Ptak et al. (2003)
detect the 6.4 keV FeK line, Armus et al. (2004) detect the
[NeV]14.3 $\mu$m line suggesting the presence of a buried AGN and
Lonsdale et al. (2003) find a very compact radio source they
interpret as having an AGN origin. Imanishi \& Maloney (2003) and
Imanishi, Dudley \& Maloney (2001) show that the absorption features
at at $3.1$ and $3.4\;\rm \mu m$ are inconsistent with well-mixed
geometry for luminosity sources and absorbers such as that expected
in a starburst, and favour an energetically dominant AGN luminosity
source.

Condon, Frayer \& Broderick (1991) classify the source as
starburst-dominated from the radio morphology, infrared spectral
index and infrared-radio flux ratio. In the mid-infrared diagnostic
diagram of Genzel et al. (1998) this source lies at a position
implying an AGN contribution of $< 25\%$.

Armus et al. (2007) find that the AGN fraction as estimated from the
line ratios $\rm Ne[V]/Ne[II]$ and $\rm O[IV]/Ne[II]$ is $\sim 7\%$
whereas if the equivalent width of PAH features or the mid-IR slope
are used estimates of $52 - 80\%$ result.

Farrah et al. find a 7\% AGN contribution is necessary for a good
fit to the optical/far-infrared SED. Our models show that an AGN
contribution of only 1 \% produces the best fit to the SED.

\smallskip

\noindent
{\bf IZw~107 (Mrk~848)}\\
Dudley (1999) classify the source as starburst dominated based on
the large equivalent width of the $11.3\;\rm \mu m$ PAH feature.

Although we fit this source with a pure starburst model, the fit at
 radio frequencies is very poor. The data suggest a very steep
radio spectral index at 8~GHz of -1.4 that is steeper than that of
our assumed synchrotron component of -0.8.
\smallskip

\noindent
{\bf IRAS~01364-1042}\\
The low value of the 25 $\mu$m IRAS flux required the highest value
of the dust optical depth of our sample ($\tau_1 \simeq 116$). The
best fit to the SED also required an AGN contribution of 11\%.
Notice, however, that the SED is poorly sampled, with only 13
data-points, while the number of free parameters is 14. The Prouton
et al. (2004) analysis required an even higher optical depth of 181,
but a lower AGN contribution of 4\%.

\smallskip

\noindent
{\bf IRAS~10173+0828}\\
Baan, Salzer \& LeWinter (1998) fail to classify this mega-maser
galaxy due to the non-detection of any strong optical emission
lines. The sub-arcsecond 15 and 22~GHz observations of Smith,
Lonsdale \& Lonsdale (1998) show an unresolved nuclear source. We
fit the SED without the need for an AGN component.

\smallskip

\noindent
{\bf Arp~299}\\
Actually a close interacting pair consisting of NGC~3690 and IC~694.
The $38\;\rm \mu m$ imaging of Charmandaris, Stacey \& Gull (2002)
shows that IC~694 emits twice as much energy as NGC~3690. However,
at $15\;\rm \mu m$ both galaxies emit similar fluxes (Gallais et al.
2004).

Soifer et al. (2001) show that the $12\;\rm \mu m$ flux measured by
IRAS originates entirely from 5 compact emission regions. These
regions are also detected at radio wavelengths (Neff et al. 2004).
Although Risaliti et al. (2000) find no evidence for the presence of
an AGN at hard X-ray wavelengths the later detection of the FeV line
in IC~694 and of Fe-K$\alpha$ in NGC~3690 (Ballo et al. 2004) and of
hard X-ray continuum emission from both (Zezas et al. 2003, Della
Ceca et al. 2002) gives evidence for the presence of AGN. However,
Neff et al. (2004) point out that the size and radio flux of source
A (nucleus of IC~694) are similar to that seen in Arp~220, which has
been resolved into individual sources. It seems likely that there is
indeed an AGN in each nucleus but that only one (source B1, nucleus
of NGC~3690) may be energetically important. The mid-infrared
results of Gallais et al. (2004) show strong PAH lines in all
positions and they derive visual extinctions of 40 mag towards
source A and 60 mag towards source B1. They conclude that only
source B1 is likely to be an AGN based on the strong hot dust
continuum at this location.

At near-infrared wavelengths there is no evidence that \emph{any} of
the compact sources are an AGN (Alonso-Herrero et al. 2000; Satyapal
et al. 1999).

This system illustrates particularly well how an infrared luminous
object, known to contain an AGN, nonetheless has its far-infrared
emission dominated by the energetic output of a starburst. In this
merger, the nucleus which harbours the AGN becomes progressively
less luminous towards far-infrared wavelengths, where most energy is
emitted. The starburst nucleus dominates at far-infrared
wavelengths.

We find a good fit to the SED of the pair with a pure SB model.

\smallskip

\noindent
{\bf UGC~4881 (Arp~055)}\\
Smith, Lonsdale \& Lonsdale (1998) find that the radio emission can
be explained with a starburst model. Our model fit to the SED needs
no AGN component.

\smallskip

\noindent
{\bf CGCG~436-30}\\
This galaxy is almost certainly interacting and is detected in HI in
both emission and absorption; the deduced atomic gas mass is greater
than $3.4 \times 10^9$ M$_\odot$ and must be concentrated near the
core in order to explain the absorption feature (Mirabel \& Sanders
1988). There is evidence of a compact radio core that may be
difficult to explain as an ensemble of young supernovae associated
with the starburst; the major contribution to the FIR and radio
emission originates on larger spatial scales (Lonsdale et al. 1993;
Smith et al. 1998).

Our SED plus IRS spectrum fit confirms the value for the MC optical
depth of $\tau_1=45$ found by Prouton et al. (2004), and requires an
AGN contribution of 5\%.

\smallskip

\noindent
{\bf IC~1623 (VV~114, Arp~236)}\\
The mid-infrared emission is dominated by the eastern member of this
interacting pair and Soifer et al. (2002) classify this source as a
starburst, based on the strength of the $11.3\;\rm \mu m$ PAH
feature. Laurent et al. (2000) however, also using mid-infrared
data, classify the nucleus as an AGN.  Le Floc'h et al. (2002) note
that only 40\% of the mid-infrared emission originates from the
compact nucleus of the eastern component with the remainder of the
emission being extended. The fact that both the radio emission
(Condon et al. 1991) and the molecular gas emission are also
extended seems to indicate that the source is starburst dominated.
Frayer et al. (1999) find the cool dust emission follows the
molecular gas distribution and argue that it will evolve into a
compact starburst. We fit the SED of this source with a pure SB
model.

\smallskip

\noindent
{\bf NGC~1614}\\
Soifer et al. (2001) find that 87\% of the $12\;\rm \mu m$ flux
originates within the $4^{\prime\prime}$ (1.2~kpc) diameter nuclear
region. Neff et al. (1990) find no evidence for an AGN from the
optical to the radio but argue that it will develop an AGN in the
future. Risaliti et al. (2000) find evidence for the presence of an
AGN from hard X-ray data.
 Alonso-Herrero et al. (2001) explain the near-infrared line fluxes using a starburst model with 2
 short duration bursts separated by 5~Myr. Shier, Rieke \& Rieke (1996) fit near-infrared data with
 a starburst model with an age of 12~Myr. Young ages are consistent with the detection of
Wolf-Rayet features in the spectrum (Armus et al. 1989, Vacca \&
Conti 1992).

Our fit to the SED includes an $18\%$ AGN contribution. The
modelled, current star formation rate of $30\;\rm M_{\odot}\,
yr^{-1}$ is similar to the value of $27\;\rm M_{\odot}\, yr^{-1}$
found by Kotilainen et al. (2001)  by
 using $\rm Br\gamma$ observations and
near-infrared photometry, if a continuous star formation is assumed.

\smallskip

\noindent
{\bf UGC~8387 (Arp~193, IC~883)}\\
Downes \& Solomon (1998) find a compact nuclear ring or disk of
molecular gas  and identify the  CO peak as an `extreme starburst
region'. Clemens \& Alexander (2004) find that free-free absorption
flattens the radio spectral index towards the centre of the
molecular gas distribution showing there to be a large region of
dense ionized gas. The strong Br$\gamma$ line and CO index (Smith et
al. 1995), strong $3.3\;\rm \mu m$ PAH emission (Dudley, 1999) and
extended $12\;\rm \mu m$ emission (Soifer et al. 2001) are further
evidence that intense star formation powers the source.

Condon, Frayer \& Broderick (1991) classify the source as
starburst-dominated from the radio morphology, infrared spectral
index and infrared-radio flux ratio.

We find a good fit to SED and the IRS spectrum with a pure starburst
model at all frequencies with the exception of the high frequency
radio data.
\smallskip

\noindent
{\bf NGC~7469}\\
This source is thought to have a Sy 1.2 nucleus and a circumnuclear
starburst. Using high resolution MIR observations, Soifer et al.
(2003) argued that the central source is an AGN rather than a
nuclear starburst because of its very high 12 $\mu$m surface
brightness. Miles et al. (1994) and Nikolic et al. (2003) both found
that PAH features in the MIR spectrum, thought to trace star
formation, are found in the ring of emission surrounding the nucleus
but not in the nucleus itself. If the hard ionizing photons from the
AGN destroy PAH molecules this supports the picture of a central AGN
surrounded by a starburst ring.

The best fit  model derived by Prouton et al. (2004) required an AGN
contribution $< 17\%$ and a MC optical depth between 20 to 45. Our
best fit, to a better sampled SED, requires an AGN contribution of
$10\%$ and a MC optical depth of 33, in agreement with the Prouton
et al. results.

\smallskip

\noindent
{\bf UGC~2369}\\
Smith, Lonsdale \& Lonsdale (1998) argue for the presence of an AGN
based on 18 cm VLBI radio continuum observations. We model the SED
of the source without the need for an AGN component.
\smallskip

\noindent
{\bf IIIZw~35}\\
Pihlstr{\"o}m et al. (2001) map both the 18 cm radio continuum and
OH maser emission at VLBI resolution and find that the radio
continuum can be explained as a combination of supernova remnants
and very luminous radio supernovae. Earlier radio work at $\sim
1^{\prime\prime}$ resolution concluded that he source was instead an
AGN (Chapman et al. 1990). A compact (100 pc) molecular disk in the
nucleus is responsible for the dynamical properties of the maser
emission (Montgomery \& Cohen, 1992). The optical spectrum has been
classified as liner by Baan et al. (1998). We model the source
without the need for an AGN component.

\smallskip

\noindent
{\bf IC~5298}\\
Wu et al. (1998) found that IC~5298 has properties that are
intermediate between HII regions and liners; it was classified as a
Seyfert 2 by Veilleux et al. (1995) although the $[O I]$ lines are
rather weak. Poggianti \& Wu (2000) classify this galaxy as e(a)
type, namely with H$\alpha$ in emission with moderate equivalent
with ($\simeq 52$ ${\AA}$) and H$\gamma$ in absorption ($\simeq 4$
${\AA}$). The best fit to the SED obtained by Prouton et al. (2004)
required and AGN contribution of 36 \% and a very high optical depth
for the molecular clouds, $\tau_1 = 181$. However, a reasonable good
fit was also obtained without the inclusion of an AGN and with a
much lower optical depth ($\tau_1 =15$).

Our best fit model to a better sampled SED is with a pure SB model
and with an optical depth of 33.
\smallskip

\noindent
{\bf Arp~148 (A~1101+41)}\\
Mouri et al. (1998) demonstrate the presence of a dust enshrouded
starburst from the detection of strong PAH features in the $6 - 12
\;\rm \mu m$ ISO spectrum.

We find a good fit to the infrared-radio SED with a pure starburst
model.
\smallskip

\noindent
{\bf NGC~2623}\\
Soifer et al. (2001) find 80\% of the $12\;\rm \mu m$ flux
originates within the $4^{\prime\prime}$ (1.5 kpc) diameter nuclear
region. The sub-arcsecond $2.2\;\rm \mu m$ imaging of Eales et al.
(1990) shows that the central source is extended but nonetheless
contains an unresolved core less than 0.3 kpc in diameter. Risaliti
et al. (2000) find no X-ray evidence for the presence of an AGN
whereas later work by Maiolino et al. (2003) argue for the presence
of a heavily obscured AGN based on Chandra X-ray data. Shier, Rieke
\& Rieke (1996) argue for the presence of both a starburst and an
AGN as long as a large fraction of the $\rm Br\gamma$ flux can come
from the AGN. Condon, Frayer \& Broderick (1991) assign a starburst
classification based on the radio morphology, infrared spectral
index and infrared-radio flux ratio.

We fit the SED plus the IRS spectrum with an AGN component
accounting for just $2\%$ of the infrared luminosity.

\smallskip

\noindent
{\bf Mrk~331}\\
Observational data show evidence of both starburst (e.g., Lancon et
al. 1996; Veilleux et al. 1995; Roche et al. 1991) and AGN (Lonsdale
et al. 1993) activity.

Prouton et al. (2004) found an upper limit to the contribution of
the AGN of about 23 \%. Their average SFR was between 72 and 84
$M_\odot/yr$ for the case with and without an AGN respectively.

Our best fit to a better sampled SED requires an AGN contribution of
5\% and a MC optical depth of 54.
\smallskip

\noindent
{\bf NGC~0034 (Mrk~938)}\\
The presence of two nuclei separated by approximately 1.2\arcsec in
the mid-infrared (Miles et al. 1996) and optical tidal tails
indicate this galaxy is undergoing a merger (e.g. Mulchaey et al.
1996). The nature of the activity has been controversial, some
authors (e.g., V\'eron-Cetty \& V\'eron 1991) claiming a Seyfert 2
nucleus, while others suggest a composite spectrum. Mulchaey et al.
(1996) emission-line images show the galaxy to be a weak emitter of
[OIII]$\lambda5007$ when compared with Seyfert galaxies in their
sample. In addition they argue its strong H$\alpha$ emission,
distributed over the entire galaxy, as indicative of a starburst.

Prouton et al. (2004) find that a fit with no AGN component is
marginally consistent with the radio data, but also that a large
(40\%) AGN contribution results in a good fit if a model with larger
MC optical depth is used (i.e. $\tau_1 =181$).

Our best fit to the SED required an AGN component of 4\% and a MC
optical depth of $\tau_1 = 42$.

\smallskip
\noindent

{\bf NGC~5256 (Mrk~266)}\\
Both Wang et al. (1997) and Condon, Frayer \& Broderick (1991)
classify the source as having both starburst and AGN components.
Mizutani, Hiroshi \& Toshinori (1994) detect the $3.3\;\rm \mu m$
PAH feature  despite the source's optical classification as a Sy2.

Our model fit to the SED is excellent from the near-infrared to the
radio with an AGN component contributing about 5\% to the IR
luminosity.

\smallskip

\noindent
{\bf UGC~6436 (IC~2810)}\\
The NVSS shows that a companion galaxy, IC~2810b, located $1\,
\farcm 2$ to the south-east has a flux of $7.7\; \rm mJy$ compared
to the $19.4\; \rm mJy$ of UGC~6436. This would be confused at IRAS
resolutions and so the IRAS fluxes have been reduced by 26\%. No
other observations were at so low a resolution as to make
contamination from this source a problem. Hattori et al. (2004) find
both extended and nuclear H$\rm \alpha$ emission.

Our purely starburst fit to the SED is the oldest of our whole
sample with age$_{\rm b} \sim 10^{8}\;\rm yr$.

\smallskip

\noindent
{\bf NGC~6286}\\
Condon, Frayer \& Broderick (1991) assign a starburst classification
based on the radio morphology and infrared-radio flux ratio.

An AGN component accounting for $5\%$ of the infrared luminosity
provides the best fit to the SED.

\section{Tables provided in electronic form}

\begin{table*}
\centering
 \caption{Observed (U)LIRG SEDs.
 This table is provided in electronic form. }
  \label{tab:seddata}
\begin{tabular}{lccc} \hline

NAME&$\lambda_{\rm{obs}}$ & Flux & Error \\
&($\mu$m)&(Jy)&(Jy)\\
 \hline
UGC08058& 1.23& 6.20E-02&1.15E-03\\
UGC08058& 1.66& 1.13E-01&2.00E-03\\
UGC08058& 2.16& 1.99E-01&3.14E-03\\
$\dots$\\
IRAS14348-1447&1.23&1.70E-03&1.00E-04\\
IRAS14348-1447&1.66&2.50E-03&3.00E-004\\
IRAS14348-1447&2.16&3.30E-003&3.00E-004\\
$\dots$\\
 \hline
\end{tabular}
\end{table*}

\begin{table*}
\centering
 \caption{Best fit models for (U)LIRG SEDs. The models include also
 the possible AGN contribution. Notice that $\lambda$ is in the rest
 frame and should be converted to observed wavelength by using
 the radial velocity provided in the second row.
 Fluxes are in the observed rest frame.  This table is provided in electronic form.}
  \label{tab:seddata}
\begin{tabular}{l|cccccccc} \hline
 NAME &UGC8058 &IR14348-1447 &  IR12112+0305  & IR05189-2524 & UGC9913   &  IR08572+3915 &   UGC8696&$\dots$  \\
 V$_{\rm{r}}$ (Km/s)& 12642.0 &24802.0&21980.0&12760.0&5400.00&17493.0& 11326.0&$\dots$ \\
 $\lambda_{\rm{rest}}$ (${\AA}$)&   Flux(Jy)& Flux(Jy)& Flux(Jy)& Flux(Jy)&  Flux(Jy)& Flux(Jy)& Flux(Jy)&$\dots$ \\
 \hline
 1.6870E+002& 3.6604E-006 &  1.2163E-007 &  1.7817E-008 &  1.0241E-004 &  3.5968E-007 &  2.5111E-007  & 5.6713E-007&$\dots$  \\
 2.0200E+002& 4.0096E-006   &1.3195E-007  & 2.1113E-008  & 1.4306E-004  & 4.2663E-007 &  2.7477E-007 &  6.5618E-007 &$\dots$ \\
 2.4200E+002 &4.2074E-006 &  1.3802E-007  & 2.6699E-008 &  1.9900E-004 &  5.4172E-007  & 2.8617E-007 &  7.4418E-007&$\dots$  \\
$\dots$&$\dots$ &$\dots$ &$\dots$ &$\dots$ &$\dots$ &$\dots$ &$\dots$ &$\dots$  \\
1.0400E+005&1.1139E+000&3.2844E-002&3.4478E-002&4.2329E-001&3.3697E-002&7.5539E-002&6.6111E-002 &$\dots$ \\
1.0500E+005&1.1514E+000&3.5322E-002&3.8008E-002&4.4170E-001&3.5531E-002 &8.1599E-002&7.2495E-002&$\dots$ \\
1.0600E+005&1.1899E+000&3.7927E-002&4.1767E-002&4.6065E-001&3.8162E-002&8.8146E-002&7.9397E-002&$\dots$ \\
$\dots$&$\dots$ &$\dots$ &$\dots$ &$\dots$ &$\dots$ &$\dots$ &$\dots$ &$\dots$  \\
 \hline
\end{tabular}
\end{table*}

\begin{table*}
\centering
 \caption{As in table \ref{tab:seddata}, but including only the starburst component
 of the best fit model.  This table is provided in electronic form.}
  \label{tab:seddata}
\begin{tabular}{l|cccccccc} \hline
 NAME &UGC8058 &IR14348-1447 &  IR12112+0305  & IR05189-2524 & UGC9913   &  IR08572+3915 &   UGC8696&$\dots$  \\
 V$_{\rm{r}}$ (Km/s)& 12642.0 &24802.0&21980.0&12760.0&5400.00&17493.0& 11326.0&$\dots$ \\
 $\lambda_{\rm{rest}}$ (${\AA}$)&   Flux(Jy)& Flux(Jy)& Flux(Jy)& Flux(Jy)&  Flux(Jy)& Flux(Jy)& Flux(Jy)&$\dots$ \\
 \hline
 1.6870E+002& 3.6604E-006 &  1.2163E-007 &  1.7817E-008 &  1.0241E-004 &  3.5968E-007 &  2.5111E-007  & 5.6713E-007&$\dots$  \\
 2.0200E+002& 4.0096E-006   &1.3195E-007  & 2.1113E-008  & 1.4306E-004  & 4.2663E-007 &  2.7477E-007 &  6.5618E-007 &$\dots$ \\
 2.4200E+002 &4.2074E-006 &  1.3802E-007  & 2.6699E-008 &  1.9900E-004 &  5.4172E-007  & 2.8617E-007 &  7.4418E-007&$\dots$  \\
$\dots$&$\dots$ &$\dots$ &$\dots$ &$\dots$ &$\dots$ &$\dots$ &$\dots$ &$\dots$  \\
1.0400E+005&4.8245E-001&1.6684E-002&3.4478E-002&2.8236E-001&3.3697E-002&8.8408E-004&3.4949E-002&$\dots$  \\
1.0500E+005&5.1720E-001&1.8587E-002&3.8008E-002&2.9908E-001&3.5531E-002&9.6866E-004&3.8960E-002&$\dots$ \\
1.0600E+005&5.5300E-001&2.0601E-002&4.1767E-002&3.1635E-001&3.8162E-002&1.0779E-003&4.3316E-002&$\dots$ \\
$\dots$&$\dots$ &$\dots$ &$\dots$ &$\dots$ &$\dots$ &$\dots$ &$\dots$ &$\dots$  \\
 \hline
\end{tabular}
\end{table*}

\end{appendix}
\end{document}